\documentclass[11pt,a4paper]{article}
\pdfoutput=1
\usepackage{jheppub}
 \usepackage{latexsym,bm,amsmath,amssymb,amsfonts}
\usepackage{slashed}

\long\def\comment#1{ }

\newcommand{\eqn}[1]{Eq.~\eqref{#1}}

\newcommand{\mcal}{\mathcal}

\newcommand{\slv}{\raise.15ex\hbox{$/$}\kern-.53em\hbox{$v$}}
\newcommand{\slF}{\raise.15ex\hbox{$/$}\kern-.53em\hbox{$F$}}
\newcommand{\slL}{\raise.15ex\hbox{$/$}\kern-.53em\hbox{$L$}}
\newcommand{\slP}{\raise.15ex\hbox{$/$}\kern-.53em\hbox{$P$}}
\newcommand{\slp}{\raise.15ex\hbox{$/$}\kern-.53em\hbox{$p$}}
\newcommand{\slq}{\raise.15ex\hbox{$/$}\kern-.53em\hbox{$q$}}
\newcommand{\slR}{\raise.15ex\hbox{$/$}\kern-.53em\hbox{$R$}}
\newcommand{\slQ}{\raise.15ex\hbox{$/$}\kern-.53em\hbox{$Q$}}
\newcommand{\slK}{\raise.15ex\hbox{$/$}\kern-.53em\hbox{$K$}}
\newcommand{\slk}{\raise.15ex\hbox{$/$}\kern-.53em\hbox{$k$}}
\newcommand{\slD}{\raise.15ex\hbox{$/$}\kern-.53em\hbox{$D$}}
\newcommand{\slC}{\raise.15ex\hbox{$/$}\kern-.53em\hbox{$C$}}
\newcommand{\slA}{\raise.15ex\hbox{$/$}\kern-.53em\hbox{$A$}}
\newcommand{\slSigma}{\raise.15ex\hbox{$/$}\kern-.53em\hbox{$\Sigma$}}
\newcommand{\slpartial}{\raise.15ex\hbox{$/$}\kern-.53em\hbox{$\partial$}}
\newcommand{\slcalP}{\raise.15ex\hbox{$/$}\kern-.63em\hbox{$\cal P$}}

\def\bs{\boldsymbol}
\def\del{\partial}

\def\p{{\boldsymbol p}}
\def\q{{\boldsymbol q}}

\def\k{{\boldsymbol k}}

\def\x{{\boldsymbol x}}
\def\y{{\boldsymbol y}}
\def\X{{\boldsymbol X}}
\def\Y{{\boldsymbol Y}}

\def\r{{\boldsymbol r}}
\def\z{{\boldsymbol z}}
\def\v{{\boldsymbol v}}
\def\u{{\boldsymbol u}}
\def\w{{\boldsymbol w}}

\def\Q{{\boldsymbol Q}}

\def\P{{\boldsymbol P}}

\def\tform{\tau_{_{\rm f}}}

\newcommand{\rmd}{{\rm d}}
\newcommand{\rme}{{\rm e}}

\newcommand{\beq}{\begin{eqnarray}}
\newcommand{\eeq}{\end{eqnarray}}
\newcommand{\be}{\begin{eqnarray*}}
\newcommand{\ee}{\end{eqnarray*}}

\newcommand{\nn}{\nonumber\\ }

\title{Medium-induced gluon branching}

\author[a]{Jean-Paul Blaizot,}

\author[a]{Fabio Dominguez,}

\author[a]{Edmond Iancu,}

\author[a]{Yacine Mehtar-Tani}

\affiliation[a]{Institut de Physique Th\'eorique, CEA Saclay, 
F-91191 Gif-sur-Yvette, France}

\emailAdd{jean-paul.blaizot@cea.fr}
\emailAdd{fabio.dominguez@cea.fr}
\emailAdd{edmond.iancu@cea.fr}
\emailAdd{yacine.mehtar-tani@cea.fr}

\abstract{We study the evolution of an energetic jet which radiates
gluons while propagating through a dense QCD medium modeled as a random
distribution of color sources.  Motivated by the heavy ion experimental
program at the LHC, we focus on the medium-induced radiation
of (relatively) soft gluons, 
which are abundantly emitted at large angles and thus can 
transport a small fraction of the jet energy far away from the jet axis.
We perform a  complete calculation of the
medium--induced gluon branching in the regime where the gluons that take part in the branching undergo multiple soft scattering with the medium. We extend the BDMPSZ theory of radiative 
energy loss by including the transverse momentum dependence in the kernel that describes the branching and by analyzing the correlations between the two offspring gluons.
We demonstrate that these gluons lose color coherence with respect to each other
over a time scale that is comparable to the duration of the branching. It follows that interference 
effects between successive emissions are suppressed, a necessary ingredient for a description of multiple emission of soft gluons by a probabilistic, branching process.
}

\dedicated{Preprint: }

\begin{document}
\maketitle

%%%%%%%%%%%%%%%%%%%%%%%%%%%%%%%%%
\section{Introduction}
%%%%%%%%%%%%%%%%%%%%%%%%%%%%%%%%%

The LHC heavy ion program has produced a wealth of remarkable results \cite{Aamodt:2010jd,Aad:2010bu,Chatrchyan:2011sx,Chatrchyan:2012ni} that motivate new theoretical efforts in the study of jet propagation in matter.  For one thing, these results confirm those obtained at RHIC \cite{Putschke:2008wn,Salur:2008hs,Lai:2009zq}, indicating that matter formed in the collisions strongly suppresses the yield of high-$p_T$
hadrons as compared to the yield that one  deduces from proton-proton collisions after proper scaling  by the number of
binary collisions in the nucleus-nucleus collision. This suppression of high-$p_T$ hadrons,  referred to as  ``jet-quenching'', is usually attributed to  the energy loss of
the leading partons caused by the radiation of  soft gluons induced by their collisions with the matter constituents. The
theory of radiative parton energy loss has been developed in late 90's. It is commonly referred to as  BDMPSZ theory, from the names of the original authors 
\cite{Baier:1996kr,Baier:1996sk,Baier:1998kq,Zakharov:1996fv,Zakharov:1997uu}.  Further developments are presented in \cite{Wiedemann:1999fq,Wiedemann:2000za,Wiedemann:2000tf,Gyulassy:2000fs,Gyulassy:2000er,Arnold:2001ba,Arnold:2001ms,Arnold:2002ja}. In this theory, the energy loss is characterized by a single parameter, a transport coefficient called $\hat q $
(the `jet quenching parameter'), which measures how much transverse momentum $\Delta k_\perp$ a given parton acquires through multiple scattering as it travels through the medium over a distance $\Delta l$: $\Delta k_\perp^2=\hat q\Delta l$.  

However, the LHC experiments provide much more detailed information about what is going on as the jet propagates through the medium, beyond the mere evidence for energy loss. There is in particular clear evidence that the jet shape is affected with a large number of soft particles being emitted at larger angles, i.e.,  outside the jet cone, and carrying a small (some ten percent or so) of the total energy
\cite{Aad:2010bu,Chatrchyan:2011sx,Chatrchyan:2012ni}. While these soft particles do not contribute much to the energy loss, they carry important information, as we shall see,  on the basic microscopic mechanisms at work. In order to fully exploit this new available information,  a more complete description of exclusive jet observables (such as
jet-shapes, particle correlations, etc) is called for. This paper  presents the first step towards such a more complete theory of jet propagation in matter, which is valid in a specific regime that we shall specify shortly. 

In order to put our work into perspective, we need first to recall a few basic features of the BDMPSZ theory \cite{Baier:1996kr,Baier:1996sk,Baier:1998kq,Zakharov:1996fv,Zakharov:1997uu}.  When propagating through a medium in which it undergoes multiple scattering, a high energy parton can radiate a gluon over a typical time scale (``formation time'') $\tform$ given by 
\beq\label{formationtime}
\frac{1}{\tform}\sim \frac{k_\perp^2}{2\omega},\eeq
where $\omega$ and $k_\perp$ are respectively  the energy of the radiated gluon and its transverse momentum (with respect to the parent gluon). 
The quantity $1/\tform$, which, as suggested by  Eq.~(\ref{formationtime}) may be read as a non relativistic energy, with $\omega$ playing the role of a mass (the relevance of this analogy will be clarified in the main text) is essentially the amount of energy that is required to put the gluon on-shell. In  the medium, such an energy is obtained from multiple scattering with the plasma constituents, each collision providing the colliding hard partons some transverse momentum.  As mentioned earlier, the  rate at which transverse momentum is accumulated by a parton along its trajectory  is given by $\hat q$, that is $\Delta k_\perp^2=\hat q\Delta t$. 
One sees therefore that emission of a gluon of a given energy $\omega$ can take place if  the transverse momentum acquired during $\tform$ matches $2\omega/\tform$, that is if $1/\tform\sim (\hat q\tform/\omega)$. This  provides a self-consistency condition that determines, as a function of $\omega$, the time scale for in-medium splitting, that we shall denote by $\tau_{_{\rm br}}$:
\beq\label{tauf}
 \tau_{_{\rm br}}(\omega)\,\sim \,\sqrt{\frac{2\omega}{\hat q}}.\eeq
Related to $\tau_{_{\rm br}}$, it is also convenient to define $k_{_{\rm br}}$,  the typical transverse momentum acquired during the time $\tau_{_{\rm br}}$: $k_{_{\rm br}}^2=\hat q \tau_{_{\rm br}}$, or $k_{_{\rm br}}(\omega)\sim
 (2\omega\hat q)^{1/4}$. Note that the time scale of the branching process is larger for harder gluons, because it requires more collisions to put a hard radiated gluon on-shell than a soft one ($k_{_{\rm br}}$ is as slowly growing function of $\omega$). Note also that the radiation is emitted with a characteristic angle $\theta_{_{\rm br}}\sim k_{_{\rm br}}/\omega\sim \left( \hat q/\omega^3  \right)^{1/4}$. Thus the softer the emission, the larger the emission angle, with all angles such that $\theta\gtrsim\theta_c$,   the minimal angle $\theta_c=\left( \hat q L^3 \right)^{-1/2}$ corresponding to $\tau_{_{\rm br}}=L$, the length of the medium.

The BDMPSZ  energy spectrum of the radiated gluons  is of the form
 \beq\label{spec}
\omega \frac{\rmd N}{\rmd \omega}
 \,\simeq\,\frac{\alpha_s N_c}{\pi}\,\sqrt{\frac{\omega_c}{\omega}}\equiv
 \bar\alpha\sqrt{\frac{\omega_c}{\omega}}, \eeq
 for $\omega\lesssim\omega_c$ (and more strongly suppressed for $\omega>\omega_c$). The frequency $\omega_c$ is that for which $\tau_{_{\rm br}}(\omega_c)\sim L$, that is,  $\omega_c=\frac{1}{2}\hat q L^2$.
The first factor in Eq.~(\ref{spec})  is the standard bremsstrahlung spectrum for
radiation by the parent gluon. The correction factor, which we can write as $\sqrt{ \omega_c/\omega }=L/\tau_{_{\rm br}}(\omega)$, increases as $\omega$ decreases, and the spectrum (\ref{spec})  should be cut-off at a minimal frequency $\omega_{_{\rm BH}}$, at which multiple
scattering ceases to be important and the radiation is produced by incoherent collisions 
(Bethe--Heitler spectrum):  $\omega_{_{\rm BH}}$ is the frequency for which the formation time is of the order of the mean free path $\ell$ between successive collisions, that is $\tau_{_{\rm br}}(\omega_{_{\rm BH}})\sim \ell$. Thus, $L/\tau_{_{\rm br}}(\omega)$ is the number of effective scattering
centers, which is maximum for $\omega\sim\omega_{_{\rm BH}}$ and is of order 1 in the vicinity of $\omega_c$. The decrease of $L/\tau_{_{\rm br}}(\omega)$ as $\omega$ increases may be understood as  a consequence of  the 
Landau--Pomeranchuk--Migdal (LPM) effect.   

This mechanism for gluon production, and the approximations used to calculate it,  require  the formation time to be much larger than the mean free path
$\ell$, but  smaller than the size $L$ of the medium, that is $ \omega_{_{\rm BH}}\ll\omega\lesssim\omega_c$. For energies within this range, there is a large number of scattering
centers, of order $\tau_{_{\rm br}}(\omega)/\ell\gg 1$, which coherently contribute
to the  emission process.  This can be rephrased in terms of the typical transverse momentum exchanged in one collision. Having in mind a picture of the medium where the typical collisions are induced by  a screened one gluon exchange,  we call that typical momentum $m_D$ (with $m_D$ the screening mass). 
Then, from the definition of $\hat q$ given earlier,  $m_D^2=\hat q \ell$, and the condition $ \omega_{_{\rm BH}}\ll\omega$ translates into $ k_{_{\rm br}}^2(\omega)\gg {m_D^2}$.  

The initial motivation for the BDMPSZ theory was to provide a framework for calculating the energy loss. An estimate of this energy loss can be obtained by integrating the spectrum (\ref{spec}):
\beq\label{enloss}
 \Delta E\,=\int_{\omega_0}^{\omega_c}
 \rmd\omega\,\omega\, \frac{\rmd N}{\rmd \omega}
 \,\sim\,\bar\alpha \omega_c  \,\sim\,\bar\alpha\,\hat q L^2. \eeq
 This  is  dominated by the upper limit $\omega=\omega_c$, 
 the maximal energy that can be taken away by a single gluon. Such emissions of hard gluons are  rather rare: they occur with a probability of order
$ \bar\alpha$. However, from the spectrum  (\ref{spec}) one can also  compute the average number of 
gluons emitted with energies larger than a given value $\omega$:
 \beq\label{number}
 \Delta N(\omega)\,=\int_{\omega}^{\omega_c}
 \rmd\omega'\, \frac{\rmd N}{\rmd \omega'}
 \,\sim\,\bar\alpha \,\frac{L}{\tau_f(\omega)}\,.\eeq
As long
as $\Delta N(\omega)\lesssim 1$ (that is, as long as $\omega\gtrsim \bar\alpha^2\,{\omega_c}$), 
it may be identified to the probability to emit one gluon with energy
$\omega'\ge \omega$. In such a case, the probability for multiple
emissions is small. This is the case of the relatively hard emissions that dominate the energy loss. But for sufficiently soft gluon
emissions, such that
 \beq\label{omegas}
  \omega\,\lesssim\,\omega_s\equiv \bar\alpha^2\,{\omega_c}\,,
  \eeq
one has $\Delta N(\omega)\gtrsim 1$ and then the multiple emissions are
clearly important. Note that these multiple emissions are `large angle' emissions, with $\theta\sim \bar\alpha^{-3/2}\theta_c\gg\theta_c$.
 Of course this can only occur if the medium is large enough, since the condition $\omega_{_{\rm BH}}\ll {\omega_s}$ implies 
$  \ell\ll \bar\alpha L$.  

To summarize, there is a regime, characterized by gluon energies in the range $\omega_{_{\rm BH}}\ll \omega\ll \omega_c$, were medium induced radiation dominates, and where multiple emissions are important. This is the regime that we explore in this paper. Note that the  conditions on $\omega$ and $L$ that characterize this regime are not very restrictive, and in
fact they are expected to be well satisfied in the LHC experiments. For a rough orientation, taking $\hat q\sim 1$GeV$^2$/fm and $L\sim 6$ fm, one finds $\omega_c\sim 100$ GeV, while the typical energy of soft particles is in the GeV range.

When multiple emissions become important, i.e., when $\Delta N(\omega)\gtrsim 1$, the whole calculationnal setting needs to be revised.  In technical terms, when 
$\bar\alpha{L}/{ \tau_{_{\rm br}}}\sim 1\,,
$ the perturbative expansion breaks down, and powers of
$\bar\alpha L/ \tau_{_{\rm br}}$ have to be resummed.  (For instance, the longitudinal 
phase-space for two independent  successive branchings goes like $L^2$ and the process  
is of order $(\bar\alpha L/\tau_{_{\rm br}})^2$.) The treatment of  multiple emission is a priori complicated by interferences between various high order processes. 
Various aspects of interference phenomena for medium--induced gluon radiation have been
recently studied,  but only for the case of a frozen configuration of the emitters (a pair of partons forming a `colour antenna') \cite{MehtarTani:2010ma,MehtarTani:2011tz,CasalderreySolana:2011rz,MehtarTani:2011gf,MehtarTani:2012cy}.
However,  in the soft regime of interest for us here, these  interferences turn out to be negligible. Indeed, as  we shall see, color correlations between the offspring gluons disappear on the same time scale $\tau_{_{\rm br}}$ at which the splitting occurs.  As a consequence, the newly formed gluons propagate independently from each other except for a relatively small period
$\tau_{_{\rm br}}\ll L$. Subsequent emissions from these gluons can interfere
with each other only if they occur during that period of order $\tau_{_{\rm br}}$
where color coherence is still present. This implies that the longitudinal phase space for 
interference effects is smaller, by a factor $\tau_{_{\rm br}}/L\ll 1$, than the corresponding
phase space for independent emissions.   Note that these features are specific to medium induced radiation\footnote{In fact, the same estimate for the color decoherence time,
that is, a time scale of order $\tau_{_{\rm br}}(\omega)$, would also follow from the
previous analyses of a quark--antiquark color antenna in Refs.~ \cite{MehtarTani:2010ma,CasalderreySolana:2011rz} provided in these analyses one identifies the
angular opening of the antenna with the formation angle 
$\theta_{_{\rm br}}\sim \left( \hat q/\omega^3  \right)^{1/4}$.}. By contrast,  in vacuum, color is conserved along the parton shower, which  implies coherence of successive parton branchings, that is responsible for angular ordering \cite{Bassetto:1982ma,Mueller:1981ex,Ermolaev:1981cm,Dokshitzer:1991wu}.

The argument of the previous paragraph suggests that multiple emissions could be treated as a probabilistic cascade of independent branchings, since this naturally resums the terms with maximal powers of  $(\bar\alpha L/\tau_{_{\rm br}})$. There is however a limit to this argument, since successive emissions may overlap with each other when the energies of the produced gluons become too soft. Consider indeed the  probability for emitting one gluon with 
frequency  $\omega \ge \omega_0$ during an interval $\Delta t$. In the regime where this probability is small, this may be estimated from the BDMPSZ spectrum, \eqn{spec}, and is (see \eqn{number})
 \beq
 P(\Delta t; \omega_0)  \,\sim\,\bar\alpha \,\frac{\Delta t}{\tau_{_{\rm br}}(\omega_0)}.
 \eeq
This becomes of order unity when $ \Delta t=\tau_{\rm rad}(\omega_0)$ with
 \beq\label{taurad}
\tau_{\rm rad}(\omega_0)\,=\,\frac{1}{\bar\alpha }
 \,\tau_{_{\rm br}}(\omega_0)=\frac{1}{\bar\alpha }\sqrt{\frac{2\omega_0}{\hat q}}
 \,.\eeq
This quantity $\tau_{\rm rad}(\omega_0)$ may be understood as the typical time interval that a gluon
with energy $\omega$ can survive without emitting any gluon with energy $\omega'$
within the interval $\omega_0<\omega'<\omega$.
Note that it is the emission of the softest allowed gluons which controls the size of 
this interval. Because of the inverse power of $\bar\alpha$ in \eqn{taurad}, $\tau_{\rm rad}(\omega_0)$ is parametrically larger than the formation time of the radiated gluon. This property would hold all along the cascade if produced gluons are emitted with comparable frequencies: successive emissions remain well separated, the phase--space for overlapping emissions is 
small and successive emissions can be treated as independent. This is what is implicitly assumed in the previous paragraph. Under these conditions, the whole branching process reduces to a classical stochastic process\footnote{This conclusion brings some support to previous phenomenological studies inspired by the BDMPSZ theory \cite{Baier:2000sb,Baier:2001yt,Jeon:2003gi} and in particular the MonteÐCarlo codes developed in Refs. \cite{Armesto:2009fj,Schenke:2009gb,Zapp:2011ya}, which have already assumed the factorization of subsequent emissions.} obtained by iterating the elementary building block corresponding to one single splitting of a gluon into two gluons. This is the process that will be studied in detail in this paper. 

However, one cannot exclude a priori the situation where a very soft gluon is emitted, with   $\tau_{\rm rad}(\omega_0)$ comparable to the formation time of the parent gluon (with energy $\omega$). The condition $\tau_{\rm rad}(\omega_0)\gtrsim \tau_{_{\rm br}}(\omega)$ holds if $\omega_0\gtrsim \bar\alpha^2\omega$. Thus, if $\omega_0$ is  parametrically smaller than $\omega$ (by two powers of $\bar\alpha$),
the interval $\tau_{\rm rad}(\omega_0)$ between two successive emissions becomes of the order of  the duration of the branching process that has produced the gluon $\omega$
in the previous step, and then the argument of independent emissions breaks down. This particular issue related to the role of very soft emissions in an actual cascade will be discussed in a forthcoming publication. 

The calculation to be presented in this paper is technically involved, but conceptually simple. We consider a high energy gluon propagating in a quark-gluon plasma. The medium is modeled by a random color field whose fluctuations account for  collisions with plasma constituents and determine $\hat q$.  In Section \ref{sec:strategy}, we present the general strategy of the calculation by discussing the well known phenomenon of momentum broadening. Then, in Section \ref{sec:branch} we discuss the general structure of the gluon splitting amplitude and probability. 
The proof of factorization of the splitting probability requires a special study of a 4-point function,
that we shall be able to explicitly compute only in the limit of a large number of colors $N_c\gg 1$. 
The corresponding analysis is presented in Section \ref{sec:twogluonprop}. 
However,  the physical argument
for the color decoherence between the offspring gluons is quite general and we expect
it to remain valid for any value of $N_c$. Section \ref{splitting} is devoted to the calculation of the splitting kernel and the completion of the formula for the splitting probability. In the conclusion we summarize the results and their range of validity. Technical material is gathered in the Appendices. Appendix \ref{app:glueprop} reviews general features of gluon propagation in a specific background field. Appendix \ref{app:pathint} is devoted to the calculation of specific path integrals that enter the $n$-point functions that are studied in the text. Finally Appendix \ref{app:color} provides details on the  color algebra that is needed to calculate the 4-point function.

%%%%%%%%%%%%%%%%%%%%%%%%%%%%%%%%%
\section{Setting up the calculation with a simple example}\label{sec:strategy}
%%%%%%%%%%%%%%%%%%%%%%%%%%%%%%%%%%%

We shall set up formalism, and fix the notation, by reviewing the well known phenomenon of 
momentum broadening of a parton traveling through a  quark-gluon plasma. This will also
give us the opportunity to present  our main results concerning the medium--induced
gluon branching, anticipating on the detailed calculations to be done in the forthcoming sections.

We consider an energetic parton  propagating through a quark--gluon plasma. For simplicity, we restrict ourselves to the case where this energetic parton is a gluon. The extension to the case where it is a quark is straightforward. The gluon is produced inside the
medium, via some hard scattering process, and then it propagates along a
distance $L$ until it escapes into the vacuum. During this propagation,
the gluon  interacts with the medium, and exchanges with it color and transverse momentum.

We assume that the energetic parton propagates at nearly the speed of light along
the $x^3$ axis and we  use light--cone coordinates, e.g.
$x^\mu=(x^+,x^-,\x)$, with\footnote{From now on, we renounce to the subscript  $\perp$
on transverse components, that is, we write e.g. $\x_\perp\equiv \x$ to alleviate 
notations.}
 \beq \label{LC}
 x^+=\frac{1}{\sqrt{2}}\,(x^0+x^3)\,,\quad
 x^-=\frac{1}{\sqrt{2}}\,(x^0-x^3)\,,\quad\x=(x^1,\,x^2)\,,
 \eeq
together with the light--cone gauge $A^+_a=0$. 
We  describe the medium as a random color field with correlation function
\begin{align}\label{2pA}
\langle {\cal A}^-_a(x^+, x^-, \x){\cal A}^-_b(y^+,x^-,\y)\rangle
\,=\,\delta_{ab}
 \delta(x^+-y^+)\,\gamma(\x-\y),
\end{align}
where the angular brackets denote the medium average.
 More general situations may be considered, e.g. we may allow $\gamma$ to depend on $x^+$, but we shall not consider such straightforward extensions here.
 
In writing \eqn{2pA}, we have made implicitly several simplifications. The coupling of the energetic gluon with the medium is described in the eikonal approximation, i.e.,  we assume that it couples only to the  component $A^-$ of the gauge field. Furthermore, this  field is probed only at small values of $x^-$ (closed to the trajectory of the gluon $x^-\simeq 0$), and we can ignore its dependence on $x^-$. In other words, we assume that the $+$ component of the gluon momentum is conserved during the propagation through the medium.  Furthermore,  because of Lorentz time dilation, the hard gluon  has a very poor resolution in $x^+$, and the medium correlations, that have a finite extent in  $x^+$,  appear to it as effectively {\em local}~:
this is the origin of  the $\delta$--function $\delta(x^+-y^+)$ in the correlator
\eqref{2pA}.  By the same token, the field of the medium, which has a finite longitudinal extent $L$,  appears frozen during the time when the hard gluon traverses it.  That implies that the average over the medium will be done after squaring the amplitudes. We take this average over the fields to be Gaussian, assuming that the corrections to this approximation are of higher order in the coupling strength.  Finally, as obvious in Eq.~(\ref{2pA}) we assume  homogeneity
 in the transverse plane, i.e., the correlation $\gamma$ is a function of 
$\x-\y$ alone.
In summary, the problem that we  address is the propagation of an energetic gluon in a random background $A^-$ field that is independent of $x^-$, with Gaussian correlations. Technical aspects of this problem are studied in detail in Appendix \ref{app:glueprop}. 
 
Although this will not enter explicitly in our derivations, it is perhaps useful to have in mind a specific model for the medium. We then briefly discuss the case where this medium is a weakly coupled quark--gluon plasma in thermal equilibrium at 
high enough temperature $T$. In this case, the medium constituents are
 quarks and gluons with energies and
momenta $p\sim T$. 
Assuming that the charge carriers are correlated over distances determined by the screening length (inverse Debye mass) $m_{_D}^{-1}$,  one can estimate the correlator of field fluctuations:
\beq\label{gamma}
 \gamma(\x-\y)\,=\,g^2 n\int \frac{\rmd^2
 \q}{(2\pi)^2}\,\rme^{i\q\cdot(\x-\y)}\,
 \frac{ 1}{(\q^2+m_{_D}^2)^2},
 \eeq
where $n\propto T^3$ is the density of (point-like) color charges (weighted with appropriate color factors), 
and $1/(\q^2+m_{_D}^2)$ is the screened Coulomb propagator.
In fact, the quantity which will  enter our analysis is not $\gamma$ itself, 
but the difference $\gamma(0)-\gamma(\r)$, 
 \beq\label{sigmar} 
 \gamma(0)-\gamma(\r)
 \,=\,g^2 n
 \,\int \frac{\rmd^2
 \q}{(2\pi)^2}\,
 \frac{1-\rme^{i\q\cdot\r}}
 {(\q^2+m_{_D}^2)^2}.
 \eeq
 The integral over $\q$ in \eqn{sigmar} is dominated by $q\lesssim 1/r$, and in the relevant case where, typically, $1/r\gg m_{_D}$, it is 
logarithmically sensitive to all the scales within the range
$m_{_D}\ll q \ll 1/r$. To leading logarithmic accuracy, 
it can be evaluated 
by expanding the complex exponential to second order. One gets
 \beq\label{sigmar2} 
 \gamma(0)-\gamma(\r)
 \,\simeq \,\frac{1}{4}g^2 n
 \,\r^2\,\int \frac{\rmd^2
 \q}{(2\pi)^2}\,
 \frac{\q^2}
 {(\q^2+m_{_D}^2)^2}\approx \frac{1}{16\pi}g^2 n
 \,\r^2\,\ln\frac{1}{r^2 m_{_D}^2}.
 \eeq
In the context of the present calculation the relevant  values of $r$ are typically 
fixed by transverse momentum broadening. In the regime dominated by multiple scattering, one often uses the \emph{harmonic approximation}, where one ignores the weak dependence on $r$ of the logarithm, and set
 \beq\label{harmonic}
g^2N_c \big[\gamma(0)-\gamma(\r)
\big]\approx \frac{1}{4}\,\hat q\,\r^2,
 \eeq
with  the jet quenching parameter $\hat q$ (evaluated at the typical scale $1/\bar r$) given by
 \beq\label{ourqhat}
 \hat q\,\approx 4\pi\alpha_s^2 N_c \,n\,\ln\frac{1}{\bar r^2 m_{_D}^2}\,.
 \eeq
Parametrically, we have, $m_{_D}\sim gT$, $\ell\sim 1/g^2T$, $\hat q\sim g^4 T^3$, and $\omega_{BH}\sim T$.

We now return to our main discussion. We assume that the energetic gluon is created inside the medium
at time $x^+=t_0$ by some local current ${\bs J}(t_0)$,   and then propagates up to
time $x^+=t_L$, when it escapes the medium. Note that,  as mentioned already in the introduction, we ignore in our calculation the `vacuum radiation' that would be associated with the possible virtuality of the energetic gluon.  (In what follows we shall often denoted 
the light--cone time
$x^+$ simply as $t$, to alleviate the notation. Note that $t_L=\sqrt{2} L$.)  
The probability amplitude for  finding the (on--shell) gluon at $t_L$, with momentum
$k^\mu=(k^-=\k^2/2k^+,k^+,\k)$,  colour $b$,
and  polarization  $\lambda$, can be obtained from a standard calculation, as explained in Appendix \ref{app:glueprop} (see also
Refs.~\cite{MehtarTani:2006xq,MehtarTani:2010ma,CasalderreySolana:2011rz}
for previous applications of a similar formalism). One gets
\beq\label{Mone}
{\cal M}^b_{\lambda}(k^+,\k)\,={\rm e}^{i\frac{\k^2}{2k^+}t_L}\, 
\int \frac{\rmd^2\p_0}{(2\pi)^2}\,(\k\, b|\,{\cal
G}(t_L,t_0)\,|\p_0\, a)\,{\bs \epsilon}_\lambda^i\,(p_0^+\,\p_0\, a|{\bs J}^i(t_0)|0),
\eeq
with $(p_0^+\,\p_0\, a|{\bs J}^i(t_0)|0)$ the matrix element of the current 
creating the gluon from the vacuum, with momentum $(p^+_0=k^+, \p_0)$ 
and color $a$. Summation over repeated discrete indices is implied.
Here, $(\k|\,{\cal G}^{ba}(t_L,t_0)\,|\p_0)=(\k\, b|\,{\cal G}(t_L,t_0)\,|\p_0\, a)$ (we use indifferently both notations) is the effective propagator that describes the (non relativistic) motion of the gluon in the transverse plane under the influence of the time dependent background field $A^-$. This propagation preserves $k^+$, as already mentioned, and  ${\cal G}$ depends on $k^+$, but we shall generally not indicate this dependence explicitly. Note also that the propagator ${\cal G}$ do not carry any Lorentz indices. This is because, as shown in Appendix \ref{app:glueprop}, the propagation in the background field preserves the (transverse) polarization of the gluon.

It is shown explicitly in Appendix \ref{app:glueprop} that ${\cal G}$ is the propagator of a Schr\" odinger equation
in  two dimensions (the transverse plane) for a non relativistic particle of mass $k^+$ 
moving in a time dependent potential $A^-(x^+,\x)$, with $x^+$ playing the role of the time. 
That is, it satisfies the equation
\beq
\left[ i{\cal D}^-+\frac{\nabla_\perp^2}{2k^+}\right]_{ac}(\x|\,{\cal G}^{cb}(x^+,y^+;k^+)\,|\y)=i\delta_{ab}\delta(x^+-y^+)\delta(\x-\y),
\eeq
with $i{\cal D}^-=i\del^-+g A^-$ and $\del^-=\del/\del x^+$. The solution  can be written as  a path integral
\beq
(\x|\,{\cal G}(x^+,y^+;k^+)\,|\y)=\int {\cal D} \r\, {\rm e}^{i\frac{k^{\!_+}}{2} \int_{y^{\!_{+}}}^{x^+}    \rmd t \, \dot\r^2   }\,\tilde U(x^+,y^+; \r),
\eeq
with $\r(y^+)=\y$, $\r(x^+)=\x$, and $\tilde U$ is a Wilson line evaluated along the path $\r(t)$ in the adjoint representation
\beq
\tilde U(x^+,y^+; \r)={\rm T}\exp\left\{ ig\int_{y^+}^{x^+} \rmd t \, A^-_a(t, \r(t))\, T^a   \right\}.
\eeq
Note that we work in a regime dominated by multiple scattering, so that the accumulated phase over a typical interval $\Delta t$ is large, i.e., $g A^- \Delta t\sim 1$. It follows that the exponential cannot be expanded in general (as done for instance in  the opacity expansion \cite{Gyulassy:2000fs,Wiedemann:2000za}).

\begin{figure}[tbp]
\begin{center}
\includegraphics[width=8cm]{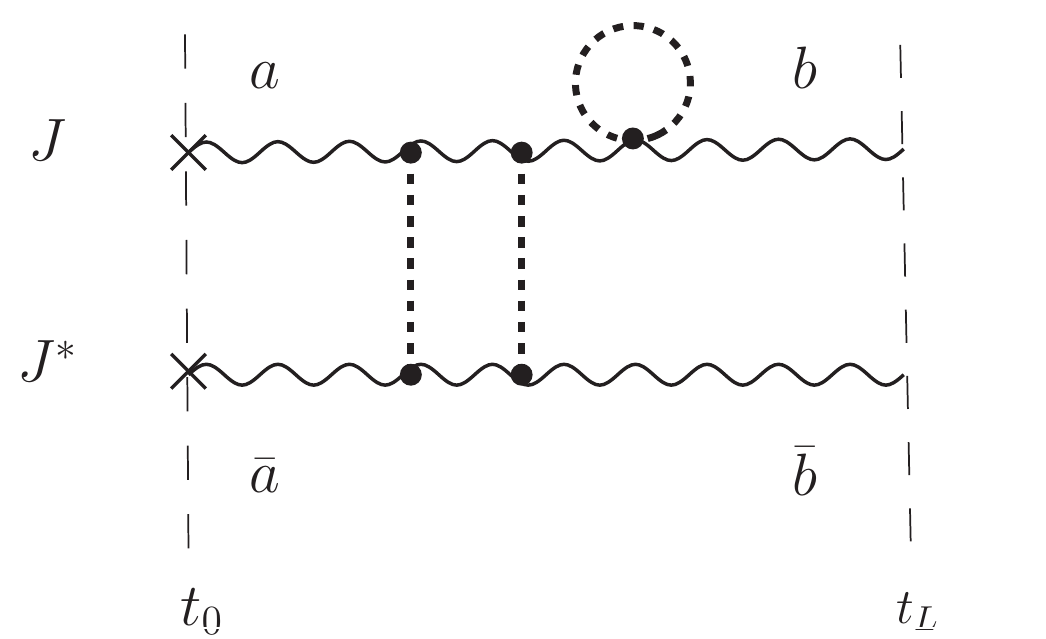}
\caption{\sl Amplitude (upper part of the diagram) and complex conjugate
amplitude (lower part) describing the propagation of a gluon in the
medium from time $t_0$ to time $t_L$ when it leaves the medium.
The horizontal wavy lines represent the gluon propagator or its complex
conjugate.  Note that we use a
bar to indicate color indices
 in the conjugate amplitude. At time $t=t_L$, $b=\bar b$. 
The (Gaussian) medium average generates instantaneous
contractions that  are depicted as vertical dotted lines, or as dotted circles.   }
\label{amplitsqonegluon}
\end{center}
\end{figure}

The probability to find the gluon with momentum $\k$ is obtained by  taking the modulus squared of the  amplitude in  \eqn{Mone}, summing over the
final color indices, and then averaging over the color background
field.  Since the distribution of gauge fields is taken to be Gaussian, cf. \eqn{2pA},
the medium averaging amounts to contracting pairs of fields from the Wilson lines in the two
propagators representing the gluon in the direct and the
complex conjugate amplitudes, respectively. These contractions can
either connect the two gluon lines, or they can be `tadpoles' with both endpoints on a same gluon
line. It is convenient to represent the amplitude together with
its complex conjugate on the same diagram, with time flowing from left to
right. (This is somewhat analogous to the close time path technique,
although we shall not explicitly use this particular formalism here.) The
ensuing diagram, shown in Fig.~\ref{amplitsqonegluon}, displays typical contractions. 
This representation makes it easier to discuss color and momentum flows.

Let us consider first the color structure. At time $t_L$, the system formed by the gluons in  the amplitude and its complex conjugate  is clearly in a color singlet state: both gluons carry the same color index $b$ (i.e., $b=\bar b$ in Fig.~\ref{amplitsqonegluon}), which is summed over since we do not observe the color of the final gluon. The field
contractions involved in the  medium averaging do not change the overall color state of the two gluon system, which remains therefore  a color singlet
 at any intermediate time between $ t_0 $ and $ t _L$. This implies that the medium
averaged squared amplitude depicted in Fig.~\ref{amplitsqonegluon} 
contains a contribution of the form (as we shall see shortly $\bar \k=\k$)
\beq\label{rIK2}
\left\langle (\k| {\cal G}^{ba}(t_L,t_0)|\p_0)(\bar\p_0|{\cal G}^{\dagger\,\bar a b} (t_0,t_L)|\bar\k)\right\rangle=\delta^{a \bar a} (\k;\bar\k|S^{(2)}(t_L,t_0)|\p_0;\bar\p_0),
\eeq
where the sum over the repeated color index $b$  is implied, and 
\beq\label{S2def}
(\k;\bar\k|\,S^{(2)}(t_L,t_0)\,|\p_0;\bar\p_0)=\frac{1}{N_c^2-1}\left\langle\,\text{Tr}\,(\k| \,{\cal G}(t_L,t_0)\,|\p_0)(\bar\p_0|\,{\cal G}^{\dagger}\, (t_0,t_L)|\bar\k)\right\rangle,
\eeq
where the trace Tr concerns color indices. The quantity $S^{(2)}$, referred to as a 2-point function, is the simplest of several $n$-point functions that we shall have to consider in this paper, and which are calculated explicitly in Appendix \ref{app:pathint}. 
Note that, in defining $S^{(2)}$, we choose to keep the time ordering given by the propagator in the amplitude,  and to put all the momentum variables at the final time in the bra,  and the ones at the initial time in the ket. The  variables corresponding to the conjugate propagator are denoted by a bar, and are separated from the variables of the  propagator by a semicolon. This notation will be used throughout  (see also Appendix \ref{app:pathint}).

Similarly, since the correlator of the background field is invariant
under translations in the transverse plane, the medium does not change the
overall transverse momentum of the pair of  gluons: whichever momentum is picked up from the medium by the gluon in the amplitude is  compensated at the same time by a similar transfer to the gluon in the complex conjugate amplitude. Thus,  if one chooses the final momentum
to be the same in the amplitude and the complex conjugate amplitude, i.e., $\k=\bar\k$ (as
we need to do if we consider the production of a gluon with a given
transverse momentum) then, at each instant of time, the momenta remain
equal in the amplitude and its  complex conjugate.  In coordinate space, as shown explicitly in Appendix \ref{app:pathint}, this corresponds to the two gluons in the amplitude and its complex conjugate propagating with  a fixed separation. The medium average of the Wilson lines in the pair of propagators is then of the form
 \beq\label{Dgg}
C_g^{(2)}(t_L-t_0;\r)\,&\equiv&\,\frac{1}{N_c^2-1}\left\langle
 \text{Tr} \,\tilde U(t_L,t_0;\x)\tilde U^\dagger(t_L,t_0;\bar\x)
 \right\rangle,\nn
 &=&\exp\left\{-g^2 N_c(t_L-t_0)\,
 \big[\gamma(0)-\gamma(\r)\big]\right\},\nn
  &=&\exp\left[-\frac{N_c n}{2}(t_L-t_0)\,\sigma(\r)\right],
 \eeq
where  $\r\equiv \x-\bar\x$ is the constant distance between the two gluons.  This
is recognized as the forward scattering amplitude for a color dipole which propagates through the medium. This is of course
a fictitious dipole, which is formed with the gluon in the direct amplitude and that 
in the complex conjugate amplitude, which in the calculation of the probability
group together in an overall color singlet state --- the `dipole'.
The expression in the third line of the above equation, which features the  dipole
cross--section $\sigma(\r)$, will often be used to simplify writing in what follows.

Altogether, the medium averaging
of the squared amplitude will therefore generate a 2--point function of
the form (see Appendix \ref{app:pathint} for details)
 \beq\label{S2P}
(\k;\k|S^{(2)}(t_L,t_0)|\p_0;\bar \p_0)
= (2\pi)^2\delta^{(2)}(\p_0-\bar\p_0)~{\cal P}(\k-\p_0,t_L-t_0),
 \eeq
where
 \beq\label{PD}
 {\cal P}(\Delta \p,\Delta t)\,=\,
 \int \rmd^2\r\,{\rm e}^{-i\Delta \p\cdot \r}\,C_g^{(2)}(\Delta t;\r),
 \eeq
can be interpreted as  the probability that a gluon acquires a transverse momentum $\Delta \p$ while traversing the medium during a time $\Delta t$.  
This is best seen by using the harmonic approximation \eqref{harmonic}, where
 \beq\label{harmonic_approx}
N_c n \sigma(\r)\approx \frac{1}{2}\,\hat q\,\r^2\,.
 \eeq
Then we get
 \beq\label{Dgg2}
 C_g^{(2)}(\Delta t;\r)=\exp\left(-\frac{\hat q\,\Delta t}{4}\,
\r^2\right), \eeq
 and
 \beq\label{Prop}
{\cal P}(\Delta\p,\Delta t)=\int \rmd^2\r \;{\rm
e}^{i\r\cdot\Delta\p-\frac{\hat q\Delta t}{4}\r^2}\,=\,\frac{4\pi}{\hat
q\Delta t}\ {\rm e}^{-\frac{(\Delta\p)^2}{\hat q\Delta t}}.
\eeq
This formula confirms the physical interpretation of $\hat q$: $\hat q\Delta t$ is the transverse
momentum squared acquired by a gluon after it has traveled through the medium
for a duration $\Delta t$.  In fact, in the harmonic approximation, ${\cal P}(\Delta\p,\Delta t)$ solves the diffusion equation with initial condition ${\cal P}(\Delta\p,\Delta t=0)=(2\pi)^2\,\delta^{(2)}(\Delta\p)$, 
with $\hat q$ playing the role of a diffusion coefficient. Note also that this probability is
independent of the gluon longitudinal momentum $k^+$. This is to be expected from the formal analogy of the present problem with the 2-dimensional diffusion of a non relativistic particle with mass $k^+$.

The exponential factor in Eq.~(\ref{Dgg2}) reflects a general feature of the interaction of color objects with the medium: the propagation of the  color dipole (formed by the gluons in the amplitude and its complex conjugate) is not affected by  the interaction as long as its size is small enough to be `viewed'  by the medium as a color singlet, but it is strongly damped as soon as the size of the dipole exceeds $2/\sqrt{\hat q\Delta t}$.

\begin{figure}[htbp]
\begin{center}
\includegraphics[width=4cm]{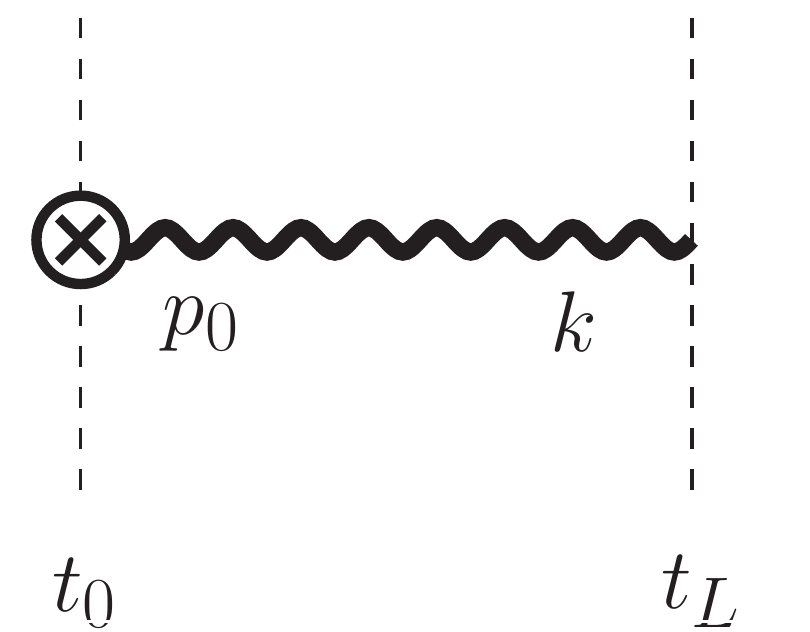}
\caption{\sl   Diagrammatic illustration for the process
in which a gluon of momentum $(p^+_0,\p_0)$ is produced
via a hard process (represented by the circled cross)
and acquires transverse momentum $\k-\p_0$ in its propagation though the
medium (the thick wavy line). The + component of the momentum is
conserved in this propagation, i.e., $k^+=p^+_0$.}\label{fat_vertex}
\end{center}
\end{figure}

We are now in  a position to calculate the differential cross section to
observe a gluon with transverse momentum $\k$ and arbitrary color and
spin at time $t_L$,  after its production through some hard process at time
$t_0$. This is given by\footnote{The subscript 0 on this cross--section
refers to the fact that the process considered here involves no gluon
branching. Similarly, the quantity $\sigma_1$ in \eqn{sigma1a} below
refers to a process which involves one branching.}
 \beq\label{sigma0} \frac{\rmd\sigma_0}{\rmd\Omega_k}\,=\,
 \int \frac{\rmd^2\p_0}{(2\pi)^2}~{\cal
P}(\k-\p_0,t_L-t_0)~\frac{\rmd\sigma_{\rm hard}}{\rmd \Omega_{p_0}},
\eeq
where $\rmd\Omega_k\equiv (2\pi)^{-3} \rmd^2\k\, 
\rmd k^+/2k^+$ is the invariant phase-space element, and 
${\rmd\sigma_{\rm hard}}/{\rmd \Omega_{p_0}}=|{\bs J}^{a}(\p_0)|^2$ is the `hard
cross section' for producing the gluon (the sum over color $a$ is understood -- we also assume that all kinematical factors necessary to build a cross section are included in the current $\bs J$). In deriving this result, we have
summed over the polarization vectors with the help of the completeness
relation $\sum_\lambda\epsilon^i_\lambda(k)\epsilon^{*j}_\lambda(k)=\delta^{ij}$. This  process, depicted in
Fig.~\ref{fat_vertex}, has a probabilistic interpretation, 
with the wavy line representing the probability \eqref{Prop}.

\begin{figure}[htbp]
\begin{center}
\includegraphics[width=7cm]{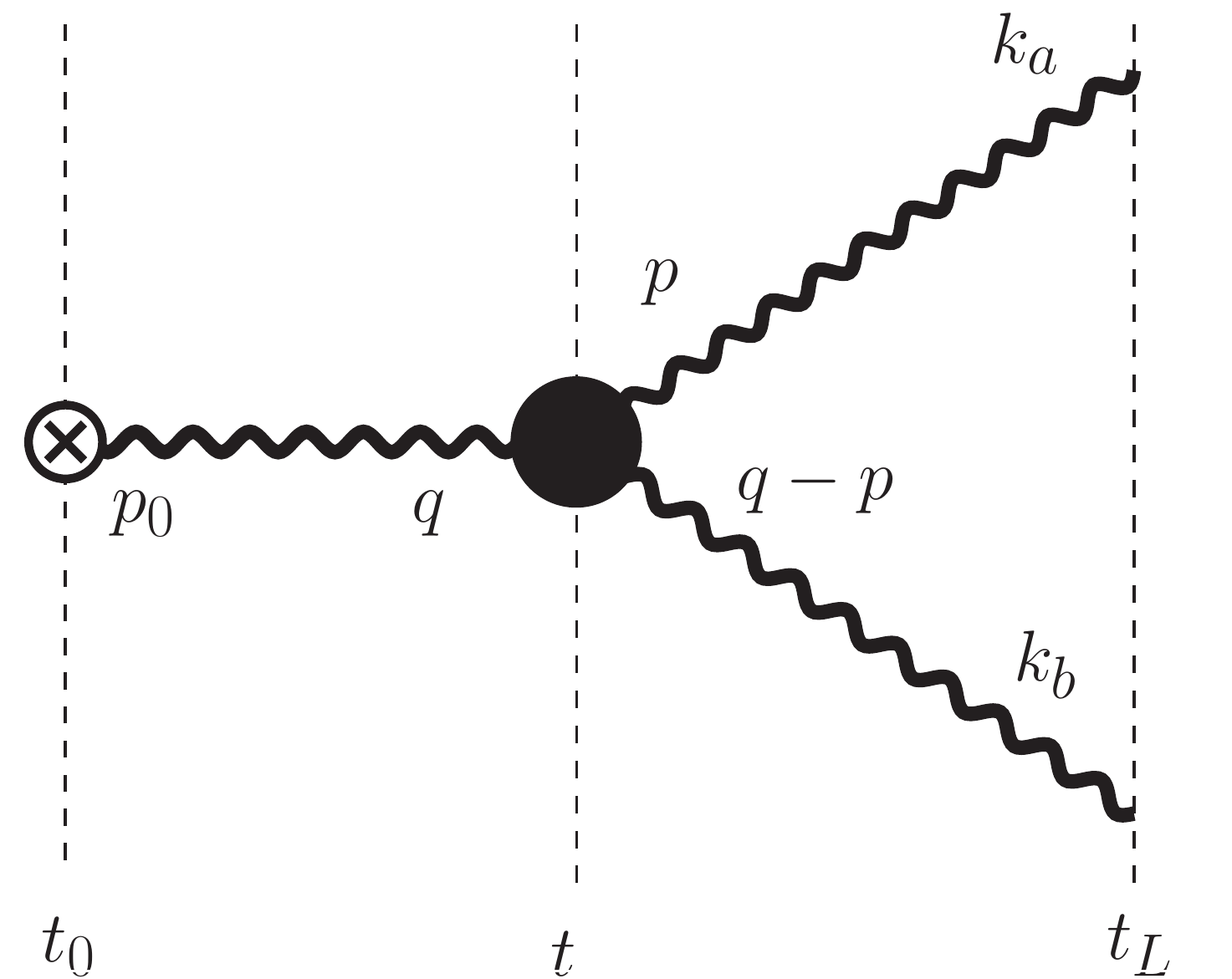}
\caption{\sl Graphical illustration of the equation (\ref{sigma1a}).
The thick wavy lines represent the probability ${\cal P}$ for transverse
momentum broadening,  the black dot is the splitting probability
${\cal K}$, and the circled cross is the cross section of the hard
process producing a gluon of momentum $p_0$. }
\label{fat_vertex2}
\end{center}
\end{figure}

It will be our main goal in this paper to show that  the cross section for the  process where
one gluon splits into two gluons under the effect of medium interactions can be given an analogous probabilistic interpretation. 
Our subsequent calculations
will lead to the following estimate for producing  soft gluons
($k^+_a,\,k^+_b\ll\omega_c$)
\beq\label{sigma1a}
\frac{\rmd^2\sigma}{\rmd\Omega_{k_a}\rmd\Omega_{k_b}}&=&2g^2z(1-z)\nn
&\times&\int_{t_0}^{t_L}\,\rmd t\,\int_{\p_0, \q,\p} \; {\cal P}(\k_a-\p,t_L-t)\,
{\cal P}(\k_b-\q+\p,t_L-t)\nn
&& \qquad \qquad \qquad \times  \,{\cal K}(\p-z\q,z,p_0^+) \,
 {\cal P}(\q-\p_0,t-t_0)\, \frac{\rmd\sigma_{hard}}{\rmd\Omega_{p_0}}\,,
\eeq
and it is understood that $z=k_a^+/p_0^+$.
This result can be interpreted as a classical branching
process, illustrated in Fig.~\ref{fat_vertex2}: after propagating from
$t_0$ to $t$, during which it acquires a transverse momentum $\q-\p_0$,
the original gluon splits into gluons $a $ and $b$ with a probability $\sim\alpha_s {\cal
K}(\p-z\q,z,q^+)$ that depends upon the longitudinal momentum $q^+$
of the parent parton, the longitudinal momentum  fraction $z=p^+/q^+$
carried by gluon $a$,  and the transverse momentum difference  $\p-z\q$. 
(The conservation of
longitudinal momentum implies of course $p_0^+=q^+=k_a^++k_b^+$ with
$k_a^+=p^+=zq^+$.) After the splitting, 
the two gluons $a$ and $b$ propagate through the medium from $t$ to $t_L$, 
and thus acquire some extra transverse momentum. 

Note that, in \eqn{sigma1a}, the splitting occurs instantaneously at time $t$, that is, 
the effective splitting vertex ${\cal K}(\p-z\q,z,q^+)$ is  local in time.
Moreover, the transverse momentum is conserved at the splitting, meaning that one neglects
the additional momentum transferred from the medium to the gluons 
during the branching process. These are of course 
approximations, which are correct so long as the duration 
of the branching  process, $\tau_{_{\rm br}}$,  is much shorter than the medium size\footnote{In
the presence of multiple emissions, the correct reference scale for $\tau_{_{\rm br}}$
is not $L$, but the typical time interval between two successive emissions. We shall
return to this issue in the concluding section.} $L$. Previously, in \eqn{tauf}, we have
estimated this time $\tau_{_{\rm br}}$ for the case of asymmetric branchings, 
where one of the daughter gluons is much softer than the other one (say $z\ll 1$). 
An estimate valid for arbitrary values of $z$ can be obtained by considering $\Delta E$,  the change in the light--cone energy  
(the minus component of the 4-momentum, or equivalently the change in the  energy in the equivalent non relativistic 2-dimensional problem), at the splitting vertex:
 \beq
\Delta E=\frac{\p^2}{2p^+}+\frac{(\q-\p)^2}{2(q^+-p^+)}-\frac{\q^2}{2q^+}
=\frac{(\p-z\q)^2}{2z(1-z)q^+}\,.
 \eeq
The inverse of this energy is the formation time, which gives access to the duration of the branching process. To get this, all what one  needs to do is, as in \eqn{tauf}, relate   $\p-z\q$ to the transverse momentum acquired by multiple scattering during $\tau_{_{\rm br}}$. That is, $(\p-z\q)^2\simeq\hat q\tau_{_{\rm br}}$, which together with 
the above expression for $\Delta E=1/\tau_{_{\rm br}}$ provides an estimate for the 
corresponding branching time:
\beq
\tau_{_{\rm br}}=\sqrt{\frac{z(1-z)p^+}{\hat q}}\,,
 \eeq
which generalizes Eq.~(\ref{tauf}) by the replacement of $\omega\equiv zp^+$ by $z(1-z)p^+$ (note that the latter quantity can be interpreted as the reduced mass for the effective non-relativistic motion of the two produced gluons in the transverse plane). This makes it clear that Eq.~(\ref{tauf}) holds  only for asymmetric splitting with  $z\ll 1$.
A slightly more precise estimate for $\tau_{_{\rm br}}$  will be given in Sect.~\ref{splitting} (Eq.~(\ref{Omega})).

%%%%%%%%%%%%%%%%%%%%%%%%%%%%%%%%%
\section{General structure of the in-medium gluon branching}\label{sec:branch}
%%%%%%%%%%%%%%%%%%%%%%%%%%%%%%%%%

In this section we  analyze the main structure of the calculation  of the gluon branching, and show how it can be conveniently divided into three main stages:  (i) production of the gluon at time $t_0$ via some local (hard) process, followed by the propagation of the gluon in the medium; (ii) the  process where the gluon splits into two gluons; (iii) finally the  propagation of the two new gluons until the end of the medium at coordinate  $x^+=t_L\equiv \sqrt{2} L$. 

\begin{figure}[htbp]
\begin{center}

\includegraphics[width=7cm]{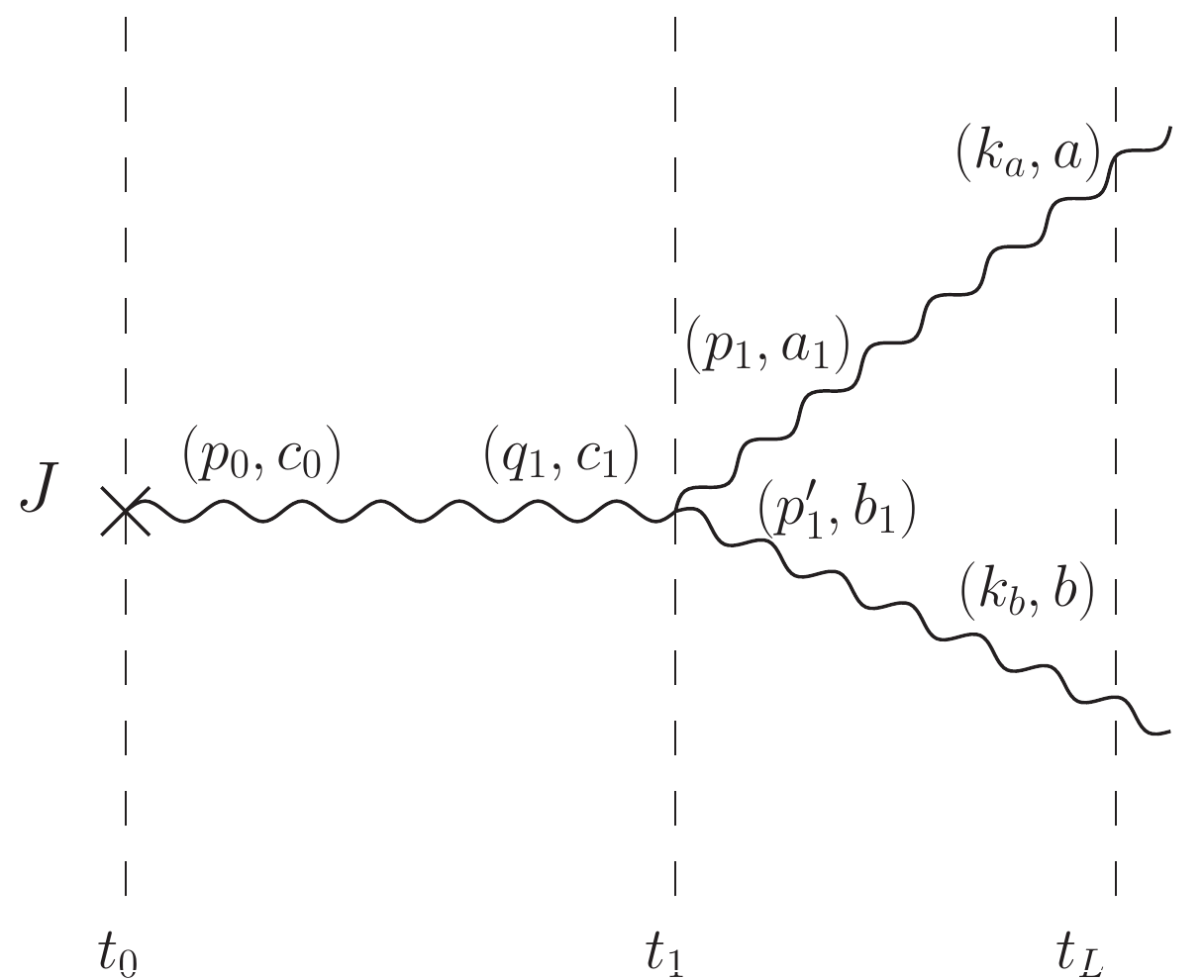}
\caption{Graphical illustration of the amplitude for one gluon splitting. The initial gluon produced at time $t_0$ with momentum $\p_0$ and color $c_0$ splits at time $t_1$ into two gluons with momenta $\p_1$ and $\p'_1$ ($=\q_1-\p_1$) respectively. Note that  the + component of the momenta is conserved both in the propagation and in the splitting, so that $k_a^+=zp_0^+$, $k^+_b=(1-z) p_0^+$.}\label{fig:bblock1}
\end{center}
\end{figure}

\subsection{The amplitude}
The amplitude for the branching process is illustrated in Fig.~\ref{fig:bblock1}: a gluon is created at time $t_0$ with color $c_0$ and momentum $(p_0^+,\p_0)$ out of the current $J^{c_0}(p_0^+,\p_0)=(\p_0\,c_0|J(t_0)|0)$; it then propagates in the medium where it acquires a transverse momentum  $\q_1-\p_0$, changes color form $c_0$ to $c_1$, and eventually splits into two gluons at time $t_1$. The newly produced gluons then propagates through the medium until the time  $t_L$, where they are observed with momenta $\k_a,\k_b$, colors $a,b$ and polarization $\lambda_a,\lambda_b$. Using the elements described in Appendix \ref{app:glueprop},
together with compact, matrix, notations similar to those already in the previous section (cf.
\eqn{Mone}) one easily gets the following expression for this amplitude:
 \beq
{\cal M}^{ab}_{\lambda_a,\lambda_b}(k_a^+\, \k_a,k^+_b\,\k_b)&=&\frac{\rme^{i(k^-_a+k^-_b)t_L}  }{2p_0^+}\,\int_{\p_0,\p_1,\q_1,\p_1'}\,\epsilon^j_{\lambda_b}(\k_b)\,\epsilon^i_{\lambda_a}(\k_a)\,\nn &\times&\ \int_{t_0}^{t_L} \rmd t_1\,(\k_a\,a;\k_b\,b|{\cal G}(t_L,t_1){\cal G}(t_L,t_1)\,| \p_1\, a_1;\p_1'\,b_1)\nn
&\times& (\p_1\,a_1;\p_1'\,b_1|\, \Gamma^{ijl}\,|\q_1\,c_1)(\q_1\,c_1|\, {\cal G}(t_1,t_0)|\p_0\,c_0)\,J^{l,c_0}(p^+_0,\p_0)\, ,\nn
\eeq
with $k^-_a=\k_a^2/(2k^+_a)$, $k^-_b=\k_a^2/(2k^+_b)$ and $z=k_a^+/p^+_0$. A summation over repeated discrete indices is implied. 
We have defined
 \begin{align}
&(\k_a\,a;\k_b\,b|\,{\cal G}(t_L,t_1){\cal G}(t_L,t_1)\,| \p_1 \,a_1;\p_1'\,b_1)
\equiv (\k_a|{\cal G}^{aa_1}(t_L,t_1)\,| \p_1 )( \k_b  |\,{\cal G}^{bb_1}(t_L,t_1)\,| \p_1')\,,\nn
& (\p_1\,a_1;\p_1'\,b_1|\, \Gamma^{ijl}\,|\q_1\,c_1)
\equiv (2\pi)^2\delta(\p_1+\p'_1-\q_1)\,2g\,f^{a_1b_1c_1}\Gamma^{ijl}(\p_1-z\q_1,z)\,,
 \end{align}
and we denote indifferently the matrix elements of the single propagator by 
$(\k_a|{\cal G}^{aa_1}(t_L,t_1)\,| \p_1 )$ or by $(\k_a\,a|{\cal G}(t_L,t_1)\,| \p_1\,a_1 )$, 
and similarly   $(\p_1\,a_1;\p_1'\,b_1|\, \Gamma^{ijl}\,|\q_1\,c_1)
=(\p_1;\p_1'|\, \Gamma^{ijl}_{a_1b_1c_1}\,|\q_1)$. 

\begin{figure}[htbp]
\begin{center}

\includegraphics[width=10cm]{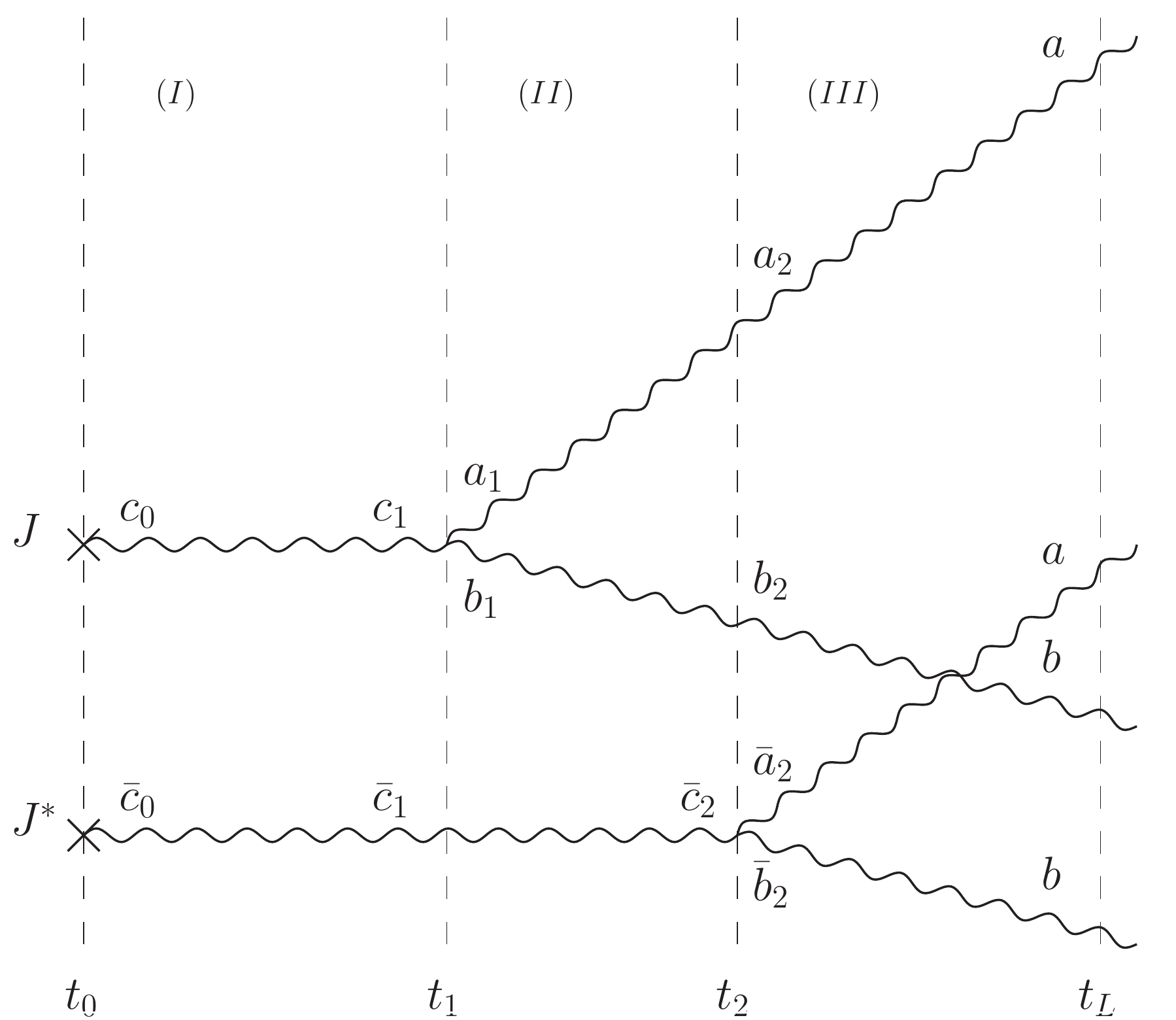}

\caption{The three regions that are involved in the description of a gluon splitting in a medium.  Time flows from left to right and the amplitude (upper part) is drawn together with the complex conjugate amplitude (lower part). The + momentum is conserved in the splitting. The various indices at each ends of the gluon lines (wavy lines) are color indices. The times separating the various regions are light-cone times: $t_1$ and $t_2$ are the light-cone times where the gluon splitting occurs in the amplitude and the complex conjugate amplitude, respectively. The indices $a,b,c$ are color indices, with each gluon carrying a given set of color indices that involve the same letter: thus `gluon a' can be seen in the various colors $a, a_1, a_2 $, etc.
 }\label{fig:bblock}
\end{center}
\end{figure}

To calculate the cross section, we have to multiply the amplitude by its complex conjugate amplitude. 
In order to analyze the effect of the medium average, it is convenient to draw the amplitude and the conjugate amplitude in the same diagram, one just below the other as shown in Fig. \ref{fig:bblock} (and in Fig.~\ref{amplitsqonegluon} of the previous section). The upper part of the diagram corresponds to the amplitude and the lower part corresponds to the conjugate amplitude, with time flowing from left to right identically in the amplitude and its complex conjugate.  Gluons which are drawn parallel to each other represent the same physical gluon (in the amplitude and conjugate amplitude, respectively), so they have the same momenta and same color at $t=t_L$. For definiteness, we have chosen the splitting to occur first in the amplitude and then in the conjugate amplitude (i.e., $t_1<t_2$), the inverse ordering 
can be accounted for by taking 2 times the real part when calculating the cross section (see Eq.~(\ref{Sigma2}) below). One may then identify three distinct time regions that correspond respectively to the propagation between $t_0$ and $t_1$ (region I), between $t_1$ and $t_2$ (region II), and between $t_2$ and $t_L$ (region III). To facilitate the grouping of terms in the different regions, we make repeated use of the identity (\ref{convprop}), writing for instance (in matrix notations)
 ${\cal G}(t_L,t_1)={\cal G}(t_L,t_2){\cal G}(t_2,t_1)$, 
 or  ${\cal G}^\dag(t_0,t_2)={\cal G}^\dag(t_0,t_1){\cal G}^\dag(t_1,t_2)$. 
Next, one introduces closure relations at appropriate times, over complete set of states 
in the transverse plane and in color space. The matrix element in  the amplitude 
can then be written as (see Figs.~\ref{fig:bblock} and \ref{fig:amp-1} for pictorial representations)
\beq
&&(\k_a;\k_b|{\cal G}^{aa_2}(t_L,t_2){\cal G}^{a_2a_1}(t_2,t_1)\,{\cal G}^{bb_2}(t_L,t_2){\cal G}^{b_2b_1}(t_2,t_1) \,\Gamma^{ijl}_{a_1b_1c_1}\,{\cal G}^{c_1c_0}(t_1,t_0)|\p_0)\nn
&&=\int_{\q_2,\p_1,\p'_1,\q'_2}(\k_a|{\cal G}^{aa_2}(t_L,t_2)|\q_2)(\q_2|{\cal G}^{a_2a_1}(t_2,t_1)|\p_1)(\k_b|{\cal G}^{bb_2}(t_L,t_2)|\q'_2)(\q'_2|    {\cal G}^{b_2b_1}(t_2,t_1)|\p'_1)     \nn
&&\times     (\p_1;\p'_1|\Gamma^{ijl}_{a_1b_1c_1}|\q_1)(\q_1|{\cal G}^{c_1c_0}(t_1,t_0)|\p_0),
\eeq
and that in the complex conjugate amplitude as
\beq
&& (\bar \p_0|{\cal G}^{\dag\,\bar c_0\bar c_1}(t_0,t_1){\cal G}^{\dag\,\bar c_1\bar c_2}(t_1,t_2)\,\Gamma^{\dagger\; \bar{i}jl}_{\bar a_2\bar b_2\bar c_2}\,{\cal G}^{\dag\,\bar a_2 a}(t_2,t_L) {\cal G}^{\dag\,\bar b_2 b}(t_2,t_L)  |\k_a;\k_b)\nn
&&=\int_{\bar\q_1,\bar \q_2,\p_2,\p'_2}(\bar p_0|{\cal G}^{\dag\,\bar c_0\bar c_1}(t_0,t_1)|\bar\q_1)(\bar\q_1|{\cal G}^{\dag\,\bar c_1\bar c_2}(t_1,t_2)|\bar\q_2)(\bar\q_2|\Gamma^{\dagger\; \bar{i}jl}_{\bar a_2\bar b_2\bar c_2}|\p_2;\p'_2)\nn
&&\times (\p_2|{\cal G}^{\dag\,\bar a_2 a}(t_2,t_L)|\k_a)(\p'_2| {\cal G}^{\dag\,\bar b_2 b}(t_2,t_L)  |\k_b).
\eeq
One can now multiply the matrix elements of the amplitude and its complex conjugate, and regroup terms in the various regions. One writes the result in the following way
\beq\label{3lines}
&&   (\k_a|{\cal G}^{aa_2}(t_L,t_2)|\q_2)  (\k_b|{\cal G}^{bb_2}(t_L,t_2)|\q'_2)(\p_2|{\cal G}^{\dagger\,\bar a_2 a}(t_2,t_L)|\k_a)(\p'_2|{\cal G}^{\dagger\,\bar b_2 b}(t_2,t_L)|\k_b)(\bar\q_2|\Gamma_{\bar a_2\bar b_2 \bar c_2}^{\dagger\; \bar{i}jl}|\p_2;\p'_2)\nn
&& \times(\q_2|  {\cal G}^{a_2a_1}(t_2,t_1)|\p_1)  (\q'_2|{\cal G}^{b_2b_1}(t_2,t_1)|\p'_1)(\bar\q_1|{\cal G}^{\dagger  \,\bar c_1\bar c_2}(t_1,t_2)|\bar\q_2)(\p_1;\p'_1|\Gamma_{a_1b_1c_1}^{ijl}|\q_1)\nn
&&\times (\q_1| {\cal G}^{c_1c_0}(t_1,t_0)|\p_0)(\bar\p_0|{\cal G}^{\dagger\,\bar c_0 \bar c_1} (t_0,t_1)|\bar\q_1),\nn
\eeq
where the three lines in the expression above correspond  to the three regions III, II, I, respectively, from top to bottom. At this point, we are ready to discuss the medium average and identify the simple, factorized, color structure of the three regions. 
%or, by introducing the $n$-point functions
%\beq
%&&  (\k_a a , \k_b b,\bar\p_2 \bar a_2 ,\bar\r_2 \bar b_2  | S^{(4)}(t_L,t_2)|\q_2 a_2, \r_2 b_2, \k_a a, \k_b b) (\bar\p_2\bar \r_2|\Gamma_{\bar a_2\bar b_2 \bar c_2}^{\dagger\;ijl}(t_2)|\bar\q_2)\nn
%&& \times(\q_2 a_2, \r_2 b_2 ,\bar\q_1\bar c_1|S^{(3)}(t_2,t_1)|\p_1 a_1,\r_1 b_1,\bar\q_2  \bar c_2)(\p_1\r_1|\Gamma_{a_1b_1c_1}^{ijl}(t_1)|\q_1)\nn
%&&\times (\q_1 c_1, \bar\p_0 c_0|S^{(2)}(t_1,t_0)|\p\bar c_0,\bar\q_1 \bar c_1)
%\eeq
%

%%%%%%%%%%%%%%%%%%%%%
\subsection{Color structure}\label{sec:colstr}
%%%%%%%%%%%%%%%%%%%%%

%For the analysis in this section, we consider only the elements that are essential to determine the color structure of the process. We will consider only the scalar part of the propagator, and retain from the  vertices,  only the color factor $f^{abc}$. We then find that the medium average we are interested in, with the appropriate color connections, takes the following form (as can be read off directly on Fig.~\ref{fig:bblock})
%\beq
%f^{a_1b_1c_1}f^{\bar a_2\bar b_2\bar c_2}&& \langle  {\cal G}_{aa_1}(Y,Z_1) {\cal G}_{bb_1}(Y,Z_1){\cal G}_{c_1c_0}(Z_1,X)\nonumber\\
%&&\times{\cal G}^\dagger_{\bar c_0\bar c_2 } (X,Z_2){\cal G}^\dagger_{\bar a_2 a}(Z_2,Y) {\cal G}^\dagger_{\bar b_2 b}(Z_2,Y) \rangle.\label{medav}
%\eeq
%While in the diagram above time goes from left to right, we use a matrix notation for the propagators. Thus, the arguments in ${\cal G}$ and in ${\cal G}^\dagger$ are read from right to left, and correspond to propagation forward in time for ${\cal G}$ and backward in time for ${\cal G}^\dagger$. That is for instance, ${\cal G}_{aa_1}(Y,Z_1)$ propagates gluon $a$ from $Z_1$ to $Y$, with color changing from $a_1$ to $a$, while ${\cal G}^\dagger_{\bar b_2 b}(Z_2,Y)$ propagates gluon $b$ from $Y$ to $Z_2$ (i.e., backward in time) with color changing along the path from $b$ to $\bar b_2$. The arguments in the propagator are those of the mixed representation, e.g., $Z_1=(t_1,\z_1)$. The $+$ components of the momenta are not indicated since these are trivially conserved: thus $k^+_c=k^+_a+k^+_b$. 

\begin{figure}[htbp]
\begin{center}
\includegraphics[width=12cm]{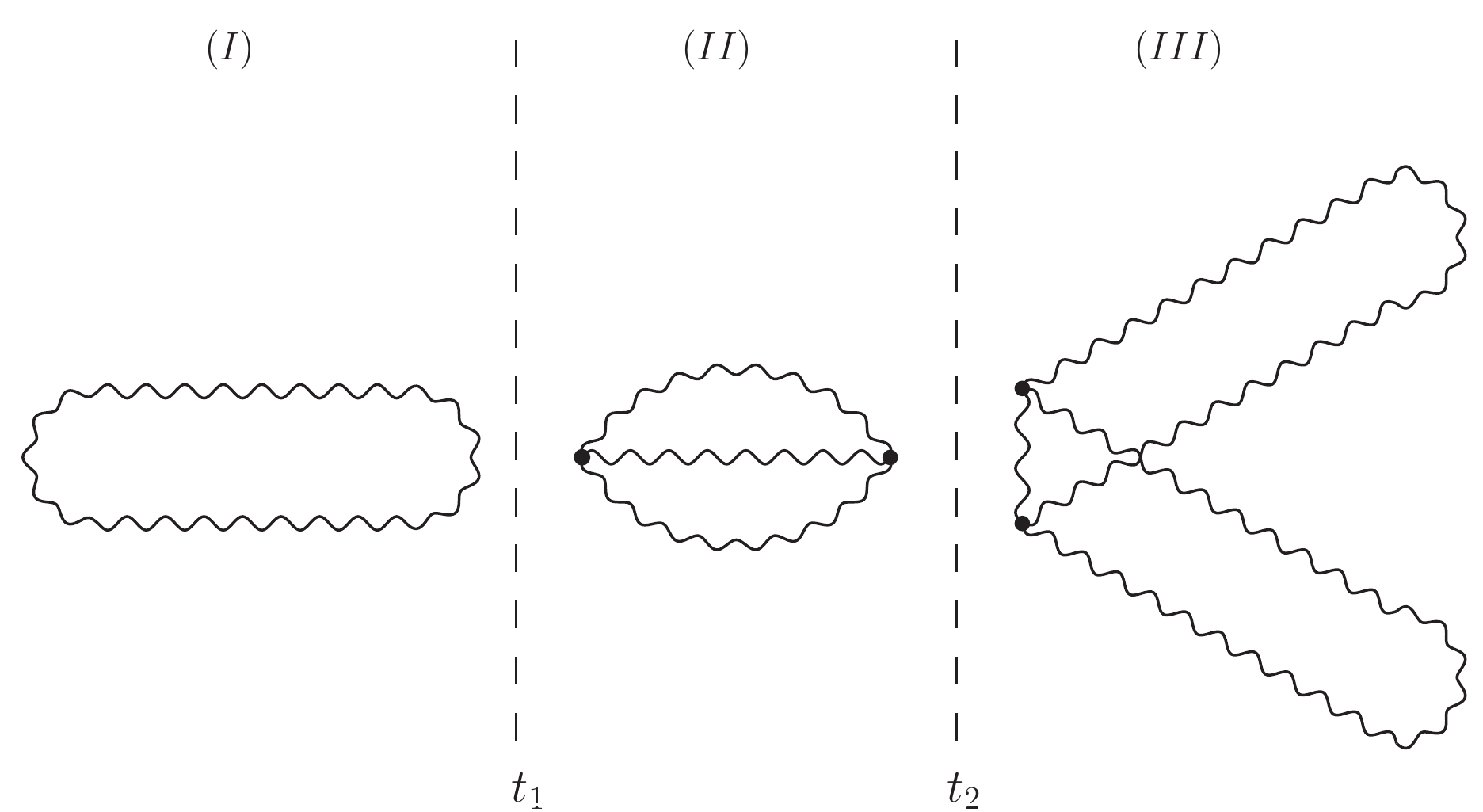}

\caption{The color connections that are responsible for making the system of gluons a color singlet in each of the  three regions. The dotted vertices represents  $f$ symbols, $f^{abc}$, while the
`vertical' connections among them represents Kronecker delta's,  $\delta_{ab}$.}\label{fig:fullsinglet}
\end{center}
\end{figure}

As already noted in the previous section, the contractions of  background fields that are involved in the medium average  preserve the overall color of the system, and since all such contractions
 are instantaneous,  the overall color state of the system comprising all gluons in the amplitude and its complex conjugate is conserved at all times. Since, as we shall verify shortly, the overall color state is a color singlet at $t=t_L$,   the gluons traversing the medium are in an overall color singlet at any given time. This  property  allows us to perform the medium averages explicitly by reducing considerably the number of color states to be considered along the calculation:  all one needs to do is to find the appropriate singlet state in which the system is at the separation points $t_1$ and $t_2$. The resulting color structure is illustrated in Fig.~\ref{fig:fullsinglet}.

That the system of gluons is in a color singlet state at $t=t_L$ is easy to see. Indeed the colors of the gluons are identical in the amplitude and in the complex conjugate amplitude, and since we do not observe the colors of the produced gluons, these colors are summed over, projecting the two pairs of gluons $a$ and $b$ onto  on color singlets. Since the system remains a color singlet at all time, it is a singlet at time $t_0$, corresponding to the beginning of region I, with which we start our detailed analysis. 

Consider then region I, that is, the third line of Eq.~(\ref{3lines}). As we have just argued, the gluon system is in a color singlet state. Hence at time $t_0$, the gluon carries  the same color index
in the direct and complex conjugate amplitude ($\bar c_0=c_0$), and then the medium averaging yields
\beq\label{rIK2}
\nn
\delta^{c_0\bar c_0}\left\langle (\q_1| {\cal G}^{c_1c_0}(t_1,t_0)|\p_0)(\bar\p_0|{\cal G}^{\dagger\,\bar c_0 \bar c_1} (t_0,t_1)|\bar\q_1)\right\rangle=\delta^{c_1 \bar c_1} (\q_1;\bar\q_1|S^{(2)}(t_1,t_0)|\p_0;\bar\p_0),
\eeq
where the quantity $S^{(2)}(t_1,t_0)$ has already been introduced in \eqn{S2def}
and evaluated in Eqs.~\eqref{S2P}--\eqref{PD}.
Note that in Eq.~(\ref{rIK2}) a sum over $c_0$ is performed. However, this sum is not free since the current also carries the same color index. However, since each term of the sum is eventually independent of $c_0$, we can correct the result by simply  dividing by $N_c^2-1$. In summary, the contribution of region I after medium average is
\beq\label{rIK3}
\frac{ \delta^{c_1\bar c_1} }{N_c^2-1} (\q_1;\bar\q_1|S^{(2)}(t_1,t_0)|\p_0;\bar\p_0).
\eeq

Now we turn our attention to region II, that is the second line of Eq.~(\ref{3lines}). We can use the Kronecker delta $\delta^{c_1\bar c_1}$ 
from Eq.~(\ref{rIK3}) and recall that the color structure of the vertex at $t_1$ is of the form $f^{a_1b_1c_1}$ to realize that the relevant correlator for this region is
\beq
f^{a_1b_1c_1}\big\langle  (\q_2|  {\cal G}^{a_2a_1}(t_2,t_1)|\p_1)  (\q'_2|{\cal G}^{b_2b_1}(t_2,t_1)|\p'_1)(\bar\q_1|{\cal G}^{\dagger  \, c_1\bar c_2}(t_1,t_2)|\bar\q_2)\big \rangle,\label{rII}
\eeq
from where it is clear that we have a color singlet state at $t_1$ by contracting the three initial color indices with an $f$ symbol. This color structure is unchanged by the medium average in this region given that there is no other way to combine these three gluons into a singlet state. It follows therefore that the color indices at $t_2$ must  also be contracted with an $f$ symbol. More specifically, this means that Eq. (\ref{rII}) takes the form
\beq
&& f^{a_1b_1c_1}\left\langle  (\q_2|  {\cal G}^{a_2a_1}(t_2,t_1)|\p_1)  (\q'_2|{\cal G}^{b_2b_1}(t_2,t_1)|\p'_1)(\bar\q_1|{\cal G}^{\dagger  \,c_1\bar c_2}(t_1,t_2)|\bar\q_2) \right\rangle\nn
&\equiv&f^{a_2b_2\bar c_2} (\q_2\q'_2;\bar\q_2|S^{(3)}(t_2,t_1)|\p_1\p'_1;\bar\q_1),\label{defs3}
\eeq
with
\beq
&&(\q_2\q'_2;\bar\q_2|S^{(3)}(t_2,t_1)|\p_1\p'_1;\bar\q_1)\nn
&=&\frac{1}{N_c(N_c^2-1)}\left\langle  (\q_2|  {\cal G}^{a_2'a_1}(t_2,t_1)|\p_1)
  (\q'_2|{\cal G}^{b_2'b_1}(t_2,t_1)|\p'_1)(\bar\q_1|{\cal G}^{\dagger  \, c_1\bar c_2'}(t_1,t_2)|\bar\q_2)\right\rangle f^{a'_2b'_2\bar c_2'}f^{a_1b_1c_1}\,.\nonumber\\
\eeq
 Due to the simple color structure of this object, one can calculate explicitly the medium average of  the Wilson lines inside the propagators,  by resumming the contractions of each pairs of gluons independently. One gets
\beq\label{3ptW}
C^{(3)}_g&=&\frac{1}{N_c(N_c^2-1)}f^{a_1b_1c_1}\langle  \tilde U_{a'_2a_1}(t_2,t_1;\r_a)  \tilde U_{b_2'b_1}(t_2,t_1;\r_b)\tilde U^\dagger_{c_1\bar c_2' } (t_1,t_2;\r_c)\rangle f^{a'_2b'_2\bar c_2'}\nn
&=&\exp\left\{-\frac{N_c\, n}{4}\int_{t_1}^{t_2} \rmd t\,  \left[\sigma(\r_b-\r_c)+\sigma(\r_b-\r_a)+\sigma(\r_a-\r_c)\right]\right\}.
\eeq
The color factor in front of the leading order dipole cross section can be calculated from the appropriate contraction of the color factors in the adjoint representation $f^{abc}f^{cde}f^{efa}=-\frac{N_c}{2}f^{bdf}$. The calculation  of the path integral needed to complete the evaluation of $S^{(3)}$  is done in Appendix \ref{app:3p}.

Turning our attention to region III, the first line  of Eq.~(\ref{3lines}), we see that the four gluons form a singlet state through the combination of two $f$'s, one coming from Eq. (\ref{defs3}) and the other from the vertex at $t_2$ (see Eq.~(\ref{3lines})). As opposed to the previous cases, there are several ways of combining four gluons to form a singlet state. The final color state is given by the identification of two pairs of gluons as the conjugate of each other as explicitly shown by setting the color indices equal for the respective propagators at $t_L$ (as already discussed, $a=\bar a$, $b=\bar b$). The relevant correlator for region III is therefore
\beq
&&f^{a_2b_2\bar c_2}f^{\bar a_2\bar b_2 \bar c_2}\langle   (\k_a|{\cal G}^{aa_2}(t_L,t_2)|\q_2)  (\k_b|{\cal G}^{bb_2}(t_L,t_2)|\q'_2)(\p_2|{\cal G}^{\dagger\,\bar a_2 a}(t_2,t_L)|\k_a)(\p'_2|{\cal G}^{\dagger\,\bar b_2 b}(t_2,t_L)|\k_b)\rangle\nn&&\equiv N_c(N_c^2-1)(\k_a\k_b;\k_a\k_b|S^{(4)}(t_L,t_2)|\q_2\q'_2;\p_2\p'_2).\label{fourpoint}
\eeq
The fact that there is no unique way to combine the four gluons into color singlets, in contrast to what happened for the 2 and 3-point functions, prevents us to give an explicit expression for the average of the Wilson lines in the propagators analogous to Eq.~(\ref{3ptW}) for the 3-point function. A more detailed analysis is required, which will be carried out in section \ref{sec:twogluonprop}.

%%%%%%%%%%%%%%%%%%%%%
\subsection{Momentum structure}\label{sec:momstr}
%%%%%%%%%%%%%%%%%%%%%

\begin{figure}
\centering
\includegraphics[width=9cm]{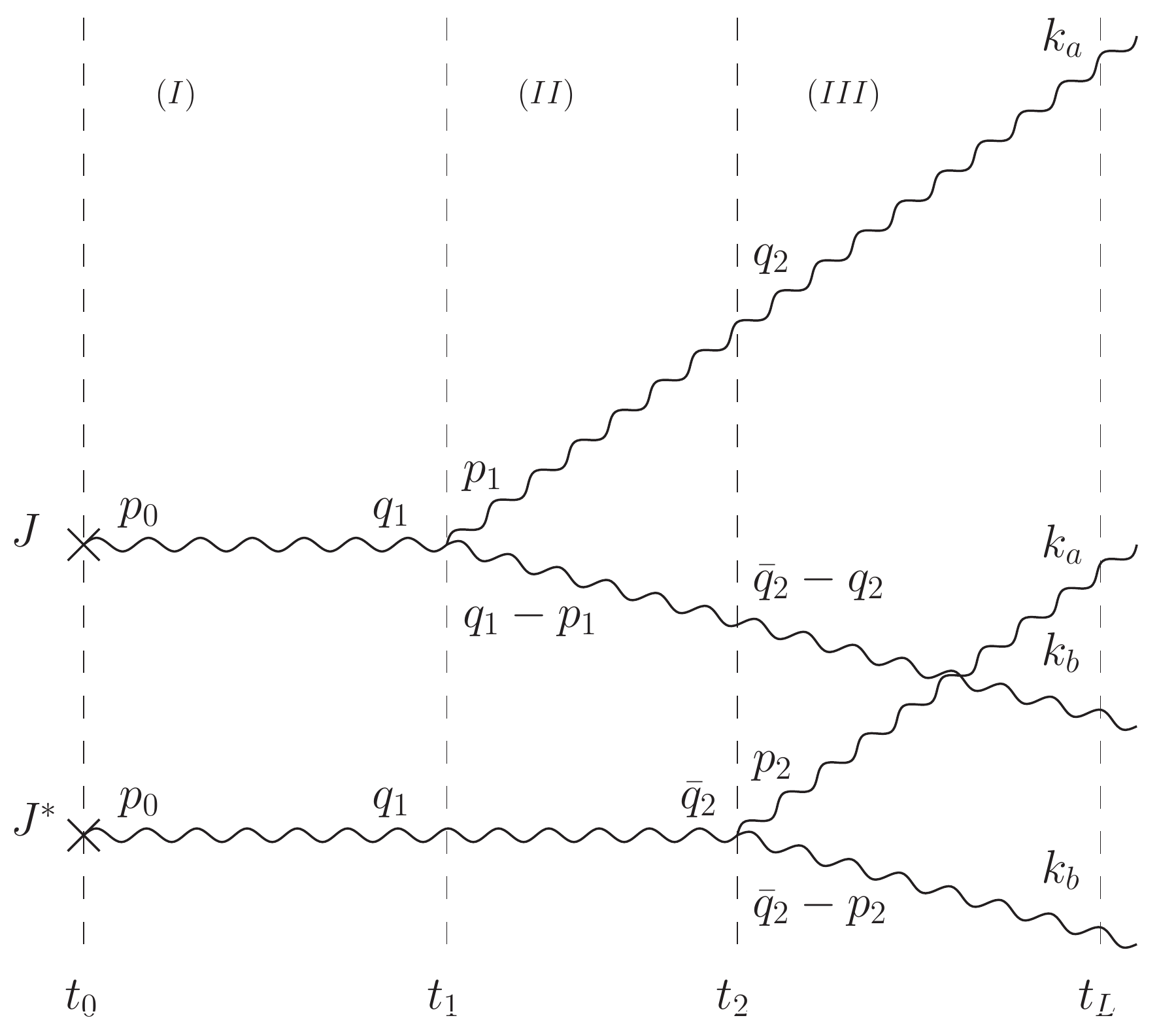}
\caption{Flow of momenta in the amplitude for one gluon splitting and its complex conjugate. At any given time, the sum of momenta in the amplitude equals the sum of momenta in the complex conjugate amplitude. The + component of the momentum is conserved, $k^+_a+k^+_b=p_0^+$, and is not indicated.}
\label{fig:amp-1}
\end{figure}

The fact that the correlator of the random background field depends only on the difference of the transverse coordinates leads to total transverse momentum conservation.  It follows that at each instant of time, the sum of momenta in the amplitude equals that of momenta in the complex conjugate amplitude, since this is the case at the final time $t_L$. As explicitly shown in Appendix \ref{app:pathint}, each of the $n$-point function contains a corresponding $\delta$-function expressing this property. There are also additional $\delta$-functions hidden in the vertices $\Gamma$ which express the conservation of transverse momentum in the local splitting. Using all these $\delta$-functions, one can then reduce considerably the number of momentum integrations when building the cross section for the branching process. The resulting flow of momenta is displayed in Fig.~\ref{fig:amp-1}, which also  lists  the independent momenta that we use. We denote with a tilde the $n$-point functions from which the delta function for momentum conservation has been factored out.

There are further simplifications that reduce the number of independent momentum variables. For instance,  the 2-point function depends only on the difference of two momenta. Similarly for the 3-point function. Also the vertex depends only on $\p-z\q$, as visible on \eqn{vertex3g}. These extra simplifications will be exploited when needed.

We are now in a position where we can write the  cross section for the two gluon production process.
After performing the medium average and summing over polarizations, one obtains
\beq\label{Sigma2}
&&\frac{\rmd^2\sigma}{\rmd\Omega_{k_a}\rmd\Omega_{k_b}}=
\frac{g^2N_c}{(2p^+_0)^2}\,2\Re e\,\int_{t_0}^{t_L}\rmd t_2\int_{t_0}^{t_2}\rmd t_1\,\int_{\p_0, \p_1\q_1\bar\q_2 \p_2\q_2} \;  \,\Gamma^{ijl}(\p_1-z\q_1,z)\,\Gamma^{\bar i jl}(\p_2-z\bar\q_2,z)\nn
&&\quad \times\,(\k_a\k_b;\k_a\k_b|\tilde S^{(4)} (t_L,t_2)|\q_2,\bar\q_2-\q_2;\p_2,\bar\q_2-\p_2)(\q_2,\bar\q_2-\q_2;\bar\q_2| \tilde S^{(3)}(t_2,t_1)|\p_1,\q_1-\p_1;\q_1)\nn
&&\quad\times(\q_1;\q_1|\tilde S^{(2)}(t_1,t_0)|\p_0;\p_0)\, J^{i,c_0}(p_0^+,\p_0) J^{*\bar i,c_0}(p_0^+,\p_0).
\eeq
In this equation, we used the compact notation $\rmd\Omega_k\equiv (2\pi)^{-3} \rmd^2\k\, 
\rmd k^+/2k^+$ for the element in phase--space and in the r.h.s. it is understood that 
$p_0^+=k_a^++k_b^+$ and $z=k_a^+/p_0^+$.

%%%%%%%%%%%%%%%%%%%%%
\subsection{Qualitative comments}\label{sec:qualcom}
%%%%%%%%%%%%%%%%%%%%%

Before getting into the details of the evaluation of the various $n$-point functions that enter the expression above, we find it useful  to make a few qualitative observations that will lead us to 
better appreciate the physical content of the calculations to be performed in subsequent sections.

As we have seen, the $n$-point functions that appear in the various regions contain a medium average of a product of Wilson lines in an overall color singlet state,  inside a path integration. The way to proceed in the explicit average evaluations is to calculate the average of a particular combination of Wilson lines with fixed paths in transverse coordinate space, then perform the integration over these paths. In general,  the medium average of the Wilson lines decays exponentially with the distance between particles constituting a color neutral state, typically as $\Delta \x^2\gtrsim 2/(\hat q\Delta t)$ (see e.g. Eq.~(\ref{Dgg2}) for the 2-point function). That is,  one gets a significant suppression factor whenever the particles are farther apart than the inverse of the typical transverse momentum acquired by multiple collisions. On the other hand, in the path integral, the gluons trajectories diffuse in transverse coordinate space, which typically increases the average distances between the gluons so that $\Delta \x^2\sim{\Delta t}/{\omega}$. 
Combining these two estimates, one gets that in order avoid a large exponential suppression factor the longitudinal extent of the medium average in consideration  should be limited to $
\Delta t\lesssim\sqrt{{2\omega}/{\hat q}}=\tau_{_{\rm br}}\ll L$.

The argument above does not always hold since it assumes that the diffusion in transverse space is independent for all the gluons, so that the mean distance between them is monotonically increasing. However, when two  gluons  are conjugate to each other their diffusions in transverse coordinate space are correlated, indeed their transverse separation stays constant (see Appendix \ref{app:pathint}). This implies that the longitudinal extent of region I is not constrained by the argument above. On the other hand, the three gluons in region II  propagate independently. This, combined with the fact that there is only one way to form a color neutral state with three gluons, implies that this region II must be short-lived, i.e.,  $t_2-t_1\lesssim \tau_{_{\rm br}}$, in order to avoid suppression factors.

In region III we run again into the case where gluons can be paired with their respective conjugates and therefore do not propagate all independently. Moreover, by pairing the gluons this way, two color neutral subsystems are formed, which can propagate independently and be arbitrarily far away from each other without introducing suppression factors. Nevertheless, the general argument concerning
the suppression of correlations by medium rescattering still holds within a narrow region bordering region II, where all four gluons are still correlated. We shall find that the longitudinal extent of that
intermediate region is of order $\tau_{_{\rm br}}$, for the same basic reasons as for region II. Accordingly, we shall conclude that  the four-point correlator obtained for the medium average in region III is factorizable into two two-point correlators except for a parametrically small ($\sim \tau_{_{\rm br}}/ L)$ region right after the second splitting.

%%%%%%%%%%%%%%%%%%%%%%%%%%%%%%%%%
\section{Factorization of two-gluon propagation}\label{sec:twogluonprop}
%%%%%%%%%%%%%%%%%%%%%%%%%%%%%%%%%

We turn now to the  detailed analysis of region III. As already anticipated in the previous section, the main goal here is to show that the propagation of the two resulting gluons can be considered as independent and correlations among them are a subleading effect. In that case, all the effects of the splitting can be included inside the splitting kernel of region II, which will be calculated in detail in the next section.

We need to calculate explicitly the 4-point function $S^{(4)}(t_L,t_2)$. Since the definition of the scalar propagators entering the average in (\ref{fourpoint}) take a simpler form in coordinate space, it is convenient to calculate first the medium average of the four propagators in coordinate space with arbitrary endpoints, perform the Fourier transform, and then impose the necessary constraints in the momentum variables.

The coordinate space 4-point function under consideration takes the explicit form
\beq
&&(\y_a\y_b;\bar\y_a\bar\y_b|S^{(4)}(t_L,t_2)|\x_a\x_b;\bar\x_a,\bar\x_b)\nn
&=&\int {\cal D}\r_a {\cal D}\r_b {\cal D}\bar\r_a {\cal D}\bar\r_b \exp\left\{\frac{i}{2}\int_{t_2}^{t_L} \rmd t\,\left(k_a^+\dot{\r}^2_a+k_b^+\dot{\r}^2_b-k_a^+\dot{\bar\r}^2_a-k_b^+\dot{\bar\r}^2_b\right)  \right\}\nn
&&\times f^{a_2b_2\bar c_2}f^{\bar a_2\bar b_2\bar c_2}\langle  \tilde U_{aa_2}(t_L,t_2;\r_a)  \tilde U_{bb_2}(t_L,t_2;\r_b)\tilde U^\dagger_{\bar a_2a} (t_2,t_L;\bar\r_a)\tilde U^\dagger_{\bar b_2b} (t_2,t_L;\bar\r_b)\rangle,\label{4pcoor}
\eeq
where $\r_{a,b}(t_2)=\x_{a,b}, \r_{a,b}(t_L)=\y_{a,b}$ and similarly for the bar coordinates.

We start by first calculating the medium average of the Wilson lines before attempting to perform the path integrations. Since for the moment we are only concerned with what occurs in region III and all the propagators have the same longitudinal extent, longitudinal coordinates will be dropped from the Wilson lines, and only the transverse trajectories will be kept explicit.

The calculation of the correlator in the last line of \eqn{4pcoor} turns out to be extremely complex
even for our simple, Gaussian, model for the random background field. So, before we start the
calculation, it is useful to make some remarks about its general structure and the strategy that will be employed for the rest of the section. Since our goal in this section is to show that, to the level of desired accuracy, one can factorize this correlator into the propagation of two independent gluons, it will be convenient to split the evaluation into a `factorizable' piece and a `non-factorizable' piece where,
\beq
S^{(4)}(t_L,t_2)=S_{\rm fac}^{(4)}(t_L,t_2)+S_{\rm nfac}^{(4)}(t_L,t_2),
\eeq
with
\beq\label{facs4}
&&(\y_a\y_b;\bar\y_a\bar\y_b|S^{(4)}_{\rm fac}(t_L,t_2)|\x_a\x_b;\bar\x_a,\bar\x_b)\nn
&=&(\y_a;\bar\y_a|S^{(2)}(t_L,t_2)|\x_a;\bar\x_a)(\y_b;\bar\y_b|S^{(2)}(t_L,t_2)|\x_b;\bar\x_b),
\eeq
or, in terms of the momentum representation entering the expression for the cross section,
\beq
&&(\k_a\k_b;\k_a\k_b|\tilde S^{(4)}_{\rm fac} (t_L,t_2)|\q_2,\bar\q_2-\q_2;\p_2,\bar\q_2-\p_2)\nn
&=&(2\pi)^2\delta^{(2)}(\p_2-\q_2){\cal P}(\k_a-\q_2,t_L-t_2){\cal P}(\k_b-\bar\q_2+\q_2,t_L-t_2).\label{S4tildefac}
\eeq
The analysis below is therefore aimed to show explicitly that the non-factorizable pieces are suppressed with respect to the factorizable pieces and therefore it is only the expression in (\ref{S4tildefac}) which enters the result for the cross section of the two-gluon production process.

The complete evaluation of the correlator in (\ref{4pcoor}) is complicated because of the many singlet states that can be constructed with four gluons, but by transforming this expression into one with only Wilson lines in the fundamental representation it is possible to gain some insight about the transitions between the different color states. In order to do that, one makes use of the following identity relating Wilson lines in the adjoint representation with Wilson lines in the fundamental representation,
\beq
\tilde U_{ab}(\r)=2\text{Tr}\left[U^\dagger(\r)t^aU(\r)t^b\right].\label{adjtofund}
\eeq
After repeated use of this identity,  one can perform the color algebra and eliminate all the explicit color factors (see details in Appendix \ref{app:color}), arriving at
\beq
&& f^{a_2b_2\bar c_2}f^{\bar a_2\bar b_2\bar c_2}\langle  \tilde U_{aa_2}(\r_a)  \tilde U_{bb_2}(\r_b)\tilde U^\dagger_{\bar a_2a} (\bar\r_a)\tilde U^\dagger_{\bar b_2b} (\bar\r_b)\rangle\nn
&=&\frac{1}{2}\left\langle\text{Tr}\left[U(\r_a)U^\dagger(\bar\r_a)\right]\text{Tr}\left[U(\bar\r_b)U^\dagger(\r_b)\right]\text{Tr}\left[U^\dagger(\r_a)U(\r_b)U^\dagger(\bar\r_b)U(\bar\r_a)\right]\right.\nonumber\\
&&\left.-\text{Tr}\left[U(\r_a)U^\dagger(\bar\r_a)U(\r_b)U^\dagger(\bar\r_b)U(\bar\r_a)U^\dagger(\r_a)U(\bar\r_b)U^\dagger(\r_b)\right] + \text{h.c.}\right\rangle.\label{4pointfund}
\eeq

\begin{figure}
\centering
\includegraphics[width=\textwidth]{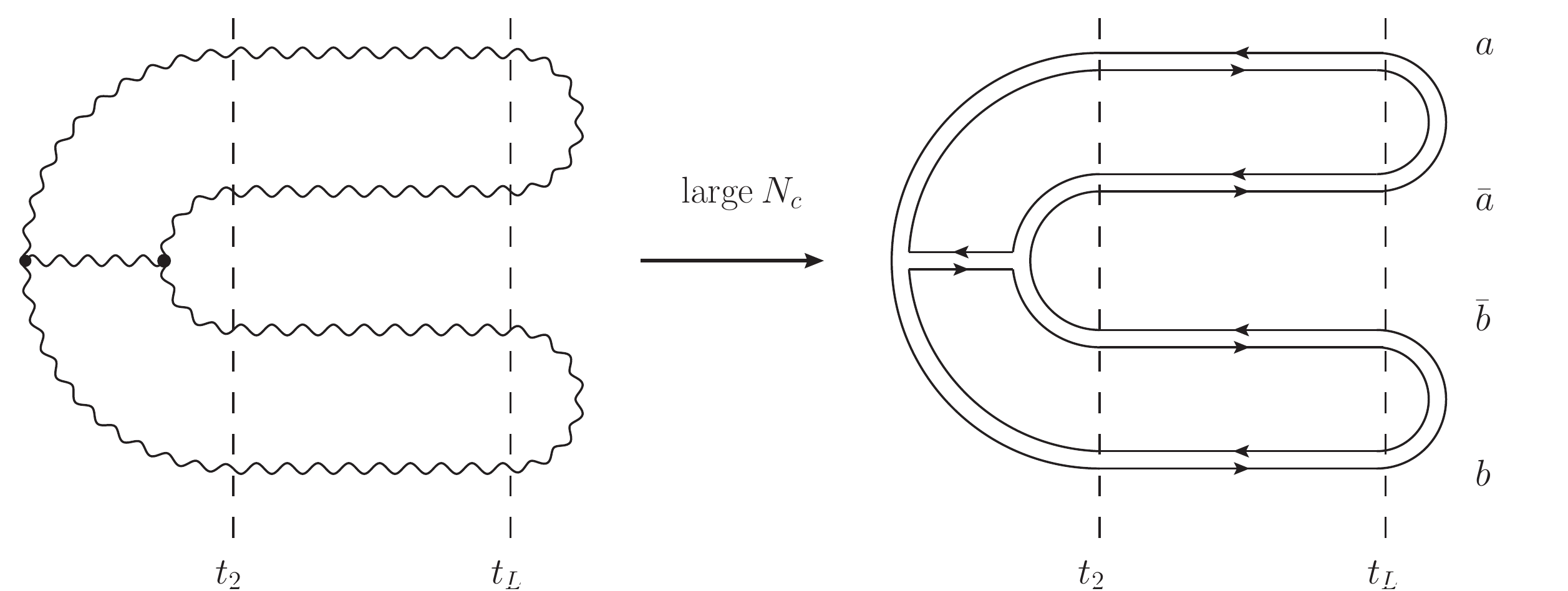}
\caption{Graphical representation of the correlator (\ref{4pointfund}) and its large-$N_c$ version (\ref{4plargenc}). The horizontal lines in between the vertical dashed lines represent the Wilson lines either in the adjoint (gluon lines) or fundamental (quark lines) representation. Lines in the right and left ends of the diagrams indicate color connections only, with gluon lines identifying adjoint color indices, quark lines identifying fundamental color indices, and three gluon vertices representing $f$ symbols.}
\label{fig:color4p}
\end{figure}

This may  not look like a real improvement: the number of Wilson lines has doubled and now we have several nontrivial terms to deal with! The advantage of this approach is that it greatly simplifies when one considers  the large-$N_c$ limit, a common approach used to simplify the calculations while capturing the essential physical behavior. In the fundamental representation, each color trace is proportional to $N_c$, from where it becomes clear that the first term will dominate over the second in the large-$N_c$ limit. Still in this limit, one can factorize the average of a product of traces as the 
product of the independent averages of each individual trace. Indeed, correlations between fields entering different traces are suppressed by inverse powers of $N_c$. By also using the fact that
all the considered correlators are real (at least, in our Gaussian model for the medium averages), 
we finally get the following expression for the dominant contribution at large $N_c$~:
\beq
&&f^{a_2b_2\bar c_2}f^{\bar a_2\bar b_2\bar c_2}\langle  \tilde U_{aa_2}(\r_a)  \tilde U_{bb_2}(\r_b)\tilde U^\dagger_{\bar a_2a} (\bar\r_a)\tilde U^\dagger_{\bar b_2b} (\bar\r_b)\rangle\nn
&=&\left\langle\text{Tr}\left[U(\r_a)U^\dagger(\bar\r_a)\right]\right\rangle\left\langle\text{Tr}\left[U(\bar\r_b)U^\dagger(\r_b)\right]\right\rangle\left\langle\text{Tr}\left[U^\dagger(\r_a)U(\r_b)U^\dagger(\bar\r_b)U(\bar\r_a)\right]\right\rangle.\label{4plargenc}
\eeq
This relation can be easily seen graphically by replacing the gluons with quark-antiquark pairs as shown in Fig. \ref{fig:color4p}, where only the planar part was kept. Each of the loops corresponds to one of the traces in (\ref{4plargenc}) and connections among them are suppressed.

The three factors in (\ref{4plargenc}) can then be evaluated independently. Two of them are quark dipoles which take the by now standard form (compare with \eqn{Dgg})
\beq
C_q^{(2)}(t_L,t_2;\r,\bar\r)\equiv\frac{1}{N_c}\left\langle\text{Tr}\left[U(\r)U^\dagger(\bar\r)\right]\right\rangle=\exp\left[-\frac{C_F n}{2}\int_{t_2}^{t_L}\rmd t\,\sigma(\r-\bar\r)\right].
\eeq

The third factor in (\ref{4plargenc}) has the form of a quadrupole amplitude. This kind of correlator has been studied in the literature, always for Wilson lines with fixed transverse coordinates while here we need to be able to evaluate it for arbitrary trajectories. A straightforward generalization, as it was
possible in the case of the dipole amplitude, is so far not available. On the other hand, the use of the large-$N_c$ limit helps greatly to further simplify the calculation and allows us to gain some insight into which terms are the important ones for the situation at hand.

There are only two independent singlet states that can be formed with two quark-antiquark pairs corresponding to the two possible ways of pairing quarks with antiquarks. Considering the way the color indices are contracted for the quadrupole amplitude, one can consider this quadrupole amplitude as the probability of switching from one singlet state to the other one, after undergoing multiple scatterings with the medium. For the  case at hand, it is clear that the natural causal way to consider the process 
is such that the two singlet states are made with the pairs of lines $a,b$ and respectively
$\bar a, \bar b$ in the initial state, and with the pairs $a,\bar a$ and $b,\bar b$ in the final state.

The two aforementioned singlet states are not orthogonal and therefore one can isolate a contribution where the system was initially in the final color state and all the interactions with the medium preserve this color state. In that case, one can consider the four-particle system as two separate quark dipoles, which do not see each other, yielding a product of two already familiar dipole amplitudes $C_q^{(2)}$. This type of contribution has the exact same structure as what was referred at the beginning of the section as the factorizable piece and the calculation below will show explicitly how one arrives at Eq. (\ref{facs4}).

\begin{figure}
\centering
\includegraphics[width=0.5\textwidth]{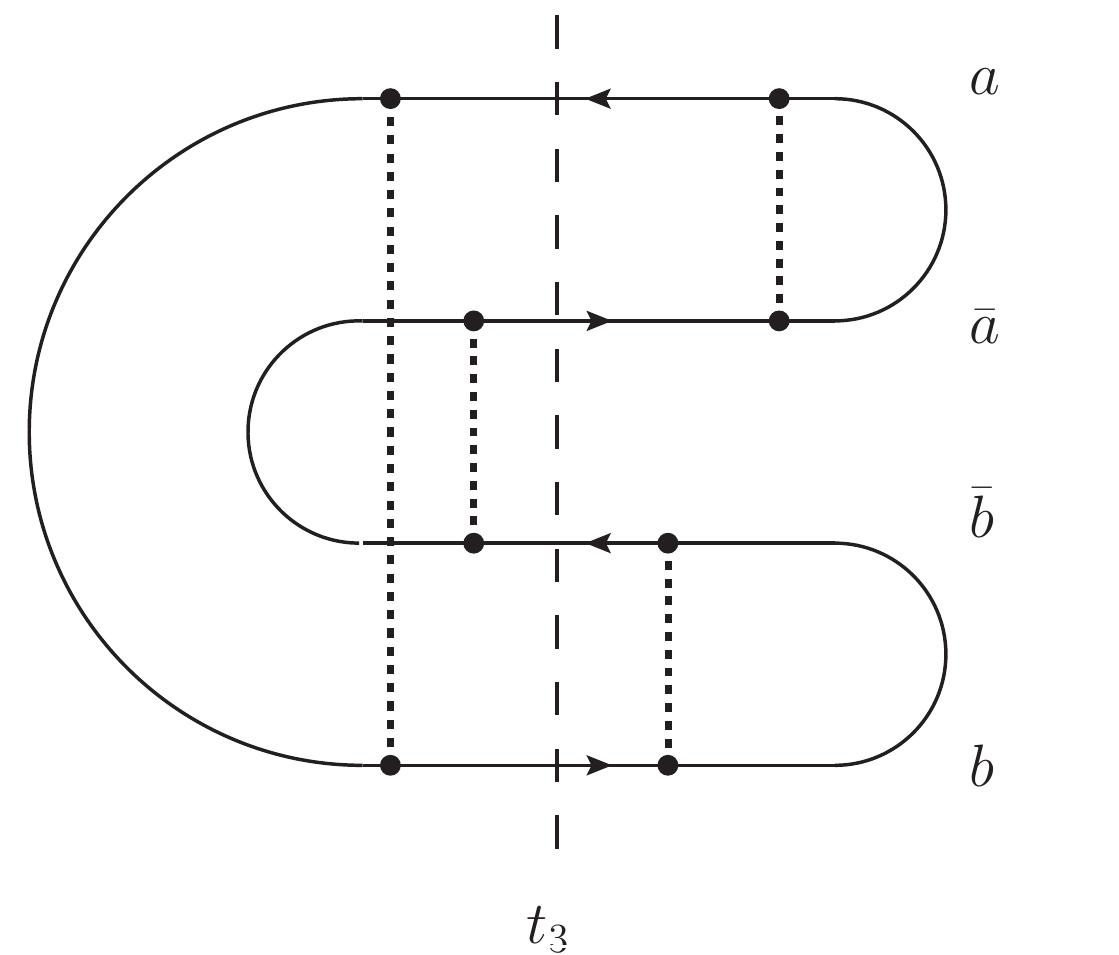}
\caption{Sample of the kind of interactions allowed at each end of the quadrupole amplitude. The dashed line in the middle represents the transition point in between the two regions where a different kind of contraction is allowed.}
\label{fig:4pfact}
\end{figure}

The non-factorizable piece consists of all the possible cases not accounted for in the factorizable piece, which are those where transitions between the color states are allowed during the interaction. In general one can have an arbitrary number of such transitions, but in the large-$N_c$ limit those are suppressed by inverse powers of $N_c$. In that case, one must take into account only contributions where there is only one transition between the two possible singlet states. All the diagrams which are not suppressed by inverse powers of $N_c$ have the general form depicted in Fig. \ref{fig:4pfact}:
 on the left side of the diagram only contractions between lines $a$ and $b$, or between lines $\bar a$ and $\bar b$ are allowed, while on the right side of the diagram only contractions between the lines $a$ and $\bar a$, or between lines $b$ and $\bar b$ are allowed.  These two regions are separated
by just one contraction performing the transition, which can connect either lines $a$ and $\bar b$, or lines $\bar a$ and $b$. It is easy to show that those are the only planar diagrams present in this calculation.

In order to find a closed expression for the sum of all these diagrams one has to choose an orientation to perform the resummation. For our purpose, it is convenient to read the diagrams from right to left. In that case, one starts with the dipole formed by lines $a,\bar a$ and $b,\bar b$ up to  $t_3$ at which point the transition contraction occurs. Specifically, at $t_3$,  a transition occurs to the singlet state where the two dipoles are formed by the pairs $a,b$ and $\bar a,\bar b$. All the contractions on both sides of the interactions can be resummed into dipole amplitudes over the corresponding longitudinal extent. This procedure assumes that there is a transition,  and therefore ignores all the diagrams with no transition, i.e., the diagrams  where all the contractions respect the singlet structure in the right side of the diagram. In order to properly account for those diagrams one must include the factorizable contribution,  given by the product of two dipole amplitudes for the respective dipoles in the right hand side of the diagram. The total result is then the following:

\begin{figure}
\centering
\includegraphics[width=\textwidth]{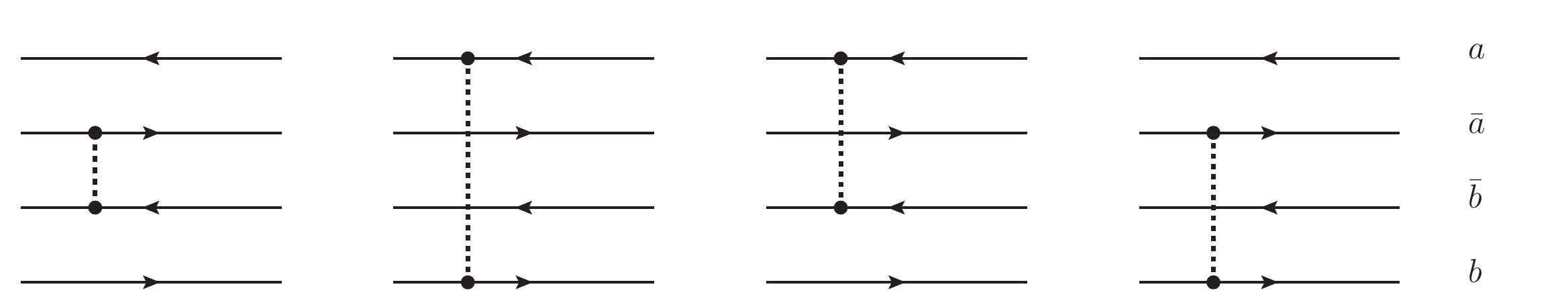}
\caption{Contractions contributing to the transition amplitude between the two singlet states.}
\label{fig:T}
\end{figure}

\beq
Q(t_L,t_2;\r_a,\r_b,\bar\r_a,\bar\r_b)&\equiv&\frac{1}{N_c}\left\langle\text{Tr}\left[U^\dagger(\r_a)U(\r_b)U^\dagger(\bar\r_b)U(\bar\r_a)\right]\right\rangle\nn
&=&C^{(2)}_q(t_L,t_2;\r_a,\bar\r_a)C^{(2)}_q(t_L,t_2;\r_b,\bar\r_b)\nn
&&+\int_{t_2}^{t_L}\rmd t_3\; C^{(2)}_q(t_L,t_3;\r_a,\bar\r_a)C^{(2)}_q(t_L,t_3;\r_b,\bar\r_b)T(t_3)\nn
&&\quad\times C^{(2)}_q(t_3,t_2;\r_a,\r_b)C^{(2)}_q(t_3,t_2;\bar\r_a,\bar\r_b),\label{quadrupole}
\eeq
where $T(t_3)$ denotes the transition between the two states. It receives contributions from the diagrams in Fig. \ref{fig:T} and is explicitly given by
\beq
T(t_3)=C_Fn\,[\sigma(\r_a-\r_b)+\sigma(\bar\r_a-\bar\r_b)-\sigma(\r_a-\bar\r_b)-\sigma(\bar\r_a-\r_b)]_{t_3},
\eeq
where all coordinates are evaluated at $t_3$.
The separation between the factorizable piece and the non-factorizable one is manifest in formula (\ref{quadrupole}). It is important to note that the same analysis can be performed by reading the diagrams from left to right. One obtains a formula which is the same as above except that the labels $b$ and $\bar a$ are interchanged. In that case the transition amplitude includes $a,\bar a$ and $b,\bar b$ contractions instead of the first two diagrams of Fig. \ref{fig:T}. It is easy to show that the two expressions are equivalent by noticing that if one substracts the two different ways of writing the non-factorizable term one can perform the $t_3$ integral explicitly and show that the resulting contribution vanishes. 

Now we are in a position to insert the medium average of the Wilson lines into the 4-point function (\ref{4plargenc}). Consider first the term corresponding to the factorizable piece of the quadrupole, one gets
\beq
&&(\y_a\y_b;\bar\y_a\bar\y_b|S^{(4)}_{\rm fac}(t_L,t_2)|\x_a\x_b;\bar\x_a,\bar\x_b)\nn
&=&\int {\cal D}\r_a {\cal D}\r_b {\cal D}\bar\r_a {\cal D}\bar\r_b \exp\left[\frac{i}{2}\int_{t_2}^{t_L} \rmd t\,\left(k_a^+\dot{\r}^2_a+k_b^+\dot{\r}^2_b-k_a^+\dot{\bar\r}^2_a-k_b^+\dot{\bar\r}^2_b\right)  \right]\nn
&&\times  \left[C^{(2)}_q(t_L,t_2;\r_a,\bar\r_a)C^{(2)}_q(t_L,t_2;\r_b,\bar\r_b)\right]^2.
\eeq
In the large-$N_c$ limit the square of a quark dipole amplitude is equivalent to a gluon dipole amplitude, therefore this factorizable piece can be rewritten as
\beq
&&(\y_a\y_b;\bar\y_a\bar\y_b|S^{(4)}_{\rm fac}(t_L,t_2)|\x_a\x_b;\bar\x_a,\bar\x_b)\nn
&=&\int {\cal D}\r_a {\cal D}\bar\r_a \exp\left\{\frac{ik_a^+}{2}\int_{t_2}^{t_L} \rmd t\,\left(\dot{\r}^2_a-\dot{\bar\r}^2_a\right)  \right\}C^{(2)}_g(t_L,t_2;\r_a,\bar\r_a)\nn
&&\times \int {\cal D}\r_b {\cal D}\bar\r_b \exp\left\{\frac{ik_b^+}{2}\int_{t_2}^{t_L} \rmd t\,\left(\dot{\r}^2_b-\dot{\bar\r}^2_b\right)  \right\}C^{(2)}_g(t_L,t_2;\r_b,\bar\r_b),\nn
&=&(\y_a;\bar\y_a|S^{(2)}(t_L,t_2)|\x_a;\bar\x_a)(\y_b;\bar\y_b|S^{(2)}(t_L,t_2)|\x_b;\bar\x_b),
\eeq
where we have identified the path--integral representations of the 2-point function
$S^{(2)}(t_L,t_2)$ according to \eqn{S2path}. We thus find the factorized structure
anticipated in \eqn{facs4}, at least within the large-$N_c$ approximation.

Now let us focus on the non-factorizable piece. The last line on (\ref{quadrupole}) has the same structure as the factorizable piece except for the fact that its starting longitudinal coordinate is at $t_3$. One can still recombine this piece with the remaining dipole amplitudes in (\ref{4plargenc}) since those amplitudes can be split in two in the following way:
\beq
C^{(2)}(t_L,t_2;\r,\bar\r)=C^{(2)}(t_L,t_3;\r,\bar\r)C^{(2)}(t_3,t_2;\r,\bar\r),
\eeq
yielding to two independent gluon dipoles for the region with longitudinal coordinate greater than $t_3$.
\beq
&&(\y_a\y_b;\bar\y_a\bar\y_b|S^{(4)}_{\rm nfac}(t_L,t_2)|\x_a\x_b;\bar\x_a,\bar\x_b)\nn
&=&\int {\cal D}\r_a {\cal D}\r_b {\cal D}\bar\r_a {\cal D}\bar\r_b \exp\left\{\frac{i}{2}\int_{t_2}^{t_L} \rmd t\,\left(k_a^+\dot{\r}^2_a+k_b^+\dot{\r}^2_b-k_a^+\dot{\bar\r}^2_a-k_b^+\dot{\bar\r}^2_b\right)  \right\}\nn
&&\times  \int_{t_2}^{t_L}dt_3\left[C^{(2)}_q(t_L,t_3;\r_a,\bar\r_a)C^{(2)}_q(t_L,t_3;\r_b,\bar\r_b)\right]^2T(t_3)\nn
&&\times C^{(2)}_q(t_3,t_2;\r_a,\r_b)C^{(2)}_q(t_3,t_2;\bar\r_a,\bar\r_b)C^{(2)}_q(t_3,t_2;\r_a,\r_b)C^{(2)}_q(t_3,t_2;\bar\r_a,\bar\r_b).
\eeq
Moreover, in the same way that the scalar propagators follow the convolution relation (\ref{convprop}), one can also split the longitudinal extent of the path integrals at $t_3$ and explicitly perform the integration in the region where the system factorizes into two independent gluons. One gets,
\beq
&&(\y_a\y_b;\bar\y_a\bar\y_b|S^{(4)}_{\rm nfac}(t_L,t_2)|\x_a\x_b;\bar\x_a,\bar\x_b)\nn
&=&\int_{t_2}^{t_L} dt_3\,\int_{\z_a\bar\z_a\z_b\bar\z_b} (\y_a;\bar\y_a|S^{(2)}(t_L,t_3)|\z_a;\bar\z_a)(\y_b;\bar\y_b|S^{(2)}(t_L,t_3)|\z_b;\bar\z_b)T(Z)\nn
&&\times\int_{t_2}^{t_3} {\cal D}\r_a {\cal D}\r_b {\cal D}\bar\r_a {\cal D}\bar\r_b \exp\left\{\frac{i}{2}\int_{t_2}^{t_3} \rmd t\,\left(k_a^+\dot{\r}^2_a+k_b^+\dot{\r}^2_b-k_a^+\dot{\bar\r}^2_a-k_b^+\dot{\bar\r}^2_b\right)  \right\}\nn
&&\times C^{(2)}_q(t_3,t_2;\r_a,\r_b)C^{(2)}_q(t_3,t_2;\bar\r_a,\bar\r_b)C^{(2)}_q(t_3,t_2;\r_a,\r_b)C^{(2)}_q(t_3,t_2;\bar\r_a,\bar\r_b),\label{S4inel}
\eeq
with $\r_{a,b}(t_3)=\z_{a,b}$ and similarly for the bar coordinates.

The transition amplitude $T$ depends explicitly on the new intermediate coordinates. One can easily see that in the harmonic approximation it takes the simple form
\beq
T(Z)=-\frac{\hat q}{2}(\z_a-\bar\z_a)\cdot(\z_b-\bar\z_b).
\eeq
When switching to momentum space, these coordinate differences can be expressed in terms of derivatives of the two-point functions going from $t_3$ to $t_L$.

After performing the Fourier transform, the non-factorizable piece can be written as
\beq
&&(\k_a\k_b;\k_a\k_b|\tilde S^{(4)}_{\rm nfac} (t_L,t_2)|\q_2,\bar\q_2-\q_2;\p_2,\bar\q_2-\p_2)\nn
&=&\int_{t_2}^L \rmd t_3\int_{\q_{3a}\q_{3b}}\frac{\hat q}{2}\;{\bs\nabla}{\cal P}(\k_a-\q_{3a},t_L-t_3)\cdot{\bs\nabla}{\cal P}(\k_b-\q_{3b},t_L-t_3)\tilde{\cal I}(Q_{3a},Q_{3b},Q_2,\bar Q_2,P_2),\nn\label{FIIIinel}
\eeq
where ${\bs\nabla}$ denotes the gradient with respect to the transverse momentum variables, and $\tilde{\cal I}$ is the Fourier transform of the path integral in (\ref{S4inel}) after removing the overall momentum conserving $\delta$-function,
\beq\label{4pointApp}
&&(2\pi)^2\delta^{(2)}(\q'_2-\bar\q_2+\q_2)\tilde{\cal I}(Q_{3a},Q_{3b},Q_2,\bar Q_2,P_2)\nn
&=&\int_{\{\x,\z\}}e^{-i[\q_{3a}\cdot(\z_a-\bar\z_a)+\q_{3b}\cdot(\z_b-\bar\z_b)-\q_2\cdot\x_a-\q'_2\cdot\x_b+\p_2\cdot\bar\x_a+(\bar\q_2-\p_2)\cdot\bar\x_b]}\nn
&&\times\int_{t_2}^{t_3} {\cal D}\r_a {\cal D}\r_b {\cal D}\bar\r_a {\cal D}\bar\r_b \exp\left\{\frac{i}{2}\int_{t_2}^{t_3} \rmd t\,\left(k_a^+\dot{\r}^2_a+k_b^+\dot{\r}^2_b-k_a^+\dot{\bar\r}^2_a-k_b^+\dot{\bar\r}^2_b\right)  \right\}\nn
&&\times C^{(2)}_q(t_3,t_2;\r_a,\r_b)C^{(2)}_q(t_3,t_2;\bar\r_a,\bar\r_b)C^{(2)}_q(t_3,t_2;\r_a,\r_b)C^{(2)}_q(t_3,t_2;\bar\r_a,\bar\r_b).
\eeq
This object is explicitly calculated in Appendix \ref{app:4p}. The main result of that calculation is that the result of the path integration is not exponentially suppressed only if its longitudinal extend is small, of order of the formation time. In terms of our result for the non-factorizable piece of the 4-point function it means that we can restrict the integration over $t_3$ to only the interval between $t_2$ and $t_2+t_f$, leading to the the fact that the region after $t_3$ is of the order of the length of the medium while the region between $t_2$ and $t_3$ is of the order of the formation time.

Using the explicit expression for the $\cal P$'s in Eq. (\ref{Prop}), one can easily see that the derivatives in (\ref{FIIIinel}) give a factor of
\beq
\frac{(\k_a-\q_{3a})\cdot(\k_b-\q_{3b})}{\hat q^2(t_L-t_3)^2}\sim\frac{1}{\hat qL}\; .
\eeq
This additional factor of $L$ in the denominator can not be compensated in any way since the integration in $t_3$ was shown to have support over a small region and therefore does not depend on $L$. Therefore, the non-factorizable piece of $S^{(4)}$ is parametrically smaller than the factorizable piece by a factor of $\tau_{_{\rm br}}/L$ and can be safely discarded for the kinematical regime under study.

%%%%%%%%%%%%%%%%%%%%%%%%%%%%%%%%%
\section{The splitting kernel and the gluon-splitting cross section}\label{splitting}
%%%%%%%%%%%%%%%%%%%%%%%%%%%%%%%%%

In this section we shall complete the calculation of the cross--section for medium--induced 
gluon branching, establish the formula (\ref{sigma1a}), and give an explicit
expression for the splitting kernel $\mcal{K}$. 

We return to Eq. (\ref{Sigma2}), and make a first simplification that exploits the main result of section \ref{sec:twogluonprop}, namely the factorization of the 4-point function. Thus, we replace  in Eq. (\ref{Sigma2}),
 $S^{(4)}(t_L,t_2)$ by $ S_{\rm fac}^{(4)}(t_L,t_2)$ given by \eqn{S4tildefac}.  We obtain then
\beq\label{Sigma2fac}
&&\frac{\rmd^2\sigma}{\rmd\Omega_{k_a}\rmd\Omega_{k_b}}=\frac{g^2N_c}{(2p^+_0)^2}\,
2\Re e\,\int_{t_0}^{t_L}\rmd t_1\int_{t_0}^{t_1}\rmd t_2\,\int_{\p_0, \p_1,\p_2,\q_1,\q_2\bar\q_2} \;\,(2\pi)^2\delta^{(2)}(\p_2-\q_2)\,\nn
&&\quad \times\Gamma^{ijl}(\p_1-z\q_1,z)\,\Gamma^{\bar i jl}(\p_2-z\bar \q_2,z)\nn
&&\quad \times {\cal P}(\k_a-\q_2,t_L-t_2){\cal P}(\k_b-\bar\q_2+\q_2,t_L-t_2)\nn
&& \quad \times  (\p_2,\bar\q_2-\q_2;\bar\q_2| \tilde S^{(3)}(t_2,t_1)|\p_1,\q_1-\p_1;\q_1) \nn
&&\quad \times{\cal P}(\q_1-\p_0,t_1-t_0) J^{i,c_0}(p_0^+,\p_0) J^{*\bar i, c_0}(p_0^+,\p_0),\eeq
where $p_0^+=k_a^++k_b^+$, $z=k_a^+/p_0^+$. (In order to  follow the flow of momenta
it may be useful to refer to Fig.~\ref{fig:amp-1}.) At this point we recognizes in \eqn{Sigma2fac} the three `classical propagators' ${\cal P}$ expressing the transverse momentum broadening of the gluons before and after the branching. These are almost the same as in  \eqn{sigma1a}, except that the momenta are those appropriate to regions I and III, while in  \eqn{sigma1a} the finite extent of region II is neglected. We shall return to this question shortly, and identify now  the splitting kernel  $\mcal{K}$ by combining 
the 3-point function $\tilde S^{(3)}(t_2,t_1)$ with the vertex factors that are
explicit in \eqn{Sigma2fac}. 

Consider first  the two vertex functions. They  combine to yield\beq
&&\frac{1}{4}\,\Gamma^{ijl}(\hat\P_1,z)\,\Gamma^{\bar i jl}(\hat\P_2,z)= \,\left[\frac{1}{z^2}+\frac{1}{(1-z)^2}\right]\,\hat\P_1\cdot\hat\P_2\,\delta^{i\bar i}+2\,\hat\P_1^i\, \hat\P_2^{\bar i}, 
\eeq
where we have set $\hat \P_1=\p_1-z\q_1$, $\hat \P_2=\p_2-z\bar\q_2$. 
For simplicity, we consider here only inclusive cross sections that are averaged over azimuthal angles. Under this assumption one can then replace
$
2\,\hat\P_1^i\, \hat\P_2^{\bar i}$ by $ (\hat\P_1\cdot\hat\P_2)\,\delta^{i\bar i},
$
and get
\beq\label{vertex}
 N_c\,\Gamma^{ijl}(\hat\P_1,z)\Gamma^{\bar i jl}(\hat\P_2,z)=
 \frac{4}{z(1-z)}P_{gg}(z)\,(\hat\P_1\cdot\hat\P_2)\,\delta^{i\bar i}\,,
\eeq
where
\be
P_{gg}(z)=N_c\left[\frac{z}{1-z}+\frac{1-z}{z}+z(1-z)\right]\,,
\ee
is the leading--order  Altarelli--Parisi splitting function \cite{Altarelli:1977zs}.

%
%Keeping in mind that our goal is to express the cross section for this process in the manner depicted in Fig. \ref{fat_vertex2}, we group together the contribution of the vertex functions that we have just considered with the 3-point function, and define the splitting kernel,
%\beq\label{kernel1}
% {\cal K}(\hat\Q_2,\hat{\bar\Q}_2,\hat\P_1,\hat\P_2,t_2-t_1,z)
%= \frac{P_{gg}(z)}{2z(1-z)\,p^{+2}}\;(\hat\P_1\cdot\hat\P_2)\,(\Q'_2|\tilde S^{(3)}(t_2-t_1)|\P'_1)\,.\nn
%\eeq

Next we consider the 3-point function $\tilde S^{(3)}(t_2,t_1)$, which is calculated in Appendix \ref{app:3p}. In the harmonic approximation it reads (cf. \eqn{eq:3-PF-FT-HO})
\begin{align}\label{S3kernel1}
& (\p_2,\bar\q_2-\p_2;\bar\q_2| \tilde S^{(3)}(t_2,t_1)|\p_1,\q_1-\p_1;\q_1)=
 \nonumber\\*[0.2cm] %\vspace*{0.6cm}
 &\qquad =\,
  \frac{8\pi[ 1+z^2+(1-z)^2]}{3\hat q\Delta t}\, {\exp}\left\{ -\frac{2[ 1+z^2+(1-z)^2](\q_1-\bar\q_2)^2}{3\hat q\Delta t}  
 \right\}\nn
&\qquad \times\,  \frac{2\pi(1+i)}{k_{_{\rm br}}^2\sinh(\Omega \Delta t) }
\exp{\left\{  -(1+i)\frac{(\hat\P_1+\hat\P_2)^2}{4k_{_{\rm br}}^2\coth(\Omega \Delta t/2)}
-(1+i)\frac{(\hat\P_1-\hat\P_2)^2}{4k_{_{\rm br}}^2\tanh(\Omega \Delta t/2)}\right\}  },
\end{align}
with $\Delta t=t_2-t_2$, and 
 \beq\label{Omega}
 \Omega\equiv\,\frac{1+i}{2\tau_{_{\rm br}}}\,,\quad 
 \tau_{_{\rm br}}\equiv\sqrt{\frac{z(1-z)p^+_0}{\hat q_{\text{eff}}}}\,,\quad
 k_{_{\rm br}}^2\equiv\hat q_{\text{eff}}\,\tau_{_{\rm br}}=\sqrt{z(1-z)p_0^+ \hat q_{\text{eff}}}\,,
 \eeq
where $\hat q_{\text{eff}}$ denotes an average, $z$--dependent, version of the jet 
quenching parameter:
 \beq
\hat q_{\text{eff}}\equiv\frac{1}{2}\hat q \left[z^2+(1-z)^2+1\right].
 \eeq
  In writing the expression above for the 3-point function, we have used the delta-function $\delta^{(2)}(\p_2-\q_2)$ in \eqn{Sigma2fac} in order to replace $\q_2$ by $\p_2$. This  allowed us in particular to express $\tilde S^{(3)}(t_2,t_1)$ in terms of $\hat\P_2$.
This expression (\ref{S3kernel1}) of  $S^{(3)}(t_2,t_1)$ makes it clear that the extent of region II is limited: indeed, since $1/\sinh(|\Omega| \Delta t) \propto \exp\{-\Delta t/\sqrt{2} \tau_{_{\rm br}}\}$ for large $\Delta t$, the  time
separation  $\Delta t=t_2-t_1$ needs to be kept smaller that $ \tau_{_{\rm br}}$ in order to avoid exponential suppression. Thus, as anticipated via simple considerations in Sect.~\ref{sec:strategy},  and in the qualitative discussion of Sect.~\ref{sec:qualcom}, one sees 
that the time scale of the branching process is indeed determined by $\tau_{_{\rm br}}$. Recall that this is also (approximately) the same time scale that controls the exponential damping of the non factorizable part of  the 4-point function (see Sect.~\ref{sec:twogluonprop}
 and appendix \ref{app:3p}, \eqn{taubranching4}). 

In order to proceed further, we shall exploit the fact that region II has limited extent and neglect in the factors $\cal P$ the fraction of momenta that can be attributed to momentum broadening within region II. Since this region is of extent $\tau_{_{\rm br}}$, this fraction is typically of order $\tau_{_{\rm br}}/L$ as compared to the momentum acquired through propagation before and after the splitting (this is in fact explicit for the dependence of  $\tilde S^{(3)}$ on $\q_1-\bar\q_2$ in Eq.~(\ref{S3kernel1}), where it can be seen that $(\q_1-\bar\q_2)^2$ is at most of order $\hat q\tau_{_{\rm br}}$). Thus we shall replace 
in \eqn{Sigma2fac}
${\cal P}(\k_b-\bar\q_2+\q_2,t_L-t_2)$ by ${\cal P}(\k_b-\q_1+\q_2,t_L-t_2)$ (ignoring the small difference $\q_1-\bar\q_2$), and use as independent variables $\q_1, \hat \P_1, \hat\P_2, \q_1-\bar\q_2$ in place of $\q_1, \p_1, \bar\q_2,\q_2$. Since the factors ${\cal P}$ do not depend on $\hat\P_1$ nor $\q_1-\bar\q_2$, the integration over these variables in \eqn{Sigma2fac} leaves these $\cal P$ factors intact. This allows us to define a kernel as follows
\beq\label{kernel5}
 {\cal K}(\hat\P_2,t_2-t_1,z)\,\equiv\,
 \frac{P_{gg}(z)}{[z(1-z)p^+_0]^2}\,  
 \Re e \int_{\q_1-\bar \q_2,\hat\P_1}
 (\hat\P_1\cdot\hat\P_2)\,  \tilde S^{(3)}(\Delta t, \hat\P_1,\hat\P_2, \q_1-\bar\q_2),
\eeq 
where we have made explicit the independent momentum variables on which $\tilde S^{(3)}$ depends.
A straightforward calculation of the gaussian integrals in \eqn{kernel5} yields
\beq\label{kernel2}
 {\cal K}(\p,\Delta t,z)= {P_{gg}(z)}\,\frac{\hat\P_2^2}{2[z(1-z)p^+_0]^2} 
\Re e \left[\left(\frac{1}{\cosh^2(\Omega \Delta t) }\right)
 \exp\left\{ -\frac{i\hat\P_2^2}{2z(1-z)p^+\Omega} \,\tanh(\Omega \Delta t)   \right\}\right].\nn
\eeq

Using this definition for ${\cal K}$, one can rewrite Eq. (\ref{Sigma2fac})  as 
\begin{align}\label{Sigma2fac-2}
\frac{\rmd^2\sigma}{\rmd\Omega_{k_a}\rmd\Omega_{k_b}}=2g^2z(1-z) &
\,\int_{t_0}^{t_L}\rmd t_2\,\int^{t_2}_{t_0} \rmd t_1\,\int_{\p_0,\hat\P_2,\q_1} \;{\cal P}(\k_a-\q_2,t_L-t_2){\cal P}(\k_b-\q_1+\q_2,t_L-t_2)\nn
 &\times  \,{\cal K}(\hat\P_2,\Delta t,z) 
 \,{\cal P}(\q_1-\p_0,t_1-t_0) \,\frac{\rmd\sigma_{hard}}{\rmd\Omega_{p_0}}\,,
\end{align}
where $\rmd\sigma_{hard}/\rmd\Omega_{p_0}=|{\bs J}(p_0^+,\p_0)|^2$. Now, as already explained,
the kernel ${\cal K}(t_2-t_1)$ effectively restricts the time integrations in \eqn{Sigma2fac-2} to
$\Delta t\equiv t_2-t_1\lesssim \tau_{_{\rm br}}$. This is a small time interval as compared to the
typical values of $t_1$ and $t_2$, of order $t_L$,
so, in line with our previous approximations,  we integrate the kernel over $\Delta t$
while neglecting the difference $t_2-t_1$ in the various factors  ${\cal P}$.
We then redefine the kernel after integration over $\Delta t$:
\eqn{kernel2}, to be later restored)
 \beq\label{kernel3}
&&\int_{0}^{t_2-t_0} \rmd\Delta t \
\frac{1}{2z(1-z)p^+_0} \left(\frac{\hat\P_2^2}{\cosh^2(\Omega \Delta t) }\right)\exp\left\{ -\frac{i\hat\P_2^2}{2z(1-z)p^+_0\Omega} \,\tanh(\Omega \Delta t) \right\}\nn
&=&i\int_{0}^{t_2-t_0} \rmd\Delta t\;\frac{\rmd}{\rmd\Delta t} \exp\left\{ -\frac{i\hat\P_2^2}{2z(1-z)p^+_0\Omega} \,\tanh(\Omega \Delta t) \right\},\nn
&=&i\left[\exp\left\{ -\frac{i\hat\P_2^2}{2z(1-z)p^+_0\Omega} \,\tanh(\Omega (t_2-t_0)) \right\}
-1\right].
\eeq
To further simplify the kernel, and again in line with our approximations,  we  neglect the region of integration $t_2\lesssim\tau_{_{\rm br}}+t_0$. When $t_2-t_0\gg \tau_{_{\rm br}}$, one can use $\tanh(\Omega (t_2-t_0))\approx 1$ and then the kernel becomes time-independent. After taking the real part in \eqn{kernel3} and restoring the proper factors from \eqn{kernel2}, we finally obtain:
\beq\label{kernel4}
&& {\cal K}(\hat\P_2,z,p_0^+) \approx \frac{2}{z(1-z)p^+_0} 
P_{gg}(z)\, \sin\left[\frac{\hat\P_2^2}{2 \k_{_{\rm br}}^2}\right]\,\exp\left[-\frac{\hat\P_2^2}{2 \k_{_{\rm br}}^2}\right].
\eeq
This kernel generalizes the result obtained in \cite{MehtarTani:2012cy} in the eikonal limit. 
 
Putting everything together, and proceeding to various  relabeling ($t_2\to t$, $\hat\P_2\to \p-z\q, \q_2\to \p, \q_1\to\q$),  one recovers the expression for the cross--section for  quasi-instantaneous medium--induced gluon branching  given in \eqn{sigma1a}

The  kernel $ {\cal K}(\p-z\q,z,p_0^+)$ describes the splitting of a gluon with longitudinal 
momentum $p_0^+$ and transverse momentum $\q$ into two gluons, one
with  longitudinal momentum fraction $z$ and transverse momentum $\p$, the
other with  longitudinal momentum fraction $1-z$ and transverse momentum $\q-\p$. Note that ${\cal K}$ does not depend on $\p$ and $\q$ separately, but only on the combination $\p-z\q$, which may be understood as the relative momentum of the effective non relativistic two-dimensional motion of the gluons in the transverse plane (with $\q$ playing the role of the `center of mass' momentum).  We may also write $\p-z\q=p^+(\v_{\p}-\v_{\q})=(p^+-q^+)(\v_{\p-\q}-\v_{\q})$,
where $\v_{\p}=\p/p^+$ etc. are transverse velocities.
% Thus  $\p-z\q$ is proportional to the relative velocity of one of the daughter gluons 
%with respect to  the parent gluon. 
Interestingly, the quantity $|\v_p-\v_q|$ is a measure of the actual emission angle in
three dimensions.  To see that, let us introduce the 
three--dimensional velocities,  $\vec v_{\q}=(v_{z\q},\v_\q)$ and $\vec v_{\p}=(v_{z\p},\v_{\p})$,
with $v_{z\q}^2+\v_{\q}^2=1$, etc.
Then,  $(\v_{\p}-\v_{\q})^2= 2(1-\cos\theta_{\p})\simeq \theta_{\p}^2$,
and similarly for $|\v_{\q-\p}-\v_\q|$.  
Since \eqn{kernel4} constrains the value of $\p-z\q$ to be of order $k_{_{\rm br}}$ at most, emission angles are constrained 
as follows
 \beq
 |\theta_{\p}|\,\simeq\,|\v_\p-\v_\q|\,\lesssim\,\frac{k_{_{\rm br}}}{zq^+}\,,\qquad
 |\theta_{\q-\p}|\,\simeq\,|\v_{\q-\p}-\v_q|\,\lesssim\,\frac{k_{_{\rm br}}}{(1-z)q^+}\,.\eeq
These formul{\ae}  make it clear that it is the softest offspring gluon which is emitted 
at the largest angle and hence that controls the geometry of the branching.
 
We may get  more precise  on  the probabilistic interpretation of the kernel ${\cal K}$ by integrating the cross section over suitable elements of phase space. 
Consider the formula  (\ref{sigma1a}) that we have just derived, 
\beq\label{sigma1abis}
\frac{\rmd^2\sigma_1}{\rmd\Omega_{k_a}\rmd\Omega_{k_b}}&=&2g^2z(1-z)\int_{t_0}^{t_L}\,\rmd t\,\int_{\p_0, \q,\p} \; {\cal P}(\k_a-\p,t_L-t)\,
{\cal P}(\k_b-\q+\p,t_L-t)\nn
&& \qquad \qquad \qquad \times  \,{\cal K}(\p-z\q,z,p_0^+) \,
 {\cal P}(\q-\p_0,t-t_0)\, \frac{\rmd\sigma_{hard}}{\rmd\Omega_{p_0}}\,,
\eeq
in which it is understood that $k_a^++k_b^+=p_0^+$. From this we may calculate 
\beq
\frac{\rmd\sigma_1}{\rmd z\rmd\Omega_0^+ }=\frac{1}{2}\int {\rmd^2\k_a}\int {\rmd^2\k_b}\,\frac{\rmd\sigma_1}{\rmd\k_a \rmd\k_b\rmd z \rmd\Omega_0^+}.
\eeq
This quantity represents the cross section for a gluon with initial + momentum in the phase space element $\rmd\Omega_0^+\equiv \rmd p_0^+/((2\pi)2p_0^+)$ to split into two gluons, one of which carries the fraction $zp_0^+$ of the initial + momentum.  The factor $1/2$ is a symmetry factor that accounts for the fact that each configuration of 2 identical gluons is counted twice in the integration over $\k_a$ and $\k_b$. The calculation is easily done 
and yields
\beq\label{sigma1abis2}
 \frac{\rmd\sigma_1}{\rmd z\rmd\Omega_0^+ }=\frac{g^2}{4\pi}\, \int_{t_0}^{t_L}\,\rmd t\,\int_{\p-z\q} 
 \,{\cal K}(\p-z\q,z,p_0^+) \,
 \frac{\rmd\sigma_{hard}}{\rmd\Omega_0^+},
\eeq
with
\beq
\frac{\rmd\sigma_{hard}}{\rmd\Omega_0^+}=\int\frac{\rmd\p_0}{(2\pi)^2} \frac{\rmd\sigma_{hard}}{\rmd\Omega_{\p_0}}.
\eeq
 From Eq.~(\ref{sigma1abis2}), one reads the probability per unit time for a gluon with energy $p_0^+$ and transverse momentum $\q$ to produce a splitting with one of the produced gluon carrying a momentum fraction $zp_0^+$ and a transverse momentum $\p$:
 \beq
 \frac{\rmd P}{\rmd\p\, \rmd z\, \rmd t}=\frac{\alpha_s }{(2\pi)^2}
 \,{\cal K}(\p-z\q,z,p_0^+).
 \eeq
 The integration over the transverse momentum yields
 \beq\label{kernel}
  { \cal K}(z,p^+)\equiv \int \frac{\rmd^2\q}{(2\pi)^2}\, {\cal K}(\q,\,z, p^+)=\frac{1}{2\pi}\,\frac{P_{gg}(z)}{\tau_{_{\rm br}}(z, p^+)}.
  \eeq
Note that the gluon spectrum
produced via a {\em single} medium--induced emission may be recovered from this formula,  by integrating
\eqn{kernel} over time up to $t_L$ and multiplying by a factor of $2z$~:
 \beq\label{BDMPSZ}
 z\,\frac{\rmd N}{\rmd z}\bigg|_{\rm one\ emission}\,=\,
 \frac{\alpha_s}{\pi} zP_{gg}(z) t_L \sqrt{\frac{\hat q_{\text{eff}}}{z(1-z)p^+}}\,.
\eeq
This is recognized as the BDMPSZ spectrum, as expected 
 \cite{Baier:1996kr,Baier:1996sk,Baier:1998kq,Zakharov:1996fv,Zakharov:1997uu}.

The kernel
in \eqn{kernel4} may turn negative for large value of the transverse momentum. This would not be a serious problem in practice since this occurs for momenta ($q\gtrsim \pi k_{_{\rm br}}$) for which the gaussian  in \eqn{kernel4} is very small, but it indicates a limitation of the approximation that has been used to arrived at  \eqn{kernel4}.
In fact, a large transverse momentum  $q \gg k_{_{\rm br}}$ cannot be acquired over a time scale $\sim \tau_{_{\rm br}}$ as a result of soft multiple scattering. Rather it must be associated with 
some rare but hard collision, which can be more accurately treated in the single
scattering approximation --- that is, by keeping only the term linear in the
dipole cross--section in the expansion of the 2-point function \eqref{Dgg} ---, 
but by keeping the logarithmic dependence upon $r$. When applied to the momentum broadening probability, this procedure yields for  the Fourier transform in \eqn{PD}, 
${\cal P}(\Delta\p,\Delta t)\sim (\hat q\Delta t)/(\Delta\p)^4$. We expect this procedure to lead to a similar behavior in $1/\q^4$ at large $\q$  for the kernel  ${\cal K}(\q,z,p^+)$. 
 We leave the detailed discussion of this particular point for a subsequent study.

%%%%%%%%%%%%
\section{Conclusions}
%%%%%%%%%%%%

In this paper, we have provided a complete calculation of medium--induced gluon
branching in the regime where the gluons that take part in the branching undergo multiple soft scattering with the medium. The kernel that describes the branching has been calculated as a function of  transverse momentum, beyond the eikonal approximation. An important  conclusion of our study  is that the offspring gluons 
lose color coherence with respect to each other on the same time scale as that of the branching process itself. Thus, as soon as they are produced,  they propagate independently from each other, a picture that holds to within corrections of order $\tau_{_{\rm br}}/L\ll1$. Our explicit proof relies on a large--$N_c$ approximation but we believe that our conclusions remain generally valid.

In the regime considered in this paper, where the energies of the emitted gluons are within the range $\omega_{_{\rm BH}}\ll\omega\ll \omega_c$, a regime dominated by multiple scattering, subsequent emissions by these gluons do not interfere with each other. Indeed, the typical duration $\tau_{_{\rm br}}(\omega)$ 
of a branching process is much smaller than the longitudinal extent $L$ of the medium.  
Since the interference effects 
between the offspring gluons are possible only during this branching time $\tau_{_{\rm br}}$, 
whereas independent emissions by these gluons can occur anywhere along the size 
$ L$ of the medium,  the longitudinal
phase--space for interference phenomena is suppressed compared to the corresponding
phase--space for independent emissions by a factor $\tau_{_{\rm br}}(\omega)/L\ll 1$.

The suppression of interference effects is a key ingredient for having independent emissions.
The other key ingredient  is that successive emissions do not overlap with
each other, a situation which, as recalled in the introduction, may occur for sufficiently soft gluons, that can be produced at a high rate. Whether a fully probabilistic description of successive branchings can be given  depends therefore on how well one can control the emission of very soft gluons (beyond the apparent infrared divergences produced by the splitting functions, and that are easily seen to cancel in the calculation of physical processes). We leave this point for a subsequent study.

There are further limitations of the present calculation that need to be emphasized. The calculation that we have presented is valid in a regime where interactions with the medium are dominated by soft multiple scattering. While this is a legitimate assumption for the leading particle
and the hardest component of its medium--induced radiation, this may not be also
the case for the very soft gluons at the end of a cascade, especially if their energies approach the Bethe--Heitler
energy. A proper treatment of this particular region would require a more accurate treatment of the 
single scattering between a gluon and the medium, which implies in particular relaxing  the eikonal approximation.
Also, we have not included in our formalism the vacuum--like emissions, that is, 
the emissions by which the energetic gluon that enters the plasma loses its initial (potentially large) virtuality. 
Such emissions  are controlled by the standard splitting functions for bremsstrahlung.
These are generally hard emissions at small angles.  In fact, all the emissions at angles much
smaller than $\theta_c$ should proceed exactly as in the vacuum with usual destructive interferences leading to angular ordering. Hard emissions at larger angles $\theta\gg\theta_c$ are possible as well,  and for them
interference effects become negligible because of medium effects \cite{MehtarTani:2010ma,MehtarTani:2011tz,MehtarTani:2011gf,CasalderreySolana:2011rz,MehtarTani:2012cy}.  
For hard emissions at intermediate angles  $\theta \sim \theta_c$ the situation 
is more complicated \cite{MehtarTani:2010ma,CasalderreySolana:2011rz}, so in order
to better understand this and also to have a unified description of the in--medium jet 
evolution, it would be very useful to extend our formalism by including vacuum--like emissions.

\noindent{\bf Acknowledgements}

We thank A. Mueller for useful discussions on various aspects of this work. This research is supported by the European Research Council under the Advanced Investigator Grant ERC-AD-267258.

\appendix

\section{Gluon dynamics in a background field}\label{app:glueprop}

In this appendix, we review briefly  properties  of the gluon propagator in a background field $A^-(x^+,\x)$ which is independent of $x^-$. To within an inessential term that can be ignored, this propagator can be entirely expressed in terms of a $2+1$ dimensional ``scalar'' propagator (independent of Lorentz indices), that describes non-relativistic propagation in the transverse plane, with $x^+$ playing the role of time. The dependence on Lorentz indices is factored out into coefficients that are used to define an effective three-gluon vertex. This vertex and the scalar propagator can be used to simplify perturbative calculations  for the effective non relativistic gluo-dynamics in the transverse plane. 

We work in the light cone gauge $A^+=0$, with covariant derivative
$ D_\mu=\del_\mu-igA_\mu$, and the only non-vanishing component of the background field is $A^-(x^+,\x)$. In this gauge, 
the free propagator reads, in momentum space,
\beq\label{freepropagator}
G_0^{\mu\nu}(p)=-G_0(p)\,d^{\mu\nu}(p)\qquad d^{\mu\nu}(p)\equiv \left[   g^{\mu\nu}-\frac{p^\mu n^\nu+p^\nu n^\mu}{n\cdot p} \right], \qquad G_0(p)\equiv \frac{-1}{p^2},
\eeq
with $n^\mu=\left( n^+,n^-,\boldsymbol{n_\perp}   \right)=\left( 0,1,\boldsymbol{0}   \right)$, so that   $n\cdot A=A^+=0$ and $n\cdot p=p^+$. Note that $G_0^{\mu\nu}$ is symmetric under the interchange of $\mu$ and $\nu$, and it vanishes if either index is $+$. The non vanishing components are 
\beq\label{freepropagator2}
G^{ij}_0(p)=G_0(p) \delta^{ij},\qquad  G_0^{-i}(p)=\frac{p^i}{p^+}G_0(p), \qquad G_0^{--}(p)= 2 \frac{p^-}{p^+} G_0(p)  .
\eeq

\begin{figure}[htbp]
\begin{center}
\includegraphics[width=4cm]{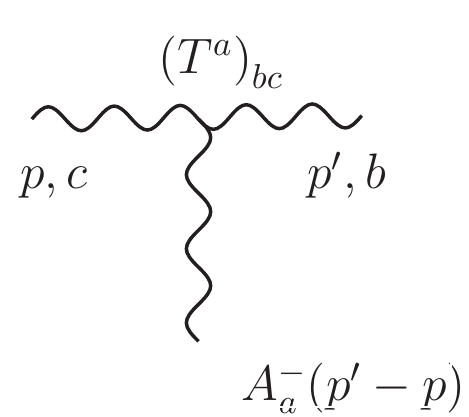}
\caption{Representation of the vertex coupling the propagating gluon to the background field $A^-_a$. The matrix $T^a$ is a matrix of the adjoint representation. The background field $A^-_a$ is independent of $x^-$, so that there is no transfer of + momentum at the vertex, that is, $p'^+=p^+$.}\label{fig:G}
\end{center}
\end{figure}

In the presence of the background field, the propagator is modified. In particular, it is no longer diagonal in momentum space (except for the + component  of the momentum). It is is also a non diagonal matrix in color space. It may be written quite generally as $G^{\mu\nu}_{ba}(p',p)$, with the convention that $p$ is incoming and $p'$ is outgoing (and similarly for the color indices). That is, one may regard $G^{\mu\nu}_{ba}(p',p)$ as the matrix element $\left({p'}\, b|G^{\mu\nu}|{p}\,a\right)$, which we shall also write $\left({p'} |G^{\mu\nu}_{ba}|{p}\right)$ when convenient. To determine the structure of $G^{\mu\nu}$, we write
\beq\label{eqntransfermatrix2}
(p'\,b|G^{\mu\nu} |p\,a)=(p'|p)\delta_{ab}\, G_0^{\mu\nu}(p)+G_0^{\mu\mu'}(p')  \,(p'\,b|{\cal T}_{\mu'\nu'}|p\,a)\, G_0^{\nu'\nu}(p),
\eeq
with $(p'|p)=(2\pi)^4\delta^{(4)}(p-p')$, and expand ${\cal T}^{\mu\nu}$  in powers of the background field. We get 
\beq\label{vertex1}
(p'|{\cal T}_{ba}^{\nu\mu} |p)&=& -i (p'|V^{\nu+\mu}_{bca}|p)\,A_c^-(p'-p)\nonumber\\
&-& (p'| V_{bed}^{\mu+\mu'}|p'')\,A_e^-(p'-p'')\, G_{0,\mu'\nu'}(p'')\,(p''|V_{dca}^{\nu'+\nu}|p)\,A_c^-(p''-p)+\cdots
\eeq
Here,  $(p'|V^{\nu+\mu}_{bca}|p)$ stands for $V^{\nu+\mu}_{bca}(p',p-p',-p)$, where $V$ is the 
 usual three-gluon vertex, defined generally as (with all momenta chosen outgoing), see Fig. \ref{fig:G},
\beq\label{vertex_definition}
V_{abc}^{\mu\nu\rho}(k,p,q)=-gf^{abc}\left[ g^{\mu\nu} (k-p)^\rho+g^{\nu\rho}(p-q)^\mu+g^{\rho\mu}(q-k)^\nu   \right].
\eeq
The fact that in Eq.~(\ref{vertex1}), or in Eq.~(\ref{eqntransfermatrix2}),  ${\cal V}^{\nu+\mu}(p',p)$ is contracted with $G_0$ propagators, puts constraints on the values of the indices $\mu$ and $\nu$ (see Eqs.~(\ref{freepropagator2})). A  simple analysis reveals that  the only relevant components of  the vertex are
\beq
V_{bca}^{j+i}(p',p-p',-p)=gf^{bca}\, g^{j i}(p+p')^+=-2igp^+ \left( T^c \right)_{ba}\, g^{j i},
\eeq
where we have used $p'^+=p^+$. It follows that  ${\cal T}_{\mu'\nu'}(p',p)$ vanishes unless the indices $\mu$ and $\nu$ are spatial indices,  and furthermore  ${\cal T}^{ij}$ is diagonal. We shall set 
\beq\label{diagonalT}
{\cal T}_{ij}(p',p)=-\delta_{ij} {\cal T}(p',p).
\eeq
Taking this property of ${\cal T}^{\mu\nu}$ into account, one can rewrite  Eq.~(\ref{eqntransfermatrix2}) as
 \beq
(p'|G_{ba}^{\nu\mu}|p)=(p'|p)\delta_{ab}\, G_0^{\nu\mu}(p)-G_0^{\nu i}(p')  \,(p'|{\cal T}_{ba}|p)\, G_0^{i\mu}(p),
\eeq
and, using (\ref{freepropagator}),  as
\beq\label{eqntransfermatrix5}
(p'|G_{ba}^{\nu\mu}|p)=-(p'|p)\delta_{ba}\, G_0(p) d^{\nu\mu}-d^{\nu i} d^{i\mu}\,G_0(p')  \,(p'|{\cal T}_{ba}|p))\, G_0(p).
\eeq
At this point, we define a `scalar propagator' $(p'|G_{ba}|p)$ 
\beq
(p'|G_{ba}|p)=(p'|p)\delta_{ba}\, G_0(p)-G_0(p')  \,(p'|{\cal T}_{ba}|p)\, G_0(p),
\eeq
and substitute this into Eq.~(\ref{eqntransfermatrix5}) to obtain
\beq\label{eqntransfermatrix3}
G_{ba}^{\mu\nu}(p',p)= (p'|p)\delta_{ba}\, G_0(p)\left[- d^{\mu\nu}(p) -d^{\mu i}(p)d^{i\nu}(p)\right]+d^{\mu i}(p')d^{i\nu}(p) G_{ba}(p',p).
\eeq
A direct calculation, using the explicit expression of $d^{\mu\nu}$ given in Eq.~(\ref{freepropagator}) reveals that
\beq
 (p'|p)\, G_0(p)\left[- d^{\mu\nu}(p) -d^{\mu i}(p)d^{i\nu}(p)\right]=- \delta^{\mu -}\delta^{\nu -} (p'|p)\,\frac{1}{(p^+)^2}.
 \eeq
We recognize the instantaneous contribution to the gluon propagator in light-cone perturbation theory. This contact term can be ignored in the present calculation. Thus we are left with
\beq
(p'|G_{ba}^{\nu\mu}|p)=d^{\nu i}(p') (p'|G_{ba}|p)d^{i\mu}(p),
\eeq
which expresses the gluon propagator as the scalar propagator multiplied by factors that carry all information about Lorentz indices. 

The Fourier transform of the scalar propagator, $(x|G|y)=G(x,y)$,  is the solution of the equation 
\beq\label{Gdefinition}
\left[ \Box_x -2ig(A^-\cdot T) \del^+\right] G(x,y)=\delta(x-y),
\eeq
an equation that  naturally arises when solving the Yang-Mills equations for a fluctuating gauge field in the presence of the $A^-$ background \cite{MehtarTani:2006xq,Blaizot:2004wu}. This is easily seen by writing this equation as $G_0^{-1}+\Sigma=1$, with $\Sigma$ the self-energy, and then identifying ${\cal T}=\Sigma-\Sigma G_0\Sigma+\cdots$. 
Eq.~(\ref{Gdefinition})  can also be written as
\beq
\left[  2\del_x^+{\cal D}^-_x -\nabla_\perp^2\right]_{ac}G_{cb}(x,y)=\delta(x-y)\delta_{ab},
\eeq 
where ${\cal D}^-=\del^--igA^-\cdot T$, and we used the fact that $\del^+A^-=0$. Since  the background field $A^-$ does not depend on $x^-$,  the propagator $G(x,y)$ depends on $x^-$ and $y^-$ only through the difference $x^--y^-$. It is then convenient to introduce a new Green's function
\beq\label{calG}
(x|G^{ab}|y)\equiv \int\frac{\rmd k^+}{2\pi}\,{\rm e}^{-ik^+(x^--y^-)}\,\frac{i}{2k^+} (\x|\,{\cal G}^{ab}(x^+,y^+;k^+)\,|\y).
\eeq
A simple calculation reveals  that ${\cal G}$ satisfies the following equation
\beq
\left[ i{\cal D}^-+\frac{\nabla_\perp^2}{2k^+}\right]_{ac}(\x|\,{\cal G}^{cb}(x^+,y^+;k^+)\,|\y)=i\delta_{ab}\delta(x^+-y^+)\delta(\x-\y),
\eeq
with $i{\cal D}^-=i\del^-+g A^-$. Thus, ${\cal G}(x^+,y^+;k^+)$ is the propagator of a Schr\" odinger equation for a non relativistic particle of mass $k^+$ moving in a time dependent potential $A^-(x^+,\x)$, with $x^+$ playing the role of the time. It can be written as  a path integral
\beq
(\x|\,{\cal G}(x^+,y^+;k^+)\,|\y)=\int {\cal D} \r\, {\rm e}^{i\frac{k^{\!_+}}{2} \int_{y^{\!_{+}}}^{x^+}    \rmd t \, \dot\r^2   }\,\tilde U(x^+,y^+; \r),
\eeq
with $\r(y^+)=\y$, $\r(x^+)=\x$, and $\tilde U$ is a Wilson line evaluated along the path $\r(t)$
\beq
\tilde U(x^+,y^+; \r)={\rm T}\exp\left\{ ig\int_{y^+}^{x^+} \rmd t \, A^-_a(t, \r(t))\, T^a   \right\}.
\eeq
From this relation, one easily deduces the following composition property (using the matrix notation)
\beq\label{convprop}
{\cal G}(x^+,y^+; k^+)={\cal G}(x^+,z^+; k^+)\,{\cal G}(z^+,y^+; k^+).
\eeq

Consider now the vertex that describes the splitting of a gluon of momentum $q$ into a gluon of momentum $p$ and a gluon of momentum $q-p$. This is given by the general expression (\ref{vertex_definition}), in which the momentum $q$ is chosen incoming while the other two momenta are outgoing:
\beq
V_{abc}^{\mu\nu\lambda}(q,p,q-p)=gf^{abc}\left[ g^{\mu\nu} (q+p)^\lambda+g^{\nu\lambda}(q-2p)^\mu-g^{\lambda\mu}(2q-p)^\nu   \right].
\eeq
A modified vertex is obtained by combining $V_{abc}^{\mu\nu\lambda}$ with the factors $d^{\mu i} $ contained in the propagators attached to the vertex. Thus we define 
\beq
\Gamma^{ijl}_{abc}=d^{i\mu}(q) d^{j\nu}(p) d^{l\nu}(q-p) V^{abc}_{\mu\nu\lambda}(q,p).
\eeq
Note that the indices on $\Gamma$ are all transverse. Taking into account that the only non vanishing  contributions of $d^{\mu i}$ are either of the form $d^{ij}$ or $d^{i-}$, one obtains easily
\beq\label{vertex3g}
\Gamma^{ijl}_{abc}(\p-z\q;z) =-2gf_{abc} \left\{-\frac{1}{1-z}(\p-z\q)^l\delta^{ij}+(\p-z\q)^i\delta^{jl}-\frac{1}{z}(\p-z\q)^j\delta^{il}   \right\},\nn
\eeq
where $ z\equiv {p^+}/{q^+}$, and we have made explicit the dependence on the single momentum  $\p-z\q$. This momentum  has a simple physical interpretation in the effective non relativistic  dynamics. It can indeed be written as $zq^+(\v_\p-\v_\q)$, where $\v_\p=\p/p^+$ and $\v_\q=\q/q^+$. That is, $\p-z\q$ is proportional to the (transverse) velocity of one of the produced gluon relative to that of the  parent gluon. 

It is convenient also to consider $\Gamma^{ijl}_{abc}(\p-z\q;z) $ as a matrix connecting single gluon states to two gluon states. We set
\beq
(\p,\,b,\,j;\k,\,c,\,l|\,\Gamma(z)\,|\q,\,a,\,i)=(\k|\q-\p)\Gamma^{ijl}_{abc}(\p-z\q;z), 
\eeq
where it is understood that $p^+=zq^+$, $k^+=(1-z)q^+$. Often, we also write  $\Gamma^{ijl}_{abc}(\p-z\q;z)=gf^{abc}~\Gamma^{ijl}(\p-z\q,z)$, isolating the coupling constant and the color factor from the momentum dependent piece. 

In summary, the dynamics of gluons  in the light cone gauge $A^+=0$, and in the presence of a background field that does not depend on $x^-$, are essentially  those of a two dimensional non-relativistic field theory. Usual perturbative calculations can be simplified by using the  propagator ${\cal G}$ and the vertex $gf^{abc}~\Gamma^{ijl}(\p-z\q,z)$ that we have discussed in this section. Note that the factor $1/2p^+$ which accompanies the propagator ${\cal G}$ (see Eq.~(\ref{calG})) is not to be included on the external lines.

%%%%%%%%%%%%%%%%%%%%%%%%%%%%%%%%%
\section{Explicit calculations of $n$-point functions}\label{app:pathint}
%%%%%%%%%%%%%%%%%%%%%%%%%%%%%%%%%
In this section we calculate explicitly the $n$-point functions that are used in the main text. These $n$-point functions are products of propagators in the amplitude and its complex conjugate, i.e. products of $\cal G$'s and ${\cal G}^\dagger$', taken between the same initial time $x^+$ and the same final time $y^+$. At these times, the transverse coordinates in the various propagators may take different values. We shall denote generically by $\X$ the set of transverse coordinates in the various propagators at time $x^+$, and by $\Y$ the set of the corresponding coordinates at time $y^+$. Often, we shall  include $x^+$ and $y^+$ in the sets  $X$ and $Y$, respectively, that is, we shall set $X=(x^+,\X)$, $Y=(y^+,\Y)$. In case of a single propagator, the same notation will be used, with $X=(x^+,\x)$ and $Y=(y^+,\y)$. Further details on the notation will be given as we proceed.

Let us recall that the free propagator reads (we set ${\cal G}^{ba}_0(Y,X)\equiv\delta^{ba}{\cal G}_0(Y,X)$)
\beq
{\cal G}_0(Y,X)=\left(\frac{\omega}{2\pi i\Delta t}    \right) {\rm e}^{i\frac{\omega(\x-\y)^2}{2\Delta t}}, \qquad \Delta t\equiv x^+-y^+,
\eeq
where we have called $\omega$ the + component of the momentum, a notation that will be used throughout this appendix.  We also write ${\cal G}_0(Y,X)$ as the matrix element $(\y|{\cal G}_0(y^+,x^-)|\x)=(\y|{\cal G}_0(\Delta t)|\x)$. The Fourier transform is 
\beq
(\q|{\cal G}_0(\Delta t)|\p)&=&\int \rmd\y\, \rmd\x\,{\rm e}^{-i\q\cdot \y}\,{\rm e}^{i\p\cdot \x}\,(\y|{\cal G}_0(\Delta t)|\x)\nn
&=& (2\pi)^2\delta(\p-\q)\,{\rm e}^{-i\frac{\p^2}{2\omega}\Delta t}.
\eeq

\begin{figure}
\centering
\includegraphics[width=0.5\textwidth]{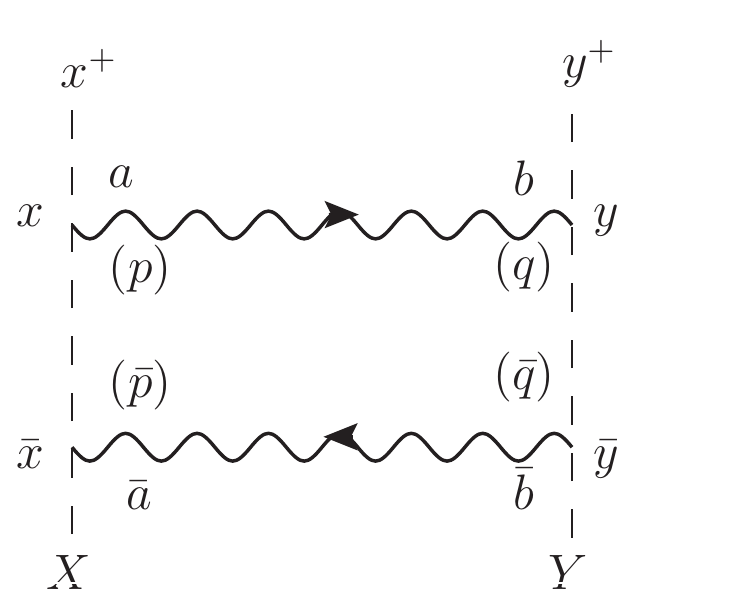}
\caption{Graphical illustration for the 2-point function $S^{(2)}(Y,X|\omega)$, where $X=(x^+,\x,\bar\x)$ and $Y=(y^+,\y,\bar \y)$. In parenthesis are given the momenta conjugate to the respective coordinates in the (mixed) Fourier representation; the + component of the momenta is denoted here by $\omega$ and is the same in the amplitude and its complex conjugate.  In the present study, a 2-point function appears typically as the product of a propagator $\cal G$ in an amplitude and a propagator ${\cal G}^\dagger$ in the complex conjugate amplitude. These propagators are represented here by oriented lines, with the arrow pointing to the right for ${\cal G}$ and to the left for ${\cal G}^\dagger$.}
\label{fig:2pf}
\end{figure}

%%%%%%%%%%%%%%%%%%%%%%%%%%%%%%%%%
\subsection{The 2-point function}\label{app:2p}
%%%%%%%%%%%%%%%%%%%%%%%%%%%%%%%%%
The scalar 2-point function $S^{(2)}(Y,X|\omega)$ is defined by
\beq\label{eq:2-point}
&&\delta^{b\bar b}\langle {\cal G}^{\dag \bar a\bar b}(\bar X,\bar Y|\omega) {\cal G}^{b a}(Y,X|\omega)\rangle =\delta^{a\bar a} S^{(2)}( Y,X|\omega).
\eeq
Note that, in order to avoid the proliferation of arguments in the functions that we discuss,  we are  using the same symbols to denote different variables, depending on the `context', that is, depending on the object where these variables appear. Thus, in the propagators  ${\cal G}(Y,X)$, $X=(x^+,\x)$, $Y=(y^+,\y)$, in  ${\cal G}^\dagger(\bar X,\bar Y)$, $\bar X=(x^+,\bar\x)$, $Y=(y^+,\bar\y)$, while in the 2-point function $S^{(2)}(Y,X|\omega)$, $X=(x^+,\x,\bar\x)$ and $Y=(y^+,\y,\bar \y)$. We shall use later a similar `contextual notation'  for the 3 and 4-point functions, where $X$ and $Y$ will represent collectively larger sets of coordinates (figures, such as Fig.~\ref{fig:2pf} will help to fix any  ambiguity that may arise). To further simplify the notation, we shall most often omit to indicate explicitly the dependence on $\omega$. Finally, whenever needed, we shall use a matrix notation, keeping separate the dependence on time arguments. Thus we shall occasionally write $S^{(2)}(Y,X)$ as the matrix element $(\Y|S^{(2)}(y^+,x^+)|\X)$ with $\X=(\x,\bar\x)$ and $\Y=(\y,\bar\y)$, with boldface letters referring to transverse coordinates only.

By using the path integral representation of the propagators ${\cal G}$ and ${\cal G}^{\dag}$ one gets, after doing the medium average, 
\beq\label{S2path}
S^{(2)}(Y,X|\omega)=\int {\cal D}\r {\cal D}\bar\r \exp\left\{\frac{i\omega}{2}\int_{x^+}^{y^+} \rmd t \left(\dot{\r}^2-\dot{\bar\r}^2\right) -\frac{N_c\, n}{2}\int_{x^+}^{y^+}  \rmd t \,\sigma(\r-\bar \r)\right\},\nonumber\\
\eeq
with  paths $\r(t)$, $\bar \r(t)$ going from $\r=\x,\bar\r=\bar\x$ at time $x^+$ to  $\r=\y, \bar\r=\bar\y$ at time $y^+$, respectively.
The change of variables
\beq\label{relativecoord2}
\u=\r-\bar\r\,,\qquad \v=\frac{1}{2}(\r+\bar\r),
\eeq
allows us to rewrite the path integral as
\beq
S^{(2)}(Y,X|\omega)=\int {\cal D}\u {\cal D}\v ~\exp\left\{{i\omega}\int_{x^+}^{y^+}  \rmd t~\dot{\u}\cdot\dot{\v}-\frac{N_c\, n}{2}\int_{x^+}^{y^+}  \rmd t\, \sigma(\u)\right\}.
\eeq
To evaluate this path integral, we first factorize the dependence on the end points, entirely contained in the value of the action for the classical paths. It is easily verified that  Euler-Lagrange equation for $\u(t)$, $\ddot\u=0$, is not modified by the interaction (even though the individual classical trajectories are no longer straight lines, as in the free case, the difference $\u(t)=\r(t)-\bar \r(t)$ remains a linear function of $t$). This is enough to calculate the  classical action. We have, in particular,
\beq
\int_{x^+}^{y^+}  \rmd t~\dot{\u}\cdot\dot{\v}=\frac{1}{2\Delta t}\left[\y-\bar\y-(\x-\bar\x)\right]\left[ \y+\bar\y-(\x+\bar\x)   \right],
\eeq
where we have used that $\dot \u$ is independent of time and set
\beq
\Delta t\equiv y^+-x^+.
\eeq
A standard calculation of the remaining quadratic path integral then yields
\beq\label{S2XY}
S^{(2)}(Y,X|\omega)&=& \left(\frac{\omega}{2\pi
\Delta t}\right)^2 \exp\left\{\frac{i\omega}{2\Delta
t}\left[(\y-\x)^2- (\bar \y-\bar \x)^2\right]-\frac{N_c\, n}{2}\int_{x^+}^{y^+} \rmd t\, \sigma(\u(t))\right\}\nn
&=&{\cal G}_0(Y,X){\cal G}^\dag_0(\bar X,\bar Y)\, \exp\left\{-\frac{N_c\, n}{2}\int_{x^+}^{y^+} \rmd t\,
\sigma(\u(t))\right\},
\eeq
where $\sigma(\u(t))$ is evaluated along  the classical path. 

By performing the Fourier transform with respect to the transverse coordinates only, we obtain the  ``mixed representation'' of the $2$-point function,  $S^{(2)}(Q,P|\omega)$ where  $Q=(y^+,\q,\bar \q)$, $P=(x^+,\p,\bar \p)$. That is, the `momentum' variables denoted by capital letters contains the conjugate to the transverse coordinates, but also the light-cone time. One may also write $S^{(2)}(Q,P|\omega)$ as the matrix element $(\Q|S^{(2)}(y^+,x^+)|\P)$, with $\P=(\p,\bar \p)$, $\Q=(\q,\bar\q)$. We have
\beq\label{FTofS2}
S^{(2)}(Q,P|\omega)=\int \rmd\x\, \rmd\bar \x\,\rmd\y\, \rmd\bar \y\, {\rm e}^{-i\q\cdot \y}\, {\rm e}^{i\p\cdot \x} \,{\rm e}^{-i\bar \p\cdot \bar \x} \,{\rm e}^{i\bar \q\cdot \bar \y}\,S^{(2)}(Y,X|\omega),
\eeq
where the signs in the various phase factors follow from our convention for the propagators ${\cal G}$ and ${\cal G}^\dagger$ that make up  $S^{(2)}$. Translational invariance of $S^{(2)}(Y,X|\omega)$ (obvious on Eq.~(\ref{S2XY})) allows us to perform the integral over the sum of coordinates, leading to a factor $(2\pi)^2\delta(\p-\bar \p-\q+\bar \q)$. To proceed further, it is convenient to change to relative coordinates. In analogy with Eq.~(\ref{relativecoord2}) we define $\u_x=\x-\bar\x$, $\v_x=(\x+\bar\x)/2$, and similarly for $\y,\bar\y$. The variable conjugate to $\q-\bar \q$ is $\v_y-\v_x$.
We are interested in the  particular Fourier  component for which the momenta at $y^+$ are the same in the amplitude and in the complex conjugate amplitude, that is $\q=\bar \q$. In this case,  the variable $\v_y-\v_x$ drops from the phase factors in Eq.~(\ref{FTofS2}), and appears only in $S^{(2)}(Y,X|\omega)$, where  $\left[(\y-\x)^2- (\bar \y-\bar \x)^2\right]/2=(\v_y-\v_x)(\u_y-\u_x)$. The  integration over $(\v_y-\v_x)$  then produces a factor $(2\pi\Delta t/\omega)^2\delta(\u_y- \u_x)$, which in turn implies that $\u(t)$ is independent of time, and also that the 2-point function depends only on the difference $\p-\q$ ($\p+\q$ being conjugate to $\u_x-\u_y$), and not on $\p$ and $\q$ separately. One finally gets
\beq\label{FTofS2b}
S^{(2)}(Q,P|\omega)&=&(2\pi)^2\delta^{(2)}(\p-\bar \p)\int \rmd\u_x
~\exp\left[i\u_x\cdot(\p-\q)-\frac{N_c\, n}{2} \,\sigma(\u_x)\,\Delta t \right],\nn
&\equiv& (2\pi)^2\delta^{(2)}(\p-\bar\p)~{\cal P}(\p-\q,\Delta t),
\eeq
where the $\delta$-function expresses momentum conservation after taking into account the condition $\q=\bar\q$. It is interesting to observe that the requirement $\q=\bar\q$ has cancelled completely the ``free'' contribution to $S^{(2)}(Q,P|\omega)$ (except for the momentum conserving $\delta$-function), leaving only in (\ref{FTofS2b}) the Fourier transform of the interaction part of $S^{(2)}(Y,X|\omega)$ (the last term in the second line of Eq.~(\ref{S2XY})). This cancellation results in the property of $S^{(2)}(Q,P|\omega)$ of being a function only of the difference of 
momenta   $\p-\q$. In the absence of interaction, ${\cal P}(\p-\q,\Delta t)$ goes over to $(2\pi)^2\delta(\p-\q)$ and the dependence on $\Delta t$ drops out.

The quantity 
\beq
{\cal P}(\Delta\p,\Delta t)=\int \rmd\u ~\exp\left[i\u\cdot\Delta\p-\frac{N_c\, n}{2}\,\sigma(\u)\, \Delta t\right].\eeq
can be interpreted as  the probability that a gluon acquires a transverse momentum $\Delta \p$ while traversing the medium during a time $\Delta t$.

%%%%%%%%%%%%%%%%%%%%%%%%%%%%%%%%%
\subsection{The 3-point function}\label{app:3p}
%%%%%%%%%%%%%%%%%%%%%%%%%%%%%%%%%
\begin{figure}
\centering
\includegraphics[width=0.5\textwidth]{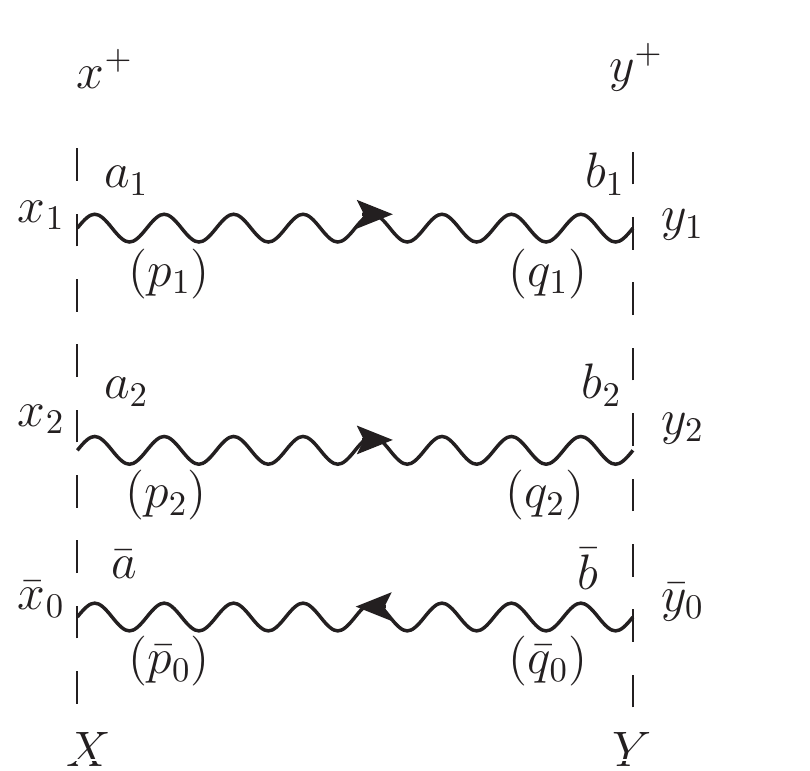}
\caption{Graphical illustration for the 3-point function $S^{(3)}(Y,X|\bs\omega)$, with $X=(x^+,\x_1,\x_2,\bar\x_0)$, $Y=(y^+,\y_1,\y_2,\bar\y_0)$ and $\bs\omega=(\omega_1,\omega_2, \omega_0)$. The typical 3-point function we deal with in this paper is the product of two ${\cal G}$ (right arrows) and one ${\cal G}^\dagger$ (left arrow). Also, the gluons 1 and 2 are  issued from gluon 0, so that conservation of the + momentum implies $\omega_0=\omega_1+\omega_2$.}
\label{fig:3-PF}
\end{figure}
The 3-point function $S^{(3)}(Y,X|\bs\omega)$ that we need is of the form
\beq\label{eq:3-point}
f^{b_1b_2\bar b_0} S^{(3)}(Y,X|\bs\omega)  = f^{a_1a_2 \bar a_0}\langle {\cal G}^{b_1 a_1}(Y,X|\omega_1){\cal G}^{b_2 a_2}(Y,X|\omega_2){\cal G}^{\dag\: \bar a_0\bar b_0}(\bar X,\bar Y|\omega_0)\rangle 
\eeq
where we use the contextual notation introduced earlier, with  the variables $X$ and $Y$ having different meanings in $S^{(3)}$ and in ${\cal G}$ or in ${\cal G}^\dagger$ (see the caption of Fig.~(\ref{fig:3-PF})).  By expressing the individual propagators in terms of path integrals, and performing the medium average of the Wilson lines, one gets
\beq
S^{(3)}(Y,X|\bs\omega)
&=&\int {\cal D}\r_2 {\cal D}\r_1 {\cal D}\r_0 \exp\left\{\frac{i}{2}\int_{x^+}^{y^+} \rmd t \left( \omega_1\dot{\r}^2_1+\omega_2 \dot{\r}^2_2-\omega_0 \dot{\r}^2_0 \right) \right\}\nn
&\times&\exp{\left\{ -\frac{N_c\, n}{4}\int_{x^+}^{y^+} \rmd t\,  \left[\sigma(\r_{1}-\r_0)+\sigma(\r_{2}-\r_0)+\sigma(\r_{2}-\r_1)\right]\right\}}.\nonumber\\
\eeq
The integral runs over paths with endpoints $\r_i(x^+)\equiv\x_i$ and $\r_i(y^+)\equiv\y_i$ ($i=0,1,2$), and the three dipole cross sections correspond to the three possible dipoles that can be formed with the three gluons. 

 We perform the following change of variables ($\v$ can be interpreted as the distance between gluon 0 and the ``center of mass'' of gluons $1$ and $2$)
\beq\label{newvariables3pts}
\u=\r_1-\r_2\,,\qquad\v=z\r_1+(1-z)\r_2-\r_0\,,\qquad \omega_1=z\omega_0\,,\qquad \omega_2=(1-z)\omega_0,\nn
\eeq
so that $\r_1-\r_0=\v-z\u$ and $\r_2-\r_0=\v+(1-z)\u$. For the endpoints, we set $\u_x=\x_1-\x_2$, $\v_x=z\x_1+(1-z)\x_2-\bar \x_0$, and similarly for the $\y$ variables. 
In the variables (\ref{newvariables3pts}), the 3-point function reads
\beq\label{3pointXY}
S^{(3)}(Y,X|\bs\omega)
&=&\int {\cal D}\u {\cal D}\v {\cal D}\r_0 \exp\left\{\frac{i\omega_0}{2}\int_{x^+}^{y^+} \rmd t\, \left[\dot{\v}^2+2\dot{\v}\cdot \dot{\r_0} +z(1-z) \dot{\u}^2\right]\right\}\nn
&\times&\exp\left\{ -\frac{N_c\, n}{4}\int_{x^+}^{y^+} \rmd t\,   \left[\sigma(\u)+\sigma(\v-z\u)+\sigma(\v+(1-z)\u)\right]\right\}.\nn
\eeq
The calculation of the path integrals over $\r_0$ and $\v$ proceeds as for the 2-point function. One identifies easily that the classical path $\v(t)$ is a straight line, that is, the center of mass of gluons 1 and 2 move with respect to gluon 0 at constant velocity. This allows us to calculate the kinetic part of the action corresponding to the variables $\r_0$ and $\v$, and to perform the remaining quadratic path integrals over $\r_0$ and $\v$. One gets
\beq
&& S^{(3)}(Y,X|\bs\omega)\equiv\left(\frac{\omega_0}{2\pi \Delta t}\right)^2\exp\left[\frac{i\omega_0}{2}\frac{\Delta \v}{\Delta t}\cdot (\Delta\v+2\Delta \r_0) \right]\nn
&\times&\int {\cal D}\u\exp\left\{\frac{i\hat \omega_0}{2}\! \int_{x^+}^{y^+} \!\!\rmd t\,\dot{\u}^2 -\frac{N_c\, n}{4}\int_{x^+}^{y^+}\! \!\rmd t\,  \left[\sigma(\u)\!+\!\sigma(\v\!-\!z\u)\!+\!\sigma(\v\!+\!(1\!-\!z)\u)\right]\right\}.\nn
\eeq
where, in the integral, $\v$ is to be taken as the classical path $\v(t)$, and we have set
\beq
 \Delta \v\equiv \v_y-\v_x,\qquad \Delta\r_0\equiv \bar\y_0-\bar\x_0, \qquad \hat\omega_0\equiv z(1-z)\omega_0,
\eeq
with $\hat\omega_0$ the reduced mass for the relative motion of gluons 1 and 2. 
Note that in the absence of interaction with the medium, $S^{(3)}(Y,X|\bs\omega)$ reduces to the product of three free propagators, as it should. We have indeed
\beq
S^{(3)}_0(Y,X|\bs\omega)&=&\left(\frac{\omega_0}{2\pi \Delta t}\right)^2{\rm e}^{\frac{i\omega_0}{2}\frac{\Delta \v}{\Delta t}\cdot (\Delta\v+2\Delta \r_0)}\times \frac{\hat\omega_0}{2\pi i \Delta t} \,{\rm e}^{i\frac{\hat\omega_0}{2\Delta t} \Delta \u^2   }\nn
&=&\sqrt{   \frac{\omega_0\omega_1\omega_2}{(2\pi)^3 i \Delta t^3}   }\;{\rm e}^{\frac{i}{2\Delta t} \left(   \omega_1\Delta \r_1^2+\omega_2\Delta \r_2^2 -\omega_0\Delta \r_0^2   \right) },
\eeq
with $\Delta \u\equiv \Delta \r_1-\Delta \r_2$, and $\Delta \r_i\equiv \y_i-\x_i$. The effect of the interaction is, as in the 2 gluon case discussed above, to produce a damping whenever the size of the dipoles exceed $1/(\hat q\Delta t)$. To analyze these effects further, it is convenient to perform a Fourier transform with respect to the transverse coordinates. 

The Fourier transform (in the mixed representation) reads
\beq\label{FTofS3}
S^{(3)}(Q,P|\bs\omega)=\int_{ \x_1\, \x_2\,\bar \x_0\,\y_1\,\y_2\, \bar \y_0}\, {\rm e}^{-i\q_1\cdot \y_1}\,{\rm e}^{-i\q_2\cdot \y_2}\, {\rm e}^{-i\bar\p_0\cdot \bar\x_0} \,{\rm e}^{i \p_1\cdot \x_1}\,{\rm e}^{i \p_2\cdot \x_2}\,{\rm e}^{i\bar \q_0\cdot \bar \y_0} \;S^{(3)}(Y,X|\bs\omega),\nn
\eeq
where  translation invariance implies $\p_1+\p_2-\q_1-\q_2=\bar\p_0-\bar\q_0$. It is convenient to introduce the following change of variables (with unit Jacobian)
\beq
\x_1=(1-z)\u_x+\v_x+\bar\x_0,\qquad\x_2=-z\u_x+\v_x+\bar\x_0,\qquad \bar\x_0=\bar\x_0,
\eeq
and similarly for the variables $\y_1,\y_2,\bar\y_0$. The phase factor in Eq.~(\ref{FTofS3}) becomes then ${\rm e}^{i\phi}$ with 
\beq
\phi&=&\bar\x_0(\p_1+\p_2-\bar\p_0)+\u_x[(1-z)\p_1-z\p_2]+\v_x(\p_1+\p_2)\\
&-& \bar\y_0(\q_1+\q_2-\bar\q_0)-\u_y[(1-z)\q_1-z\q_2]-\v_y(\q_1+\q_2).
\eeq
As obvious from its explicit expression (\ref{3pointXY}),  the 3-point function $S^{(3)}(Y,X|\bs\omega)$ depends on $\bar\x_0$ and  $\bar\y_0$ only through the difference $\bar\x_0-\bar\y_0$, which is conjugate to the variable $[\p_1+\p_2-\bar\p_0+\q_1+\q_2-\bar\q_0]/2=\q_1+\q_2-\bar\q_0$. For the particular Fourier component which we are interested in, that for which $\q_1+\q_2=\bar\q_0$, this variable $\bar\x_0-\bar\y_0$ drops from the phase factors in Eq.~(\ref{FTofS3}). One can then perform the corresponding  integration, which  yields a factor $(2\pi\Delta t/\omega_0)^2\delta(\Delta\v)$. 
The $\delta$-function $\delta(\Delta\v)=\delta(\v_x-\v_y)$ implies that  $\v(t)$ is independent of time. It also implies that, besides the momentum conserving $\delta$-function, $S^{(3)}(Q,P|\bs\omega)$ depends only on the difference $\bar\p_0-\bar\q_0$ rather than on $\bar\p_0$ and $\bar\q_0$ separately.  We are then left with 
\beq\label{eq:3-PF-FT}
 S^{(3)}(Q,P|\bs\omega)&=& (2\pi)^2\delta^{(2)}(\p_1\!+\!\p_2\!-\!\bar\p_0) \! \int \rmd\u_x \rmd\u_y \rmd\v_x \;{\rm e}^{i\u_x\cdot (\p_1-z\bar\p_0)-i\u_y\cdot(\q_1-z\bar\q_0)+i\v_x\cdot(\bar\p_0-\bar\q_0)}\nn
&&\int {\cal D}\u\exp\left\{\frac{i\hat\omega_0}{2} \int \rmd t ~\dot{\u}^2 -\frac{N_c\, n}{4}\int \rmd t\,   \left[\sigma(\u)\!+\!\sigma(\v\!-\!z\u)\!+\!\sigma(\v\!+\!(1\!-\!z)\u)\right]\right\},\nn
\eeq
where the $\delta$-function is that of momentum conservation, after taking into account that $\q_1+\q_2=\bar\q_0$.

To proceed further, we  evaluate Eq. (\ref{eq:3-PF-FT}) in the harmonic  approximation, \eqn{harmonic_approx}.  Then the path integral in the second line of Eq. (\ref{eq:3-PF-FT}) reads simply
\beq
{\rm e}^{-3\v_x^2\hat q\Delta t/8[1+z^2+(1-z)^2]}\;\int {\cal D}\u\exp\left\{   i\hat\omega_0  \int_{x^+}^{y^+} \rmd t ~\left(\frac{ \dot{\u}^2}{2}+i\omega_{_{\rm br}}^2 \u^2 \right) \right\},
\eeq
where we have set
\be
\tau_{_{\rm br}}\equiv \sqrt{\frac{\hat\omega_0}{\hat q_{\text{eff}}}},\qquad \omega_{_{\rm br}}^2\equiv \frac{1}{4\tau_{_{\rm br}}^2},\qquad \hat q_{\text{eff}}\equiv\frac{1}{2}\hat q \left[(1-z)^2+z^2+1\right].
\ee
The calculation of the quadratic   integral 
\beq\label{integralJ}
{\cal J}(\u_x,\u_y)=\int {\cal D}\u\exp\left\{   i\hat \omega_0  \int_{x^+}^{y^+} \rmd t ~\left( \frac{\dot{\u}^2}{2}+i\omega_{_{\rm br}}^2 \u^2 \right) \right\}.
\eeq
is standard, and yields
\beq
{\cal J}(\u_x,\u_y)&=& \frac{k_{_{\rm br}}^2 (1-i) }{4\pi \sinh \Omega\Delta t}\exp\left\{   \frac{(i-1)k_{_{\rm br}}^2}{4\sinh \Omega\Delta t} \left[  ( \u_x^2+\u_y^2)\cosh \Omega\Delta t -2 \u_x\cdot \u_y  \right] \right\}\nn
&=& \frac{k_{_{\rm br}}^2 (1-i) }{4\pi \sinh \Omega\Delta t}\exp\left\{   \frac{(i-1)k_{_{\rm br}}^2}{2\sinh \Omega\Delta t} \left[ \sinh^2\frac{\Omega\Delta t}{2} (\u_s+\u_y)^2+\cosh^2\frac{\Omega\Delta t}{2} (\u_s-\u_y)^2\right]   \right\},\nn
\eeq
with $\Omega\equiv (1+i)\omega_{_{\rm br}}$ and $k_{_{\rm br}}^2\equiv \hat q_{\text{eff}}\tau_{_{\rm br}}$.

The Fourier transform can  then be completed. We get
\beq\label{eq:3-PF-FT-HO}
 &&S^{(3)}(Q,P|\bs\omega)= (2\pi)^2\delta^{(2)}(\p_1\!+\!\p_2\!-\!\bar\p_0)\,\frac{8\pi[ 1+z^2+(1-z)^2]}{3\hat q\Delta t}\,  {\rm e}^{ -\frac{2[ 1+z^2+(1-z)^2](\bar\p_0-\bar\q_0)^2}{3\hat q\Delta t}  }\nn
 &\times&  \frac{2\pi(1+i)}{k_{_{\rm br}}^2\sinh(\Omega \Delta t) }
\exp{\left\{  -(1+i)\frac{(\hat\P_1+\hat\Q_1)^2}{4k_{_{\rm br}}^2\coth(\Omega \Delta t/2)}
-(1+i)\frac{(\hat\P_1-\hat\Q_1)^2}{4k_{_{\rm br}}^2\tanh(\Omega \Delta t/2)}\right\}  },\nn
\eeq
where we have set $\hat \P_1\equiv\p_1-z\bar\p_0$ and $\hat\Q_1\equiv\q_1-z\bar\q_0$. Setting $S^{(3)}(Q,P|\bs\omega)= (2\pi)^2\delta^{(2)}(\p_1\!+\!\p_2\!-\!\bar\p_0)\,\tilde S^{(3)}(Q,P|\bs\omega)$, we note that $\tilde S^{(3)}(Q,P|\bs\omega)$ is a function of just three variables, $\bar\p_0-\bar\q_0$, and the variables $\hat \P_1$, $\hat \Q_1$ just defined, and which may be interpreted as the relative momenta of the 2-dimension motion at $x^+$ and $y^+$, with $\p_0$ and $\bar\q_0$ playing the role of respective `center of mass' momenta.

%%%%%%%%%%%%%%%%%%%%%%%%%%%%%%%%%
\subsection{The path integral needed in the evaluation of the 4-point function}
\label{app:4p}
%%%%%%%%%%%%%%%%%%%%%%%%%%%%%%%%%

\begin{figure}
\centering
\includegraphics[width=0.5\textwidth]{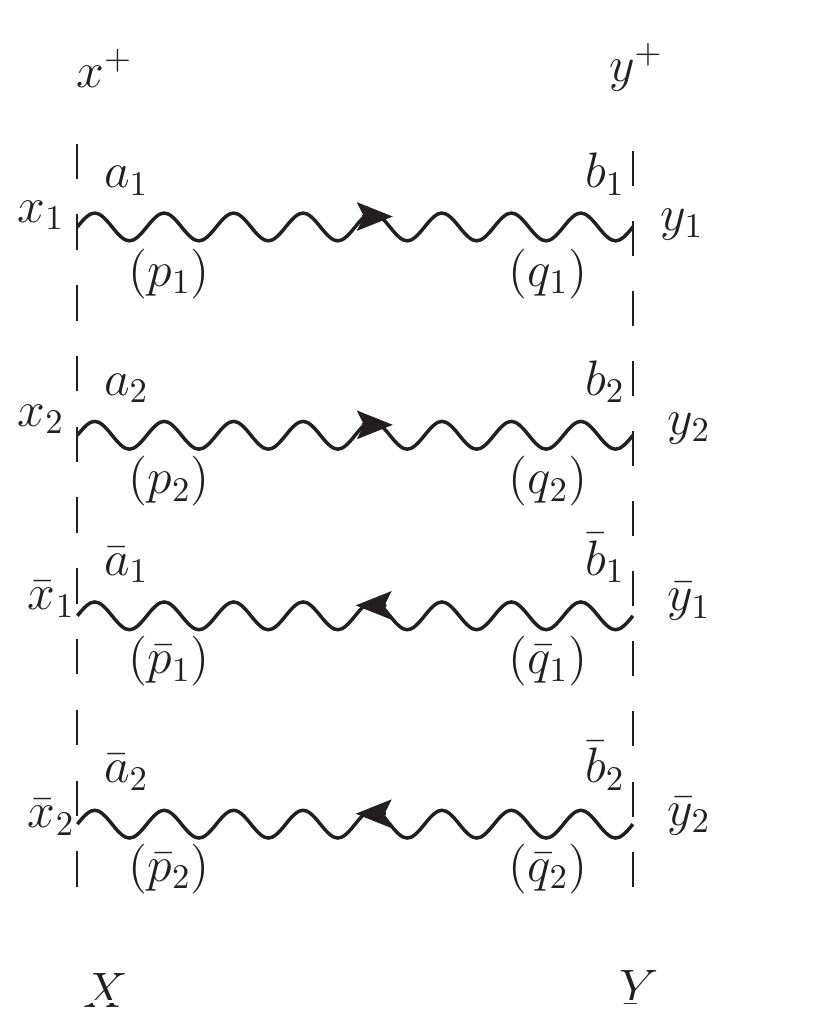}
\caption{Graphical illustration for the path integral ${\cal I}(Y,X|\bs\omega)$, with $X=(x^+,\x_1,\x_2,\bar\x_1,\bar \x_2)$, $Y=(y^+,\y_1,\y_2,\bar\y_1,\bar \y_2)$ and $\bs\omega=(\omega_1,\omega_2)$. The 4-point function is the product of two propagators ${\cal G} $ (for gluons 1 and 2, with + momenta $\omega_1$ and $\omega_2$,  and two propagators ${\cal G}^\dagger$ corresponding to the same gluons in the complex conjugate amplitude. The  notation is as in the previous two figures.}
\label{fig:4-PF}
\end{figure}

In this appendix we evaluate the path integral entering the evaluation of the non-factorizable piece of the 4-point function (see Eq.~(\ref{4pointApp})), which in coordinate space takes the form (see Fig. \ref{fig:4-PF} for notation)
\beq
{\cal I}(Y,X|\bs\omega)
&=&\int {\cal D}\r_1 {\cal D}\r_2 {\cal D}\bar \r_1 {\cal D}\bar \r_2 \exp\left\{\frac{i}{2}\int_{x^+}^{y^+} \rmd t \left( \omega_1\dot{\r}^2_1+\omega_2 \dot{\r}^2_2-\omega_1 \dot{\bar\r}^2_1-\omega_2 \dot{\bar\r}^2_2 \right) \right\}\nn
&\times&\exp{\left\{ -\frac{C_F\, n}{2}\int_{x^+}^{y^+} \rmd t\,  \left[\sigma(\r_{1}-\bar\r_1)+\sigma(\r_{2}-\bar \r_2)+\sigma(\r_{1}-\r_2)+\sigma(\bar\r_{1}-\bar\r_2)\right]\right\}}.\nonumber\\
\eeq
The integral runs over paths with endpoints $\r_i(x^+)\equiv\x_i$ and $\r_i(y^+)\equiv\y_i$ ($i=1,2$), and similarly for $\bar \r_i$. The factor $C_F$ in the second exponential of the formula above originates from the fact that the average of the Wilson lines that  leads to this expression involves quark dipoles $C^{(2)}_q$ (see \eqn{4pointApp}).

 We perform the following change of variables (with unit Jacobian)
\beq\label{newcoordinates4a}
\u=\r_1-\r_2,\quad \bar \u=\bar \r_1-\bar \r_2,\quad\v&=&z(\r_1-\bar \r_1)+(1-z)(\r_2-\bar\r_2),
\nn
\w&=& z\frac{\r_1+\bar\r_1}{2}+(1-z)\frac{\r_2+\bar\r_2}{2}.
\eeq
with as before $\omega_1=z\omega_0,  \omega_2=(1-z)\omega_0$, and $\hat\omega_0=z(1-z)\omega_0$. The inverse transformation reads
\beq\label{newcoordinates4b}
&&\r_1=\w+(1-z)\u+\frac{\v}{2}, \qquad \r_2=\w-z\u+\frac{\v}{2},\nn
&&\bar \r_1=\w+(1-z)\bar\u-\frac{\v}{2}, \qquad\bar \r_2=\w-z\bar\u-\frac{\v}{2},\
\eeq
We  also define
\beq\label{newcoordinates4c}
&&\x_1=\w_x+(1-z)\u_x+\frac{\v_x}{2}, \qquad \x_2=\w_x-z\u_x+\frac{\v_x}{2},\nn
&&\bar \x_1=\w_x+(1-z)\bar\u_x-\frac{\v_x}{2}, \qquad\bar \x_2=\w_x-z\bar\u_x-\frac{\v_x}{2},\
\eeq
and similarly for the $\y$-variables.
In the variables (\ref{newcoordinates4a}), the path integral reads
\beq
{\cal I}(Y,X|\bs\omega)
&=&\int {\cal D}\u   {\cal D}\bar \u {\cal D}\v {\cal D}\w\,\exp\left\{\frac{i\hat\omega_0}{2}\int_{x^+}^{y^+} \rmd t \left[(\dot \u^2-\dot{\bar\u}^2)+2\dot \v\cdot\dot\w \right] \right\}\nn
&\times&{\rm e}^{\left\{ -\frac{C_F\, n}{2}\int_{x^+}^{y^+} \rmd t\,   \left[\sigma((1-z)(\u-\bar \u)+\v)+\sigma(-z(\u-\bar\u)+\v)+\sigma(\u)+\sigma(\bar\u)\right]\right\}}.\nonumber\\
\eeq
The Euler Lagrange equation for $\w$ yields $\ddot \v=0$, that is, the classical path $\v(t)$ is a linear function of time. This allows us to calculate the integral over $\v$ and $\w$, and obtain
\beq\label{S4YX}
{\cal I}(Y,X|\bs\omega)
&=&\left(\frac{\omega_0}{2\pi \Delta t } \right)^2\,{\rm e}^{i\omega_0\frac{\Delta \v}{\Delta t}\Delta\w}\int {\cal D}\u   {\cal D}\bar \u\,{\rm e}^{\frac{i\hat\omega_0}{2}\int_{x^+}^{y^+} \rmd t\, (\dot \u^2-\dot\bar\u^2)  }\nn
&\times&{\rm e}^{\left\{ -\frac{C_F\, n}{2}\int_{x^+}^{y^+} \rmd t\,  \left[\sigma((1-z)(\u-\bar \u)+\v)+\sigma(-z(\u-\bar\u)+\v)+\sigma(\u)+\sigma(\bar\u)\right]\right\}},\nonumber\\
\eeq
where $\Delta\w=\w_y-\w_x$, $\Delta\v=\v_y-\v_x$ and $\v(t)=\v_x+(\Delta\v/\Delta t) (t-x^+)$. Note that if one ignores the second line of Eq.~(\ref{S4YX}), one recovers ${\cal I}(Y,X|\bs\omega)$ as a product of four free propagators, expressed in the variables (\ref{newcoordinates4a}).

At this point, it is convenient to go to the mixed representation, and take a Fourier transform with respect to the transverse coordinates. We get
\beq\label{FTofS4}
{\cal I}(Q,P)=\int_{\{\x,\y\}}\, {\rm e}^{-i\q_1\cdot \y_1}\,{\rm e}^{-i\q_2\cdot \y_2}\,{\rm e}^{i\bar \q_1\cdot \bar \y_1}\,{\rm e}^{i\bar \q_2\cdot \bar \y_2} \,{\rm e}^{i \p_1\cdot \x_1}\,{\rm e}^{i \p_2\cdot \x_2} \, {\rm e}^{-i\bar\p_1\cdot \bar\x_1}\, {\rm e}^{-i\bar\p_2\cdot \bar\x_2}\;{\cal I}(Y,X),\nn
\eeq
where $\int_{\{\x,\y\}}$ denotes the integration over the $2\times 8$ transverse coordinates, and the variables $P,Q$ are those of the mixed representation, e.g., $P=(x^+, \p_1,\p_2,\bar\p_1,\bar\p_2)$.
Translational invariance (one can easily verify that ${\cal I}(Y,X)$ is unchanged in a constant shift of all the coordinates) implies $\p_1+\p_2-\bar\p_1-\bar\p_2=\q_1+\q_2-\bar\q_1-\bar\q_2$. After changing to the variables (\ref{newcoordinates4c}), and taking into account momentum conservation, one finds that $\q_1+\q_2-\bar\q_1-\bar\q_2$ is conjugate to $\w_x-\w_y$. Now, we are interested in the particular Fourier component for which $\q_1+\q_2=\bar\q_1+\bar\q_2$. For this component,  the dependence of the phase factor in Eq.~(\ref{FTofS4}) on $\Delta\w=\w_y-\w_x$ drops out. The integration over $\Delta\w$ can then be performed, leading to a factor $(2\pi\Delta t/\omega_0)^2\delta(\Delta v)$, while the integration over $\w_y+\w_x$ leaves a factor $(2\pi)^2\delta(\p_1+\p_2-\bar\p_1-\bar\p_2)$. We are then left with
\beq
&&\tilde{\cal I}(Q,P)=  \int_{ \v_x u_x\u_y\bar \u_x\bar \u_y}\, {\rm e}^{i\v_x\cdot(\p_1+\p_2-\q_1-\q_2)}\nn
&\times&{\rm e}^{i\u_x\cdot((1-z)\p_1-z\p_2)}\,{\rm e}^{-i\u_y((1-z)\q_1-z\q_2)}\,{\rm e}^{-i\bar \u_x\cdot((1-z)\bar\p_1-z\bar\p_2)} \,{\rm e}^{i \bar\u_y\cdot ((1-z)\bar\q_1-z\bar\q_2)} \nn
&\times&\int {\cal D}\u   {\cal D}\bar \u\;\; {\rm e}^{  i\frac{\hat\omega_0 }{2}\int_{x^+}^{y^+} \rmd t \, (\dot \u^2-\dot{\bar\u}^2)   }\nn
&\times&{\rm e}^{  -\frac{C_F\, n}{2}\int_{x^+}^{y^+} \rmd t\,   \left[\sigma((1-z)(\u-\bar \u)+\v)+\sigma(-z(\u-\bar\u)+\v)+\sigma(\u)+\sigma(\bar\u)\right]  }.
\eeq
In the path integral in the expression above, the endpoints of the paths are $\u_x,\u_y$ and $\bar\u_x,\bar\u_y$, respectively, and $\v=\v_x$ is a constant.

We shall evaluate the path integrals over $\u$ and $\bar\u$ in the harmonic approximation, i.e, set $ N_cn\, \sigma(\u)\approx \hat q \, \u^2/2$. We write the path integral as
\beq
\int {\cal D}\u   {\cal D}\bar \u\;\; {\rm e}^{  \int_{x^+}^{y^+} \rmd t {\cal L}(\u,\bar u) },\qquad  {\cal L}={\cal L}_0+{\cal L}_1,\qquad {\cal L}_0= i \frac{\hat\omega_0 }{2} (\dot{\bar\u}^2-\dot \u^2).
\eeq
In the harmonic approximation we get
\beq
{\cal L}_1\approx  -\frac{C_F}{4N_c} \hat q \left\{A(\u^2+\bar\u^2)-2\,(A-1)\u\cdot\bar\u+2B(\u-\bar\u)\cdot \v +2\,\v^2\right\},
\eeq
with
\beq
A=1+z^2+(1-z)^2,\qquad B= 1-2z.
\eeq
The shift of variables 
\beq
\u\to \u-\frac{B}{2A-1}\,\v\qquad \bar\u\to\bar \u+\frac{B}{2A-1}\,\v\,,
\eeq
where $\v$ is a constant vector, leaves the measure of the path integral unchanged, as well as ${\cal L}_0$, and reduces ${\cal L}_1$ to 
\beq
{\cal L}_1=-\frac{C_F}{4N_c} \hat q \left\{A(\u^2+\bar\u^2)-2\,(A-1)\u\cdot\bar\u\right\}\,.
\eeq
 At this point, we perform a boost-like transformation, 
 \beq
\left(
\begin{array}{c}
  \u   \\
  \bar\u  
\end{array}
\right)
=\gamma \left(
\begin{array}{cc}
  1& \beta     \\
 \beta & 1     
\end{array}
\right)
\left(
\begin{array}{c}
  \u'   \\
    \bar \u'
\end{array}
\right),\qquad  \gamma=\frac{1}{\sqrt{1-\beta^2} },
\eeq
which again  leaves ${\cal L}_0$  unchanged, that is, ${\cal L}_0(\u,\bar\u)={\cal L}_0(\u',\bar\u')$. This leaves also the measure of the path integral unchanged. By choosing $\beta$ as a solution of the equation 
\beq\label{eqnforbeta}
\beta^2+1-\frac{2A}{A-1}\beta=0,\qquad \beta=\frac{A\pm \sqrt{2A-1}}{A-1},
\eeq
one can write ${\cal L}_1$ in the factorized form
\beq
&&{\cal L}_1= -\frac{C_F}{4N_c} \hat q\gamma^2 \left[A(1+\beta^2)-2(A-1)\,\beta\right]\,(\u'^2+\bar\u'^2).
\eeq
Simple algebra yields
\beq
C(z)\equiv\frac{A(1+\beta^2)-2(A-1)\,\beta}{1-\beta^2}=\pm \sqrt{2A-1}=\pm\sqrt{2z^2+2(1-z)^2+1}\,,
\eeq
where the two signs correspond to the two possible solution of Eq.~(\ref{eqnforbeta}). One can then write the path integral as
\beq
\int {\cal D}\u'{\cal D}\bar\u' \;{\rm e}^{  i \frac{\hat\omega_0 }{2}\int_{x^+}^{y^+} \rmd t  (\dot \u'^2-\dot{\bar\u}'^2)    }\;{\rm e}^{ -\frac{C_F}{4N_c}  C(z)\int_{x^+}^{y^+} \rmd t \hat q (\u'^2+\bar\u'^2)  },
\eeq
and the convergence  imposes the choice of the positive sign for $C(z)$. 
This is the product of two integral of the type of Eq.~(\ref{integralJ}):
\beq
{\cal J}(\u'_x,\u_y')\,\bar{\cal J}(\bar \u'_x,\bar \u'_y)
\eeq
with here
\beq
&&{\cal J}(\u'_x,\u'_y)=\frac{\hat\omega_0\Omega}{2\pi i \sinh(\Omega \Delta t)}\, \exp\left\{i\frac{\hat\omega_0\Omega}{2\sinh(\Omega\Delta t)}\left[\cosh(\Omega\Delta t)(\u'^2_x+\u'^2_y)-2\u'_x\cdot \u'_y\right]\right\}\,,\nn
%&&\bar{\cal K}(\Delta t,{\bar\u}')=\frac{z(1-z)\omega_0\bar\Omega}{2\pi i \sin(\bar\Omega \Delta t)}\, \exp\left\{i\frac{z(1-z)(1-\beta^2)\omega_0\bar\Omega}{2\sin(\bar\Omega\Delta t)}\left[\cos(\bar\Omega\Delta t)({\bar\u}'^2_x+\bar\u'^2_y)-2{\bar\u}'_x\cdot {\bar\u}'_y\right]\right\}\,\nn
\eeq
and similarly for $\bar {\cal J}$ with
\beq
&&\Omega=\frac{1+i}{2}\sqrt{\frac{\hat q\, C_FC (z) }{\hat\omega_0\,N_c}}\,,\qquad\bar \Omega=\frac{1-i}{2}\sqrt{\frac{\hat q\, C_FC (z) }{\hat\omega_0\,N_c}}\,,\nn
\eeq
and the values of $\u'_x$, $\u'_y$, $\bar\u'_x$, $\bar\u'_y$ can be obtained from the inverse transformation
 \beq
\left(
\begin{array}{c}
  \u'   \\
  \bar\u'  
\end{array}
\right)
=\gamma \left(
\begin{array}{cc}
  1& -\beta     \\
 -\beta & 1     
\end{array}
\right)
\left(
\begin{array}{c}
  \u +\frac{B}{2(A-1)}\v \\
    \bar \u-\frac{B}{2(A-1)}\v
\end{array}
\right).
\eeq
At this point, it is  clear that  the 4-point function that correlates gluons $a$ and $b$  is exponentially damped on a time scale comparable to that of the 3-point function, more specifically as 
\be\label{taubranching4}
\Delta t \gtrsim  \sqrt{\frac{\hat\omega_0}{\hat q\,C (z) } \frac{N_c}{C_F}}\approx \tau_{_{\rm br}}.
\ee

%%%%%%%%%%%%%%%%%%%%%%%%%%%%%%%%%%
\section{Color algebra}\label{app:color}
%%%%%%%%%%%%%%%%%%%%%%%%%%%%%%%%%%

In this appendix we review the color algebra necessary to justify Eq. (\ref{4pointfund}) by replacing adjoint Wilson lines by fundamental ones. This manipulation can be made in several different ways, here we choose to first replace all adjoint Wilson lines by fundamental ones by means of the identity in Eq. (\ref{adjtofund}) and then remove all explicit color matrices $t$'s and $f$'s by using the Fierz identity $t^a_{ij}t^a_{kl}=\frac{1}{2}\delta_{il}\delta_{jk}-\frac{1}{2N_c}\delta_{ij}\delta_{kl}$ and the Lie algebra relation $[t^a,t^b]=if^{abc}t^c$. In order to have shorter expressions we will use the shorthand notation $U_1\equiv U(\r_1)$, $U(\bar\r_1)\equiv U_{\bar 1}$ and similarly for the other coordinates and adjoint Wilson lines.
\beq
&&f^{mn\bar e}f^{\bar c \bar d \bar e}\tilde{U}_{1am}\tilde{U}_{2bn}\tilde{U}^\dagger_{\bar 1\bar c a}\tilde{U}^\dagger_{\bar 2\bar d b}\nn
&=&16f^{mn\bar e}f^{\bar c \bar d \bar e}\text{Tr}\left(U_1^\dagger t^aU_1t^m\right)\text{Tr}\left(U_2^\dagger t^bU_2t^n\right)\text{Tr}\left(U_{\bar 1}^\dagger t^aU_{\bar 2}t^{\bar c}\right)\text{Tr}\left(U_{\bar 2}^\dagger t^bU_{\bar 2}t^{\bar d}\right).
\eeq
Here we can directly apply the Fierz identity to the pairs of matrices $t^a$ and $t^b$. One can easily see that the second term of the identity would remove one of the color matrices form the traces, in which case the Wilson lines cancel out and one is left with a trace of a fundamental color matrix, which is zero. One therefore obtains
\beq
4f^{mn\bar e}f^{\bar c \bar d \bar e}\text{Tr}\left[U_1^\dagger U_{\bar 1}t^{\bar c}U_{\bar 1}^\dagger U_1t^m\right]\text{Tr}\left[U_2^\dagger U_{\bar 2}t^{\bar d}U_{\bar 2}^\dagger U_2t^n\right].
\eeq
Now all $t$ matrices have different color indices. In order to proceed we need to get rid of the $f$ symbols, arriving at
\beq
4\text{Tr}\left(U_1^\dagger U_{\bar 1}\left[t^{\bar d},t^{\bar e}\right]U_{\bar 1}^\dagger U_1t^m\right)\text{Tr}\left(U_2^\dagger U_{\bar 2}t^{\bar d}U_{\bar 2}^\dagger U_2\left[t^m,t^{\bar e}\right]\right).
\eeq
Now we have three pairs of $t$ matrices with the same color index and therefore we can make use of the Fierz identity three more times. Notice that the contribution from the second term of the Fierz identity is always zero since it would eliminate one of the matrices inside a commutator. By successively eliminating matrices with the indices $\bar d$, $m$, and $\bar e$, we arrive at
\beq
&&2\left[\text{Tr}\left(U_1^\dagger U_{\bar 1}U_{\bar 2}^\dagger U_2\left[t^m,t^{\bar e}\right]U_2^\dagger U_{\bar 2}t^{\bar e}U_{\bar 1}^\dagger U_1t^m\right) - \text{Tr}\left(U_1^\dagger U_{\bar 1}t^{\bar e}U_{\bar 2}^\dagger U_2\left[t^m,t^{\bar e}\right]U_2^\dagger U_{\bar 2}U_{\bar 1}^\dagger U_1t^m\right)\right]\nn
&=&\text{Tr}\left(U_1^\dagger U_{\bar 1}U_{\bar 2}^\dagger U_2\right)\text{Tr}\left(t^{\bar e}U_2^\dagger U_{\bar 2}t^{\bar e}U_{\bar 1}^\dagger U_1\right)-\text{Tr}\left(U_1^\dagger U_{\bar 1}U_{\bar 2}^\dagger U_2t^{\bar e}\right)\text{Tr}\left(U_2^\dagger U_{\bar 2}t^{\bar e}U_{\bar 1}^\dagger U_1\right)\nn
&&-\text{Tr}\left(U_1^\dagger U_{\bar 1}t^{\bar e}U_{\bar 2}^\dagger U_2\right)\text{Tr}\left(t^{\bar e}U_2^\dagger U_{\bar 2}U_{\bar 1}^\dagger U_1\right)+\text{Tr}\left(U_1^\dagger U_{\bar 1}t^{\bar e}U_{\bar 2}^\dagger U_2t^{\bar e}\right)\text{Tr}\left(U_2^\dagger U_{\bar 2}U_{\bar 1}^\dagger U_1\right)\nn
&=&\frac{1}{2}\left[\text{Tr}\left(U_1^\dagger U_{\bar 1}U_{\bar 2}^\dagger U_2\right)\text{Tr}\left(U_{\bar 1}^\dagger U_1\right)\text{Tr}\left(U_2^\dagger U_{\bar 2}\right)-\text{Tr}\left(U_1^\dagger U_{\bar 1}U_{\bar 2}^\dagger U_2U_{\bar 1}^\dagger U_1U_2^\dagger U_{\bar 2}\right)\right.\nn
&&\left.-\text{Tr}\left(U_1^\dagger U_{\bar 1}U_2^\dagger U_{\bar 2}U_{\bar 1}^\dagger U_1U_{\bar 2}^\dagger U_2\right)+\text{Tr}\left(U_1^\dagger U_{\bar 1}\right)\text{Tr}\left(U_{\bar 2}^\dagger U_2\right)\text{Tr}\left(U_2^\dagger U_{\bar 2}U_{\bar 1}^\dagger U_1\right)\right].
\eeq
The last line above is exactly the expression in (\ref{4pointfund}).

%\bibliography{mybib}{}

\begin{thebibliography}{10}

\bibitem{Aamodt:2010jd}
{\bf ALICE} Collaboration, K.~Aamodt and C.~A. Loizides, {\it {Suppression of
  Charged Particle Production at Large Transverse Momentum in Central Pb--Pb
  Collisions at $\sqrt{S_{_{Nn}}} = 2.76$ TeV}},  {\em Phys. Lett.} {\bf B696}
  (2011) 30--39, [\href{http://xxx.lanl.gov/abs/1012.1004}{{\tt
  arXiv:1012.1004}}].

\bibitem{Aad:2010bu}
{\bf Atlas} Collaboration, G.~Aad {\em et.~al.}, {\it {Observation of a
  Centrality-Dependent Dijet Asymmetry in Lead-Lead Collisions at Sqrt(S(Nn))=
  2.76 TeV with the Atlas Detector at the Lhc}},  {\em Phys. Rev. Lett.} {\bf
  105} (2010) 252303, [\href{http://xxx.lanl.gov/abs/1011.6182}{{\tt
  arXiv:1011.6182}}].

\bibitem{Chatrchyan:2011sx}
{\bf CMS} Collaboration, S.~Chatrchyan {\em et.~al.}, {\it {Observation and
  Studies of Jet Quenching in Pbpb Collisions at Nucleon-Nucleon Center-Of-Mass
  Energy = 2.76 TeV}},  {\em Phys. Rev.} {\bf C84} (2011) 024906,
  [\href{http://xxx.lanl.gov/abs/1102.1957}{{\tt arXiv:1102.1957}}].

\bibitem{Chatrchyan:2012ni}
{\bf CMS} Collaboration, S.~Chatrchyan {\em et.~al.}, {\it {Jet Momentum
  Dependence of Jet Quenching in Pbpb Collisions at Sqrt(Snn)=2.76 TeV}},
  \href{http://xxx.lanl.gov/abs/1202.5022}{{\tt arXiv:1202.5022}}.

\bibitem{Putschke:2008wn}
{\bf STAR} Collaboration, J.~Putschke, {\it {First Fragmentation Function
  Measurements from Full Jet Reconstruction in Heavy-Ion Collisions at
  $\sqrt{S_{_{\rm Nn}}}=200$ GeV by Star}},  {\em Eur. Phys. J.} {\bf C61}
  (2009) 629--635, [\href{http://xxx.lanl.gov/abs/0809.1419}{{\tt
  arXiv:0809.1419}}].

\bibitem{Salur:2008hs}
{\bf STAR} Collaboration, S.~Salur, {\it {First Direct Measurement of Jets in
  $\sqrt{S_{Nn}}=200$ GeV Heavy Ion Collisions by Star}},  {\em Eur. Phys. J.}
  {\bf C61} (2009) 761--767, [\href{http://xxx.lanl.gov/abs/0809.1609}{{\tt
  arXiv:0809.1609}}].

\bibitem{Lai:2009zq}
{\bf PHENIX} Collaboration, Y.-S. Lai, {\it {Probing Medium-Induced Energy Loss
  with Direct Jet Reconstruction in P+P and Cu+Cu Collisions at Phenix}},  {\em
  Nucl. Phys.} {\bf A830} (2009) 251c--254c,
  [\href{http://xxx.lanl.gov/abs/0907.4725}{{\tt arXiv:0907.4725}}].

\bibitem{Baier:1996kr}
R.~Baier, Y.~L. Dokshitzer, A.~H. Mueller, S.~Peigne, and D.~Schiff, {\it
  {Radiative Energy Loss of High Energy Quarks and Gluons in a Finite-Volume
  Quark-Gluon Plasma}},  {\em Nucl. Phys.} {\bf B483} (1997) 291--320,
  [\href{http://xxx.lanl.gov/abs/hep-ph/9607355}{{\tt hep-ph/9607355}}].

\bibitem{Baier:1996sk}
R.~Baier, Y.~L. Dokshitzer, A.~H. Mueller, S.~Peigne, and D.~Schiff, {\it
  {Radiative Energy Loss and P(T)-Broadening of High Energy Partons in
  Nuclei}},  {\em Nucl. Phys.} {\bf B484} (1997) 265--282,
  [\href{http://xxx.lanl.gov/abs/hep-ph/9608322}{{\tt hep-ph/9608322}}].

\bibitem{Baier:1998kq}
R.~Baier, Y.~L. Dokshitzer, A.~H. Mueller, and D.~Schiff, {\it {Medium-Induced
  Radiative Energy Loss: Equivalence Between the Bdmps and Zakharov
  Formalisms}},  {\em Nucl. Phys.} {\bf B531} (1998) 403--425,
  [\href{http://xxx.lanl.gov/abs/hep-ph/9804212}{{\tt hep-ph/9804212}}].

\bibitem{Zakharov:1996fv}
B.~G. Zakharov, {\it {Fully Quantum Treatment of the Landau-Pomeranchuk-Migdal
  Effect in Qed and QCD}},  {\em JETP Lett.} {\bf 63} (1996) 952--957,
  [\href{http://xxx.lanl.gov/abs/hep-ph/9607440}{{\tt hep-ph/9607440}}].

\bibitem{Zakharov:1997uu}
B.~G. Zakharov, {\it {Radiative Energy Loss of High Energy Quarks in
  Finite-Size Nuclear Matter and Quark-Gluon Plasma}},  {\em JETP Lett.} {\bf
  65} (1997) 615--620, [\href{http://xxx.lanl.gov/abs/hep-ph/9704255}{{\tt
  hep-ph/9704255}}].

\bibitem{Wiedemann:1999fq}
U.~A. Wiedemann and M.~Gyulassy, {\it {Transverse Momentum Dependence of the
  Landau-Pomeranchuk- Migdal Effect}},  {\em Nucl. Phys.} {\bf B560} (1999)
  345--382, [\href{http://xxx.lanl.gov/abs/hep-ph/9906257}{{\tt
  hep-ph/9906257}}].

\bibitem{Wiedemann:2000za}
U.~A. Wiedemann, {\it {Gluon Radiation Off Hard Quarks in a Nuclear
  Environment: Opacity Expansion}},  {\em Nucl. Phys.} {\bf B588} (2000)
  303--344, [\href{http://xxx.lanl.gov/abs/hep-ph/0005129}{{\tt
  hep-ph/0005129}}].

\bibitem{Wiedemann:2000tf}
U.~A. Wiedemann, {\it {Jet Quenching Versus Jet Enhancement: a Quantitative
  Study of the Bdmps-Z Gluon Radiation Spectrum}},  {\em Nucl. Phys.} {\bf
  A690} (2001) 731--751, [\href{http://xxx.lanl.gov/abs/hep-ph/0008241}{{\tt
  hep-ph/0008241}}].

\bibitem{Gyulassy:2000fs}
M.~Gyulassy, P.~Levai, and I.~Vitev, {\it {Non-Abelian Energy Loss at Finite
  Opacity}},  {\em Phys. Rev. Lett.} {\bf 85} (2000) 5535--5538,
  [\href{http://xxx.lanl.gov/abs/nucl-th/0005032}{{\tt nucl-th/0005032}}].

\bibitem{Gyulassy:2000er}
M.~Gyulassy, P.~Levai, and I.~Vitev, {\it {Reaction Operator Approach to
  Non-Abelian Energy Loss}},  {\em Nucl. Phys.} {\bf B594} (2001) 371--419,
  [\href{http://xxx.lanl.gov/abs/nucl-th/0006010}{{\tt nucl-th/0006010}}].

\bibitem{Arnold:2001ba}
P.~B. Arnold, G.~D. Moore, and L.~G. Yaffe, {\it {Photon Emission from
  Ultrarelativistic Plasmas}},  {\em JHEP} {\bf 11} (2001) 057,
  [\href{http://xxx.lanl.gov/abs/hep-ph/0109064}{{\tt hep-ph/0109064}}].

\bibitem{Arnold:2001ms}
P.~B. Arnold, G.~D. Moore, and L.~G. Yaffe, {\it {Photon Emission from Quark
  Gluon Plasma: Complete Leading Order Results}},  {\em JHEP} {\bf 12} (2001)
  009, [\href{http://xxx.lanl.gov/abs/hep-ph/0111107}{{\tt hep-ph/0111107}}].

\bibitem{Arnold:2002ja}
P.~B. Arnold, G.~D. Moore, and L.~G. Yaffe, {\it {Photon and Gluon Emission in
  Relativistic Plasmas}},  {\em JHEP} {\bf 06} (2002) 030,
  [\href{http://xxx.lanl.gov/abs/hep-ph/0204343}{{\tt hep-ph/0204343}}].

\bibitem{MehtarTani:2010ma}
Y.~Mehtar-Tani, C.~A. Salgado, and K.~Tywoniuk, {\it {Antiangular Ordering of
  Gluon Radiation in QCD Media}},  {\em Phys. Rev. Lett.} {\bf 106} (2011)
  122002, [\href{http://xxx.lanl.gov/abs/1009.2965}{{\tt arXiv:1009.2965}}].

\bibitem{MehtarTani:2011tz}
Y.~Mehtar-Tani, C.~A. Salgado, and K.~Tywoniuk, {\it {Jets in QCD Media: from
  Color Coherence to Decoherence}},  {\em Phys. Lett.} {\bf B707} (2012)
  156--159, [\href{http://xxx.lanl.gov/abs/1102.4317}{{\tt arXiv:1102.4317}}].

\bibitem{CasalderreySolana:2011rz}
J.~Casalderrey-Solana and E.~Iancu, {\it {Interference Effects in
  Medium-Induced Gluon Radiation}},  {\em JHEP} {\bf 08} (2011) 015,
  [\href{http://xxx.lanl.gov/abs/1105.1760}{{\tt arXiv:1105.1760}}].

\bibitem{MehtarTani:2011gf}
Y.~Mehtar-Tani, C.~A. Salgado, and K.~Tywoniuk, {\it {The Radiation Pattern of
  a QCD Antenna in a Dilute Medium}},  {\em JHEP} {\bf 04} (2012) 064,
  [\href{http://xxx.lanl.gov/abs/1112.5031}{{\tt arXiv:1112.5031}}].

\bibitem{MehtarTani:2012cy}
Y.~Mehtar-Tani, C.~A. Salgado, and K.~Tywoniuk, {\it {The Radiation pattern of
  a QCD antenna in a dense medium}},
  \href{http://xxx.lanl.gov/abs/1205.5739}{{\tt arXiv:1205.5739}}.

\bibitem{Bassetto:1982ma}
A.~Bassetto, M.~Ciafaloni, G.~Marchesini, and A.~H. Mueller, {\it {Jet
  multiplicity and soft gluon factorization}},  {\em Nucl. Phys.} {\bf B207}
  (1982) 189.

\bibitem{Mueller:1981ex}
A.~H. Mueller, {\it {On the Multiplicity of Hadrons in QCD Jets}},  {\em
  Phys.Lett.} {\bf B104} (1981) 161--164.

\bibitem{Ermolaev:1981cm}
B.~Ermolaev and V.~S. Fadin, {\it {Log - Log Asymptotic Form of Exclusive
  Cross-Sections in Quantum Chromodynamics}},  {\em JETP Lett.} {\bf 33} (1981)
  269--272.

\bibitem{Dokshitzer:1991wu}
Y.~L. Dokshitzer, V.~A. Khoze, A.~H. Mueller, and S.~I. Troian, {\it {Basics of
  Perturbative QCD}}, . Gif-sur-Yvette, France: Ed. Frontieres (1991) 274 p.
  (Basics of).

\bibitem{Baier:2000sb}
R.~Baier, A.~H. Mueller, D.~Schiff, and D.~Son, {\it {'Bottom up'
  thermalization in heavy ion collisions}},  {\em Phys.Lett.} {\bf B502} (2001)
  51--58, [\href{http://xxx.lanl.gov/abs/hep-ph/0009237}{{\tt
  hep-ph/0009237}}].

\bibitem{Baier:2001yt}
R.~Baier, Y.~L. Dokshitzer, A.~H. Mueller, and D.~Schiff, {\it {Quenching of
  hadron spectra in media}},  {\em JHEP} {\bf 0109} (2001) 033,
  [\href{http://xxx.lanl.gov/abs/hep-ph/0106347}{{\tt hep-ph/0106347}}].

\bibitem{Jeon:2003gi}
S.~Jeon and G.~D. Moore, {\it {Energy loss of leading partons in a thermal QCD
  medium}},  {\em Phys.Rev.} {\bf C71} (2005) 034901,
  [\href{http://xxx.lanl.gov/abs/hep-ph/0309332}{{\tt hep-ph/0309332}}].

\bibitem{Armesto:2009fj}
N.~Armesto, L.~Cunqueiro, and C.~A. Salgado, {\it {Q-PYTHIA: A Medium-modified
  implementation of final state radiation}},  {\em Eur.Phys.J.} {\bf C63}
  (2009) 679--690, [\href{http://xxx.lanl.gov/abs/0907.1014}{{\tt
  arXiv:0907.1014}}].

\bibitem{Schenke:2009gb}
B.~Schenke, C.~Gale, and S.~Jeon, {\it {Martini: an Event Generator for
  Relativistic Heavy-Ion Collisions}},  {\em Phys. Rev.} {\bf C80} (2009)
  054913, [\href{http://xxx.lanl.gov/abs/0909.2037}{{\tt arXiv:0909.2037}}].

\bibitem{Zapp:2011ya}
K.~C. Zapp, J.~Stachel, and U.~A. Wiedemann, {\it {A local Monte Carlo
  framework for coherent QCD parton energy loss}},  {\em JHEP} {\bf 1107}
  (2011) 118, [\href{http://xxx.lanl.gov/abs/1103.6252}{{\tt
  arXiv:1103.6252}}].

\bibitem{MehtarTani:2006xq}
Y.~Mehtar-Tani, {\it {Relating the Description of Gluon Production in Pa
  Collisions and Parton Energy Loss in Aa Collisions}},  {\em Phys. Rev.} {\bf
  C75} (2007) 034908, [\href{http://xxx.lanl.gov/abs/hep-ph/0606236}{{\tt
  hep-ph/0606236}}].

\bibitem{Altarelli:1977zs}
G.~Altarelli and G.~Parisi, {\it {Asymptotic Freedom in Parton Language}},
  {\em Nucl.Phys.} {\bf B126} (1977) 298.

\bibitem{Blaizot:2004wu}
J.~P. Blaizot, F.~Gelis, and R.~Venugopalan, {\it {High-energy pA collisions in
  the color glass condensate approach. 1. Gluon production and the Cronin
  effect}},  {\em Nucl.Phys.} {\bf A743} (2004) 13--56,
  [\href{http://xxx.lanl.gov/abs/hep-ph/0402256}{{\tt hep-ph/0402256}}].

\end{thebibliography}
%\bibliographystyle{jhep}

\end{document}